\documentclass[twocolumn, twocolappendix, times]{aastex63} 
\usepackage{amsmath}
\usepackage{soul}
\usepackage{natbib}
\usepackage{mathtools}
\graphicspath{{./HyperExs/}{./}{./Pictures/}{./Pictures/ChannelMaps/}}
\allowdisplaybreaks 

\shorttitle{An ALMA Survey of Chemistry in Disks around M4-M5 Stars}
\shortauthors{J. Pegues et al.}

\begin{document}

\title{An ALMA Survey of Chemistry in Disks around M4-M5 Stars}


\author{Jamila Pegues}
\affiliation{
Center for Astrophysics $\mid$ Harvard \& Smithsonian,  Cambridge, MA 02138, USA}

\author{Karin I. \"Oberg}
\affiliation{
Center for Astrophysics $\mid$ Harvard \& Smithsonian,  Cambridge, MA 02138, USA}

\author{Jennifer B. Bergner}
\altaffiliation{NASA Hubble Fellowship Program Sagan Fellow}
\affiliation{
Department of Geophysical Sciences, University of Chicago, Chicago, IL 60637, USA}
\affiliation{
Center for Astrophysics $\mid$ Harvard \& Smithsonian,  Cambridge, MA 02138, USA}

\author{Jane Huang}
\altaffiliation{NASA Hubble Fellowship Program Sagan Fellow}
\affiliation{
Department of Astronomy, University of Michigan, 323 West Hall, 1085 S. University Avenue, Ann Arbor, MI 48109, USA}
\affiliation{
Center for Astrophysics $\mid$ Harvard \& Smithsonian,  Cambridge, MA 02138, USA}

\author{Ilaria Pascucci}
\affiliation{
Lunar and Planetary Laboratory, The University of Arizona, Tucson, AZ 85721, USA}
\affiliation{
Earths in Other Solar Systems Team, NASA Nexus for Exoplanet System Science}

\author{Richard Teague}
\affiliation{
Center for Astrophysics $\mid$ Harvard \& Smithsonian,  Cambridge, MA 02138, USA}

\author{Sean M. Andrews}
\affiliation{
Center for Astrophysics $\mid$ Harvard \& Smithsonian,  Cambridge, MA 02138, USA}

\author{Edwin A. Bergin}
\affiliation{
Department of Astronomy, University of Michigan, 1085 S. University Ave, Ann Arbor, MI 48109}

\author{L. Ilsedore Cleeves}
\affiliation{
Astronomy Department, University of Virginia, Charlottesville, VA 22904, USA}

\author{Viviana V. Guzm\'an}
\affiliation{Instituto de Astrof{\'i}sica, Ponticia Universidad Cat{\'o}lica de Chile, Av.~Vicu{\~n}a Mackenna 4860, 7820436 Macul, Santiago, Chile}

\author{Feng Long}
\affiliation{
Center for Astrophysics $\mid$ Harvard \& Smithsonian,  Cambridge, MA 02138, USA}

\author{Chunhua Qi}
\affiliation{
Center for Astrophysics $\mid$ Harvard \& Smithsonian,  Cambridge, MA 02138, USA}

\author{David J. Wilner}
\affiliation{
Center for Astrophysics $\mid$ Harvard \& Smithsonian,  Cambridge, MA 02138, USA}

\begin{abstract}

M-stars are the most common hosts of planetary systems in the Galaxy.  Protoplanetary disks around M-stars thus offer a prime opportunity to study the chemistry of planet-forming environments.  We present an ALMA survey of molecular line emission toward a sample of five protoplanetary disks around M4-M5 stars (FP Tau, J0432+1827, J1100-7619, J1545-3417, and Sz 69).  These observations can resolve chemical structures down to tens of AU.  Molecular lines of $^{12}$CO, $^{13}$CO, C$^{18}$O, C$_2$H, and HCN are detected toward all five disks.  Lines of H$_2$CO and DCN are detected toward 2/5 and 1/5 disks, respectively.  For disks with resolved C$^{18}$O, C$_2$H, HCN, and H$_2$CO emission, we observe substructures similar to those previously found in disks around solar-type stars (e.g., rings, holes, and plateaus).  C$_2$H and HCN excitation conditions estimated interior to the pebble disk edge for the bright disk J1100-7619 are consistent with previous measurements around solar-type stars.  The correlation previously found between C$_2$H and HCN fluxes for solar-type disks extends to our M4-M5 disk sample, but the typical C$_2$H/HCN ratio is higher for the M4-M5 disk sample.  This latter finding is reminiscent of the hydrocarbon enhancements found by previous observational infrared surveys in the innermost ($<$10AU) regions of M-star disks, which is intriguing since our disk-averaged fluxes are heavily influenced by flux levels in the outermost disk, exterior to the pebble disk edge.  Overall, most of the observable chemistry at 10-100AU appears similar for solar-type and M4-M5 disks, but hydrocarbons may be more abundant around the cooler stars.
\end{abstract}


\section{Introduction}
\label{sec_introduction}

The exoplanetary family within the local Galaxy is dominated by M-stars.  M-stars are not only the most common stars in the local Galaxy, but are also the most common hosts of planetary systems~\citep[e.g.,][]{cite_henryetal2006, cite_dressingetal2015, cite_muldersetal2015, cite_henryetal2016}.  The study of protoplanetary disk chemistry is crucial for modeling and predicting the chemistry of comets, planetesimals, and planets around these common cool stars.

To date, disk chemistry around low-mass M-stars (stellar masses $<$0.5M$_\Sun$, spectral types typically from $\sim$M4-M9) has barely been explored beyond the inner $\sim$10 AU.  The majority of observational chemistry surveys of low-mass M-star disks have been of infrared molecular lines, which probe disk scales of $<$10 AU~\citep[e.g.,][]{cite_pascuccietal2009, cite_pontoppidanetal2010, cite_harveyetal2012, cite_pascuccietal2013, cite_bulgeretal2014, cite_hendleretal2017}.  At millimeter wavelengths, CO and dust observations exist toward samples of low-mass M-star disks~\citep[e.g.,][]{cite_riccietal2014, cite_ciezaetal2015, cite_pascuccietal2016, cite_vanderplasetal2016, cite_ansdelletal2017, cite_longetal2017, cite_andrewsetal2018}.  However, reported emission from other millimeter wavelength molecular lines is scarce: CN, SO, H$_2$CO, and CO fluxes have been observed toward one low-mass M-star disk in $\rho$ Ophiuchi~\citep{cite_reboussinetal2015}; CN fluxes have been observed toward low-mass M-star disks in Lupus~\citep{cite_vanterwisgaetal2019}; and bright C$_2$H emission has been resolved toward three disks with stellar masses $<$0.5M$_\Sun$ in Lupus~\citep{cite_miotelloetal2019}.

The existing molecular line observations provide important insights into the chemistry of low-mass M-star disks.  Detection rates for infrared lines of small molecular species (including H$_2$O, C$_2$H$_2$, HCN, and CO$_2$) in the inner disk regions generally increase with decreasing spectral type~\citep[from A-stars down to M-stars; e.g.,][]{cite_pontoppidanetal2010}.  Disks around low-mass M-stars and brown dwarfs (spectral types $\leq$M6) also have higher infrared C$_2$H$_2$/HCN flux and column density ratios and higher HNC/H$_2$O flux ratios relative to disks around solar-type stars~\citep{cite_pascuccietal2009, cite_pascuccietal2013}.  These infrared studies have attributed these enhancements to higher C/O ratios in the inner disks around low-mass stars.

Theoretically, models have predicted that the differences in stellar properties between low-mass and solar-type stars affect some aspects of disk chemistry, while others appear insensitive to the details of the stellar radiation field.  Models comparing disk chemistry within $\leq$10AU around an M-dwarf and T Tauri star have predicted that M-dwarf disks host more carbon-rich atmospheres, with relatively high abundances of small organic molecules like C$_2$H$_2$ and HCN~\citep{cite_walshetal2015}.  A study comparing a grid of thermochemical brown dwarf disk models and a T Tauri disk model predicted that brown dwarf disks and T Tauri disks host similar physical and chemical processes overall, but found model evidence in support of the enhanced hydrocarbon content suggested by previous infrared surveys~\citep{cite_greenwoodetal2017}.  While these theoretical efforts provide some guidance, we emphasize the lack of predictions for molecular column densities and abundances beyond 10AU in disks around low-mass M-stars.

In this study, we present an Atacama Large Millimeter/submillimeter Array (ALMA) exploratory survey of molecules toward a sample of five disks around M4-M5 stars, and we draw preliminary conclusions on chemistry across the disks.  In Section~\ref{sec_survey}, we describe the disks and molecular lines in the survey sample, the ALMA observations, and the data reduction process.  In Section~\ref{sec_analysis}, we discuss the tools used to analyze the image products, to characterize the emission, and to measure molecular column densities, excitation temperatures, and optical depths.  In Section~\ref{sec_results}, we present the detections, emission morphologies, relative fluxes and flux correlations, and estimates of molecular column densities, excitation temperatures, and optical depths.  In Section~\ref{sec_discussion}, we discuss the results, and we compare M4-M5 disk chemistry from our sample with solar-type counterparts from previous molecular surveys~\citep{cite_huangetal2017, cite_bergneretal2019, cite_bergneretal2020, cite_peguesetal2020}.  In Section~\ref{sec_summary}, we summarize the key findings of our survey.

\section{Observations}
\label{sec_survey}

%
\begin{deluxetable*}{lccccccccc}
\tablecaption{Stellar and Disk Characteristics of the Sample. \label{table_char}}
\tablehead{
Disk                                   & Spectral  & R.A.$^{[0]}$ & Decl.$^{[0]}$  & Region       & Distance$^{[0]}$  & $t_*$  & $L_*$ & $M_*$             & $T_{\mathrm{eff}}$ \\
    & Type      & {(}J2000{)}   & {(}J2000{)}   &   &  {(}pc{)} &  {(}Myr{)}    & ($L_\Sun$)    & ($M_\Sun$)    &  (K) }
 \colnumbers \startdata
\hline
FP Tau                         & M4$^{[1]}$    & 04:14:47.31 & 26:46:26.06  & Taurus       & 128.5 (140$^{[2]}$) & 1.1$^{[2]}$       & 0.32$^{[2]}$  & 0.23$^{[2]}$ & 3270$^{[2]}$ \\
J04322210+1827426$^*$ & M4.75$^{[3]}$ & 04:32:22.12 & 18:27:42.36  & Taurus       & 141.9 (140$^{[3]}$) & $\sim$1$^{[3]}$   & 0.11$^{[3]}$  & 0.14$^{[3]}$ & 3027$^{[3]}$ \\
J11004022-7619280$^*$ & M4$^{[4]}$    & 11:00:40.14 & -76:19:28.00 & Chamaeleon I & 191.5 (160$^{[4]}$) & $\sim$2-3$^{[4]}$ & 0.10$^{[4]}$  & 0.23$^{[4]}$ & 3270$^{[4]}$ \\
J15450887-3417333$^*$ & M5.5$^{[5]}$  & 15:45:08.86 & -34:17:33.80 & Lupus        & 155.0 (150$^{[5]}$) & $\sim$3$^{[5]}$   & 0.058$^{[5]}$ & 0.14$^{[5]}$ & 3060$^{[5]}$ \\
Sz 69                          & M4.5$^{[5]}$  & 15:45:17.39 & -34:18:28.64 & Lupus        & 154.5 (150$^{[5]}$) & $\sim$3$^{[5]}$   & 0.088$^{[5]}$ & 0.20$^{[5]}$ & 3197$^{[5]}$
\enddata
\tablecomments{Right ascension (R.A.) and declination (decl.) coordinates and distances are from \textit{Gaia}~\citep[e.g.,][]{cite_gaia2016, cite_gaia2018b}.  The stellar ages ($t_*$), stellar luminosities ($L_*$), stellar masses ($M_*$), and stellar effective temperatures ($T_\mathrm{eff}$) were taken from the literature, where they were derived from continuum photometry, spectral energy distribution (SED) fits, scaling relations, and/or stellar evolutionary models.  The distances in parentheses are the distances assumed in the literature for these disks.  We use the values in parentheses throughout this paper to be consistent with the derived stellar characteristics.  $*$: J04322210+1827426, J11004022-7619280, and J15450887-3417333 are referred to as J0432+1827, J1100-7619, and J1545-3417 in subsequent figures, tables, and text.  \textit{References: [0]~\cite{cite_gaia2016,cite_gaia2018b}, [1]~\cite{cite_luhmanetal2010}, [2]~\cite{cite_andrewsetal2013}, [3]~\cite{cite_wardduongetal2018}, [4]~\cite{cite_manaraetal2017}, [5]~\cite{cite_alcalaetal2017}.}}
\end{deluxetable*}
%

%
\begin{deluxetable*}{lcccccc}
\tablecaption{Molecular Lines in the Sample. \label{table_mol}}
\tablehead{
Molecule  & Transition              & Frequency     & $E_\mathrm{u}$     & $S\mu^2$ & $S\mu^2/S_\mathrm{m}\mu^2$ & $R_\mathrm{i}$   \\
          &                         & {(}GHz{)}  & {(}K{)}      & {(}Debye$^2${)}  & &
}
 \colnumbers \startdata
\hline
$^{12}$CO & J=2-1                   & 230.53800 & 16.596 & 0.024227        & ---              & ---          \\
$^{13}$CO & J=2-1                   & 220.39868 & 15.866 & 0.048753        & ---              & ---          \\
C$^{18}$O & J=2-1                   & 219.56035 & 15.806 & 0.024401        & ---              & ---          \\
C$_2$H    & N=3-2, J=7/2-5/2, F=4-3 & 262.00426 & 25.149 & 2.2809          & 1.0           & 0.572   \\
          & N=3-2, J=7/2-5/2, F=3-2 & 262.00648 & 25.148 & 1.7065          & 0.74817       & 0.428   \\
          & N=3-2, J=5/2-3/2, F=3-2 & 262.06499 & 25.159 & 1.6290          & 0.71419       & 0.605   \\
          & N=3-2, J=5/2-3/2, F=2-1 & 262.06747 & 25.160 & 1.0644          & 0.46666       & 0.395   \\
DCN       & J=3-2                   & 217.23854 & 20.852 & 80.507          & ---              & ---           \\
HCN       & J=3-2, F=4-3            & 265.88650 & 25.521 & 34.369          & 1.0           & 0.429   \\
          & J=3-2, F=3-3            & 265.88489 & 25.521 & 2.9702          & 0.086418      & 0.0370  \\
          & J=3-2, F=3-2            & 265.88643 & 25.521 & 23.761          & 0.69135       & 0.296   \\
          & J=3-2, F=2-3            & 265.88698 & 25.521 & 0.084870        & 0.0024693     & 0.00111 \\
          & J=3-2, F=2-2            & 265.88852 & 25.521 & 2.9708          & 0.086436      & 0.0370  \\
          & J=3-2, F=2-1            & 265.88619 & 25.521 & 16.039          & 0.46666       & 0.200   \\
H$_2$CO   & 3$_{03}$-2$_{02}$       & 218.22219 & 20.956 & 16.308          & ---              & ---          
\enddata
\tablecomments{All frequencies, upper energy levels ($E_\mathrm{u}$), and line intensities ($S\mu^2$) were obtained directly from the Cologne Database for Molecular Spectroscopy~\citep[CDMS;][]{cite_cdms2016}, with the exception of the HCN lines, for which these values were obtained from CDMS via Splatalogue~\citep{cite_splatalogue2016}.  The line intensities of each hyperfine transition relative to the main hyperfine transition ($S\mu^2$/$S_\mathrm{m}\mu^2$) are given in column 6.  The line intensities of each hyperfine transition relative to all same-level J transitions, denoted as $R_\mathrm{i}$, are given in column 7.}
\end{deluxetable*}
%

%
\begin{deluxetable*}{lcccccccccc}
\tablecaption{ALMA Project Code 2017.1.01107.S. \label{table_obs}}
\tablehead{
Observed Disks   & Target &  Date  & Total Time    & \# of & Baseline & Ang.       & Max Ang.       & Bandpass & Flux & Phase \\
    & Freq. &   & per Source & Ant.$^*$ & Range & Res.$^*$ & Scale$^*$ & Calibrator & Calibrator & Calibrator \\
  & (GHz)          &          &  (min)         &        &    (m)        &  (")  &   (")     &           & & 
}
\colnumbers \startdata
\hline
FP Tau, J0432+1827 & 230.5                          & 12/31/2017 & 25.2, 22.7                      & 46                 & 15-2517   & 0.13              & 2.49                           & J0510+1800    & J0510+1800      & J0426+2327       \\
                     &                                & 1/22/2018  & 25.2, 22.7                      & 44                 & 15-1398   & 0.25              & 3.74                           & J0510+1800    & J0510+1800      & J0426+2327       \\
                     &                                & 1/23/2018  & 25.2, 22.7                      & 43                 & 15-1398   & 0.25              & 3.68                           & J0510+1800    & J0510+1800      & J0426+2327       \\
 & 262.1                          & 9/14/2018  & 18.7, 16.6                      & 44                 & 15-1261   & 0.24              & 3.74                           & J0510+1800    & J0510+1800      & J0426+2327       \\
                     &                                & 9/15/2018  & 18.7, 16.6                      & 44                 & 15-1261   & 0.24              & 3.46                           & J0510+1800    & J0510+1800      & J0426+2327       \\
                     \hline
J1100-7619         & 230.5                          & 1/24/2018  & 36.3                            & 44                 & 15-1398   & 0.25              & 3.79                           & J1427-4206    & J1427-4206      & J1058-8003       \\
                     &                                & 9/15/2018  & 36.3                            & 42                 & 15-1261   & 0.27              & 3.94                           & J1037-2934    & J1037-2934      & J1058-8003       \\
                     &                                & 9/16/2018  & 36.3                            & 43                 & 15-1261   & 0.26              & 3.77                           & J0635-7516    & J0635-7516      & J1058-8003       \\
                     &                                & 9/18/2018  & 36.3                            & 44                 & 15-1398   & 0.25              & 3.56                           & J0635-7516    & J0635-7516      & J1058-8003       \\
         & 262.1                          & 12/26/2017 & 39.8                            & 46                 & 15-2517   & 0.12              & 2.35                           & J1427-4206    & J1427-4206      & J1058-8003       \\
\hline
J1545-3417, Sz 69  & 230.5                          & 1/17/2018  & 19.7, 19.7                      & 44                 & 15-1398   & 0.25              & 3.77                           & J1517-2422    & J1517-2422      & J1610-3958       \\
                     &                                & 1/18/2018  & 19.7, 19.7                      & 45                 & 15-1398   & 0.25              & 3.77                           & J1517-2422    & J1517-2422      & J1610-3958       \\
                     &                                & 1/18/2018  & 19.7, 19.7                      & 45                 & 15-1398   & 0.25              & 3.77                           & J1517-2422    & J1517-2422      & J1610-3958       \\
  & 262.1                          & 1/22/2018  & 17.1, 17.1                      & 45                 & 15-1398   & 0.22              & 3.29                           & J1517-2422    & J1517-2422      & J1534-3526       \\
                     &                                & 3/10/2018  & 17.1, 17.1                      & 41                 & 15-1241   & 0.29              & 4.23                           & J1517-2422    & J1517-2422      & J1610-3958      
\enddata
\tablecomments{Columns 5, 7, and 8 are abbreviations of \textit{Number of Antennas}, \textit{Angular Resolution}, and \textit{Maximum Angular Scale}, respectively.}
\end{deluxetable*}
%

\begin{deluxetable*}{lcccccc}
\tablecaption{Dust Continuum Emission. \label{table_contflux}}
\tablehead{
Disk      & $\lambda$     & 90\% Em. Radius    & Total Em.       & Peak Em.       & rms      & Beam Size             \\
          & (mm) & {(}AU{)}           & {(}mJy{)} & {(}mJy beam$^{-1}${)} & {(}mJy beam$^{-1}${)} &  (P.A.) 
}
\colnumbers \startdata
\hline
FP Tau     & 1.1 & 44                       & 11 $\pm$ 0.019  & 8.7            & 0.04           & 0.40" x 0.26" (-173.8$^\circ$) \\
     & 1.3 & 36                       & 8.4 $\pm$ 0.015 & 6.4            & 0.03           & 0.29" x 0.23" (-8.4$^\circ$)   \\ \hline
J0432+1827 & 1.1 & 51                       & 29 $\pm$ 0.077  & 12             & 0.09           & 0.35" x 0.25" (14.6$^\circ$)   \\ 
 & 1.3 & 42                       & 25 $\pm$ 0.051  & 8.6            & 0.05           & 0.24" x 0.20" (-16.1$^\circ$)  \\ \hline
J1100-7619 & 1.1 & 46                       & 35 $\pm$ 0.077  & 5.8            & 0.04           & 0.19" x 0.14" (-11.4$^\circ$)  \\
 & 1.3 & 63                       & 25 $\pm$ 0.056  & 8.6            & 0.07           & 0.39" x 0.25" (-3.2$^\circ$)   \\ \hline
J1545-3417 & 1.1 & 42                       & 24 $\pm$ 0.28   & 18             & 0.04           & 0.29" x 0.24" (90.0$^\circ$)   \\
 & 1.3 & 41                       & 20 $\pm$ 0.16   & 15             & 0.04           & 0.27" x 0.24" (76.4$^\circ$)   \\ \hline
Sz 69      & 1.1 & 30                       & 9.4 $\pm$ 0.023 & 9.0            & 0.04           & 0.29" x 0.24" (-89.8$^\circ$)  \\
      & 1.3 & 30                       & 8.0 $\pm$ 0.015 & 7.2            & 0.03           & 0.27" x 0.24" (77.7$^\circ$)  
\enddata
\tablecomments{\textit{Em.} in columns 3, 4, and 5 is an abbreviation of \textit{Emission}.  The pebble disk size (column 3) is represented as the radius containing 90\% of the dust continuum emission.  Total and peak emission for the 1.1mm (262GHz) and 1.3mm (231GHz) dust continuum were measured within the bounds of the $^{12}$CO and HCN Keplerian masks, respectively (Appendix~\ref{sec_appendix_kep}).  Note the difference in unit between the two quantities.  The error in the total emission was estimated across 1000 Keplerian-masked random samples extracted away from the source center.  The rms was also estimated across 1000 random samples, extracted in $2"\times2"$ regions away from the source center.  Uncertainties do not include $\sim$15\% systematic flux calibration uncertainties.}
\end{deluxetable*}


\subsection{Disk Sample}

\begin{figure}
\centering
\resizebox{0.99\hsize}{!}{
    \includegraphics[trim=10pt 10pt 0pt 0pt, clip]{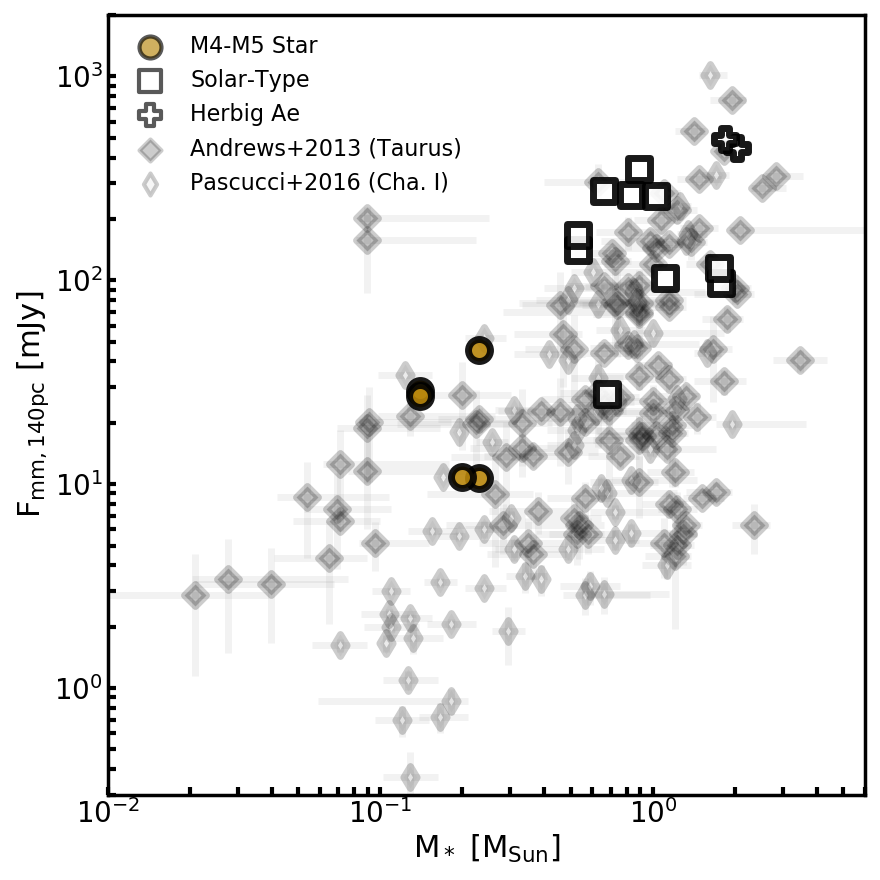}}
\caption{The dust continuum fluxes for the M4-M5 disk sample as a function of stellar mass, compared to disks compiled from the literature.  The dark gold circles are 1.1mm continuum fluxes for M4-M5 disk detections from this work.  The white points are ALMA-observed 1.1mm or 1.3mm continuum flux detections (whichever is available) for solar-type disks (marked with squares) and Herbig Ae disks (marked with crosses).  These disks were compiled from~\cite{cite_huangetal2017},~\cite{cite_bergneretal2019},~\cite{cite_bergneretal2020}, and~\cite{cite_peguesetal2020}.  This is the same compilation used later in Section~\ref{sec_results_relflux}.  The thick and thin diamonds are disks detected in 0.89mm continuum fluxes from~\cite{cite_andrewsetal2013} and~\cite{cite_pascuccietal2016}, which surveyed the Taurus and Chamaeleon I star-forming regions, respectively.  All fluxes were scaled from their original wavelengths to 1.1mm using $c=\lambda \nu$ and $F_\nu \propto \nu^\alpha$, where $c$ is the speed of light, $\lambda$ is the wavelength, $\nu$ is the frequency, and $F_\nu$ is the dust continuum flux.  We assume empirically that $\alpha=2.2$~\citep[review by][]{cite_andrewsetal2020}.  All fluxes have also been scaled to 140pc.
\label{fig_disksample}}
\end{figure}

Table~\ref{table_char} presents the stellar characteristics of the survey sample, which consists of five disks around M4-M5 stars.  The disks were chosen from multiple star-forming regions, avoiding bias toward any one molecular cloud.  Two of the disks are from the Taurus region, two from the Lupus region, and one from the Chamaeleon I region.  All disks are within 128-192pc~\citep{cite_gaia2016, cite_gaia2018b}.  Millimeter dust disk sizes are $\sim$30-63AU~\citep[where the size denotes the radius that contains $\sim$90\% of the emission, similar to the methods of][]{cite_ansdelletal2018}.  We refer to the millimeter dust disk as the ``pebble disk" throughout the remainder of the paper.
The host stars of the disks are $\sim$1-3Myr in age and range in spectral type, stellar luminosity, estimated stellar mass, and stellar effective temperature from M4-M5.5, $\sim$0.05-0.32L$_\Sun$, $\sim$0.14-0.23M$_\Sun$, and $\sim$3000-3300K, respectively.  All stellar characteristics were derived in the literature from continuum photometry, spectral energy distribution (SED) fits, scaling relations, and/or stellar evolutionary models\footnote{See~\cite{cite_peguesetal2021a} for new dynamical measurements of the stellar masses for FP Tau, J0432+1827, and J1100-7619.}~\citep{cite_luhmanetal2010, cite_andrewsetal2013, cite_wardduongetal2018, cite_manaraetal2017, cite_alcalaetal2017}.

Figure~\ref{fig_disksample} compares the dust continuum fluxes of our M4-M5 disk sample to dust continuum fluxes of solar-type and Herbig Ae disks~\citep[compiled from the chemistry surveys of][discussed further in Section~\ref{sec_results_relflux}]{cite_huangetal2017, cite_bergneretal2019, cite_bergneretal2020, cite_peguesetal2020}.  Figure~\ref{fig_disksample} also compares our sample to the dust continuum flux detections from~\cite{cite_andrewsetal2013} and from~\cite{cite_pascuccietal2016}, which were extracted from the Taurus and Chamaeleon I star-forming regions, respectively.  The M4-M5 disks in our sample allow us to probe a lower stellar mass regime than the previous chemistry surveys dominated by solar-type disks.  The majority of our M4-M5 disks fall within the brighter regime of the dust continuum flux vs. stellar mass relationship.  Our sample thus provides an initial rather than representative look into resolved disk chemistry around M4-M5 stars and in the low-mass M-star regime.

We note that the five disks in the M4-M5 disk sample were selected specifically for their bright $^{12}$CO (J=3--2) and $^{13}$CO (J=3--2) detections in existing ALMA disk surveys~\citep[e.g.,][van der Plas et al. in prep.]{cite_ansdelletal2016, cite_longetal2017}.   These selection criteria are similar to those used by early exploratory chemistry surveys of solar-type and Herbig Ae disks~\citep[e.g., the DISCS survey,][]{cite_obergetal2010, cite_obergetal2011}, and so our M4-M5 disk sample can reasonably be compared to disks around more massive stars from these previous surveys.  The solar-type and Herbig Ae disks form the core of our literature sample, which is overplotted in Figure~\ref{fig_disksample}.  We compare our M4-M5 disk sample to this literature sample in Section~\ref{sec_results_relflux}.


\subsection{Molecular Line Sample}

We survey the overall chemistry of these disks by targeting 1.1mm and 1.3mm lines of CO isotopologues and precursor organic molecules.  Our primary target molecular lines are $^{12}$CO (J=2--1), $^{13}$CO (J=2--1), C$^{18}$O (J=2--1), C$_2$H (N=3--2, J=7/2--5/2), DCN (J=3--2), HCN (J=3--2), and H$_2$CO 3$_{03}$--2$_{02}$\footnote{These lines are abbreviated as $^{12}$CO 2--1, $^{13}$CO 2--1, C$^{18}$O 2--1, C$_2$H 3--2, DCN 3--2, HCN 3--2, and H$_2$CO 3--2, respectively, in subsequent figures, tables, and text.}.  Together, these lines provide constraints on the C/N/O chemistry and organic inventories in the emitting layers of the disks~\citep[e.g.,][]{cite_berginetal2016, cite_cleevesetal2018, cite_miotelloetal2019}.  These lines have also been previously observed toward disks around solar-type stars, allowing direct comparison of the disk chemistry between these two types of stars.  We additionally use observations of the hyperfine structure of the C$_2$H and HCN lines to estimate C$_2$H and HCN column densities, excitation temperatures, and optical depths.  The molecular characteristics of the target lines are summarized in Table~\ref{table_mol}.

\subsection{Data and Data Reduction}

Table~\ref{table_obs} presents the observational characteristics of the sample.  All five disks were observed with ALMA during Project 2017.1.01107.S from December 2017 through September 2018.  Each individual observation (i.e., each execution) used a total of 41-46 antennas, providing a minimum baseline of 15m and maximum baselines from 1241-2517m.  Individual on-source observation times spanned 16-40min.  The angular resolution and maximum angular scales ranged from 0.12-0.29" and 2.35-4.23", respectively.

Initial calibration (including flux, phase, and bandpass calibration) was performed by ALMA/NAASC using standard procedures.  We then used the \texttt{Common Astronomy Software Applications} package (\texttt{CASA}) version 4.7.2 to self-calibrate each observation and image each molecular line.  We self-calibrated each observation using the line-free continuum combined from all spectral windows.  We uniformly used Briggs weighting and a robust value of 0.5.  We performed two rounds of phase calibration and one round of amplitude calibration per observation when able.  Solution intervals for the disks J1545-3417 and Sz 69 ranged from 10-100 seconds for each observation.  These intervals were chosen as the minimum values in this range that still maximized the number of solutions with signal-to-noise $\geq$2.  Due to low continuum fluxes, we used solution intervals from 70-210 seconds for the five observations toward FP Tau.  We were unable to self-calibrate on the continuum for J0432+1827 and J1100-7619.

We used \texttt{CASA}'s \texttt{uvcontsub} and \texttt{clean} functions to (1) subtract the continuum from each spectral window and (2) image each molecular line.  We created \texttt{clean}ing masks by hand to cover the molecular emission and little else in each channel.  For fainter lines, we recycled \texttt{clean}ing masks from brighter lines and \texttt{clean}ed down to 3$\sigma$ rather than the deeper 2$\sigma$, where $\sigma$ is the average rms across non-emission channels, to avoid creating image artifacts.  In order to balance sensitivity and resolution, we imaged each line using Briggs weighting and a robust value of 0.5.  The resulting synthesized beams range from 0.18" to 0.47" in size.  We imaged the dust continuum and molecular line emission toward the disk observed with the highest resolution (J1100-7619) with a pixel size of 0.02", while we imaged the emission toward the other four disks in the sample (FP Tau, J0432+1827, J1545-3417, and Sz 69) with a pixel size of 0.04".

We imaged the following line+disk combinations at a channel resolution of 0.2km/s: all line emission toward the bright disk J1100-7619; the $^{12}$CO 2--1, $^{13}$CO 2--1, and HCN 3--2 lines toward FP Tau and J0432+1827; and the $^{12}$CO 2--1 and $^{13}$CO 2--1 lines toward J1545-3417 and Sz 69.  All other line+disk combinations were imaged at a channel resolution of 0.4km/s to increase the signal-to-noise ratio.  We analyze velocity channels for each disk+line only within a specific velocity range.  We select these ranges based on visual inspection of the Keplerian-masked spectra (Section~\ref{sec_analysis_image}) to encompass the visible emission.  These ranges also include additional "buffer" channels to ensure that we have incorporated all emission.  Note that the velocity ranges for the C$_2$H 3--2 and HCN 3--2 lines are extended to include all hyperfine components.

\section{Analysis}
\label{sec_analysis}

\subsection{Image Analysis}
\label{sec_analysis_image}

We estimated the disk centers using two-dimensional Gaussian fits to the dust continuum emission.  We used Keplerian masks~\citep[e.g.,][]{cite_rosenfeldetal2013a, cite_yenetal2016, cite_salinasetal2017} to extract emission for each disk~\citep[software:][]{cite_kepmask} across all emission channels for each molecular line.  Keplerian masks decrease the amount of noise that is incorporated into the integrated emission.  Mask parameters used for this survey are given in Appendix~\ref{sec_appendix_kep}.  We used the Keplerian-masked channels to then generate spectra, velocity-integrated emission maps, radial profiles, and integrated fluxes.

We took the average rms across 1000 $2"\times2"$ random samples of non-emission channels to be the channel rms.  We estimated the noise for the integrated fluxes via bootstrapping over 1000 Keplerian-masked random samples of non-emission channels.  For the velocity-integrated emission maps, we used the median of 1000 random rms maps~\citep[$\sigma_\mathrm{map}$;][]{cite_bergneretal2018} as a representation of the noise.  For the radial profiles, we adapted the approach of~\cite{cite_bergneretal2018} and estimated the noise per ring as ($\sigma_\mathrm{map}/\sqrt{N}$).  $N$ is the number of independent measurements in the ring, assumed to be $N$=\textit{(\# pixels within each ring's width)/(\# pixels in the beam area)}.  Once the count per ring is less than the circumference of the beam, $N$ is fixed to be the value of $N$ using the circumference.

\subsection{Pixel-by-Pixel Hyperfine Fits}
\label{sec_analysis_hyperfine}

The C$_2$H (N=3--2, J=7/2--5/2), C$_2$H (N=3--2, J=5/2--3/2), and HCN 3--2 lines exhibit hyperfine structure, and so the emission spectra can be decomposed into individual hyperfine emission components (e.g., the 262.00426GHz and 262.00648GHz lines given in rows 4 and 5 of Table~\ref{table_mol} are hyperfine components for the C$_2$H (N=3--2, J=7/2--5/2) line).  Since the disk J1100-7619 is bright and well-resolved, we are able to fit models to the resolved hyperfine structure~\citep[e.g.,][]{cite_hilybrantetal2013, cite_estallelaetal2017, cite_gildas2018}.  We can then extract excitation temperatures, column densities, and optical depths from the parameters of the fitted models, \textit{through} each pixel in the image.  By fitting the spectrum per pixel rather than across a larger disk area, we reduce the line width, and therefore line blending, of the hyperfine components in the spectrum.  This in turn permits tighter constraints on the model fit.  We use a hyperfine pixel-by-pixel procedure based on the work of~\cite{cite_bergneretal2019}, which in turn was adapted from the fitting procedure of~\cite{cite_estallelaetal2017} and the calculations of~\cite{cite_mangumetal2015}.  See Appendix~\ref{sec_appendix_hyperfine} for the model equations, the fitting procedure, examples of model fits, and important discussion of intrinsic uncertainties.
%

\subsection{Column Density Estimates}
\label{sec_analysis_coldens}

For lines where we cannot use the hyperfine fitting method to measure excitation temperatures, optical depths, and column densities directly, we can still estimate the column densities over a given disk region, using assumptions of the excitation temperature and total optical depth (e.g., from pixel-by-pixel hyperfine fits to the brightest disk), as well as a measurement of the total emission across all hyperfine components within that given disk region.  This methodology is described in detail in Appendix~\ref{sec_appendix_Ntot}.  Uncertainties are discussed in Appendix~\ref{sec_appendix_hyperfine}.

\section{Results}
\label{sec_results}

\subsection{Detections}
\label{sec_results_detections}

\begin{deluxetable*}{lcccccccc}
\tablewidth{0.99\textwidth}
\tablecaption{Fluxes and Detection Upper Limits for Target Molecular Lines. \label{table_emflux}}
\tablehead{
Disk   & 90\% Em.   & Integrated      & Peak Flux           & Integrated    & Kep. Mask & Channel & Channel      & Beam \\
 & Radius & Flux & {(}mJy beam$^{-1}$ & Velocity Range & Extent & Width & rms & Size   \\
 & (AU) & {(}mJy km s$^{-1}${)} & $\times$ km s$^{-1}${)} & {(}km s$^{-1}${)}  & (") & {(}km s$^{-1}${)} & {(}mJy beam$^{-1}${)} & (P.A.)  
}
\colnumbers \startdata
\hline
\multicolumn{9}{c}{$^{12}$CO (J=2--1)}                                                                                      \\ \hline
FP Tau     & 55                & 781 $\pm$ 11   & 247 $\pm$ 4.3       & 2.0 - 14.6       & 1.24                & 0.20       & 3.4            & 0.30" x 0.23" (-6.3$^{\circ}$)   \\
J0432+1827 & 126               & 1847 $\pm$ 13  & 129 $\pm$ 2.9       & 0.6 - 10.6       & 2.16                & 0.20       & 3.3            & 0.24" x 0.20" (-14.8$^{\circ}$)  \\
J1100-7619 & 186               & 1306 $\pm$ 6.1 & 69 $\pm$ 1.4        & 2.1 - 7.3        & 2.04                & 0.20       & 2.5            & 0.40" x 0.25" (-1.1$^{\circ}$)   \\
J1545-3417 & 98                & 738 $\pm$ 10   & 141 $\pm$ 3.3       & -1.0 - 10.0      & 2.00                 & 0.20       & 3.7            & 0.27" x 0.24" (71.0$^{\circ}$)   \\
Sz 69      & 138               & 2001 $\pm$ 15  & 192 $\pm$ 3.0       & 1.2 - 9.4        & 2.68                & 0.20       & 3.8            & 0.27" x 0.24" (73.0$^{\circ}$)   \\
\hline
\multicolumn{9}{c}{$^{13}$CO (J=2--1)}                                                                                      \\ \hline
FP Tau     & 46                & 232 $\pm$ 8.3  & 100 $\pm$ 3.5       & 2.0 - 14.6       & 0.68                & 0.20       & 3.2            & 0.30" x 0.24" (-7.2$^{\circ}$)   \\
J0432+1827 & 81                & 537 $\pm$ 9.4  & 81 $\pm$ 2.9        & 0.6 - 10.6       & 1.32                & 0.20       & 3.0            & 0.25" x 0.21" (-13.4$^{\circ}$)  \\
J1100-7619 & 136               & 417 $\pm$ 6.0  & 38 $\pm$ 1.4        & 2.2 - 7.4        & 1.92                & 0.20       & 2.7            & 0.41" x 0.26" (-4.8$^{\circ}$)   \\
J1545-3417 & 80                & 218 $\pm$ 7.7  & 57 $\pm$ 3.1        & -1.0 - 10.0      & 1.32                & 0.20       & 3.3            & 0.28" x 0.25" (71.4$^{\circ}$)   \\
Sz 69      & 136               & 345 $\pm$ 13   & 60 $\pm$ 2.6        & 1.2 - 9.4        & 2.68                & 0.20       & 3.5            & 0.28" x 0.25" (72.7$^{\circ}$)   \\
\hline
\multicolumn{9}{c}{C$^{18}$O (J=2--1)}                                                                                      \\ \hline
FP Tau     & 44                & 91 $\pm$ 6.7   & 54 $\pm$ 3.3        & 1.9 - 14.7       & 0.56                & 0.40       & 2.1            & 0.30" x 0.24" (-7.1$^{\circ}$)   \\
J0432+1827 & 58                & 148 $\pm$ 9.0  & 36 $\pm$ 2.4        & 0.4 - 10.8       & 1.32                & 0.40       & 2.0            & 0.25" x 0.21" (-13.5$^{\circ}$)  \\
J1100-7619 & 116               & 71 $\pm$ 3.6   & 12 $\pm$ 1.2        & 2.2 - 7.4        & 1.36                & 0.20       & 2.1            & 0.42" x 0.27" (-2.3$^{\circ}$)   \\
J1545-3417 & 123               & 89 $\pm$ 9.5   & 21 $\pm$ 2.9        & -1.1 - 10.1      & 1.96                & 0.40       & 2.3            & 0.28" x 0.25" (69.1$^{\circ}$)   \\
Sz 69      & 37                & 38 $\pm$ 6.6   & 22 $\pm$ 2.5        & 1.4 - 9.4        & 0.96                & 0.40       & 2.2            & 0.29" x 0.25" (71.0$^{\circ}$)   \\
\hline
\multicolumn{9}{c}{C$_2$H (N=3--2, J=$\frac{7}{2}$--$\frac{5}{2}$)}                                                          \\ \hline
FP Tau     & 63                & 140 $\pm$ 18   & 81 $\pm$ 4.7        & 1.1 - 14.7       & 1.44                & 0.40       & 3.2            & 0.44" x 0.34" (-176.5$^{\circ}$) \\
J0432+1827 & 104               & 288 $\pm$ 19   & 39 $\pm$ 4.3        & -1.6 - 10.0      & 1.32                & 0.40       & 3.0            & 0.34" x 0.24" (14.4$^{\circ}$)   \\
J1100-7619 & 167               & 1180 $\pm$ 17  & 22 $\pm$ 1.9        & -0.4 - 7.4       & 1.48                & 0.20       & 2.9            & 0.18" x 0.14" (-9.5$^{\circ}$)   \\
J1545-3417 & 56                & 61 $\pm$ 12    & 47 $\pm$ 3.0        & -1.9 - 9.7       & 1.16                & 0.40       & 2.6            & 0.37" x 0.33" (72.4$^{\circ}$)   \\
Sz 69      & 39                & 28 $\pm$ 8.1   & 25 $\pm$ 3.7        & -2.6 - 9.0       & 0.40                 & 0.40       & 2.8            & 0.37" x 0.32" (72.9$^{\circ}$)   \\
\hline
\multicolumn{9}{c}{HCN (J=3--2)}                                                                                            \\ \hline
FP Tau     & 48                & 272 $\pm$ 17   & 164 $\pm$ 4.4       & 2.0 - 14.6       & 1.20                 & 0.20       & 4.2            & 0.38" x 0.25" (-173.8$^{\circ}$) \\
J0432+1827 & 85                & 600 $\pm$ 18   & 125 $\pm$ 2.9       & -0.2 - 11.4      & 1.36                & 0.20       & 3.8            & 0.34" x 0.24" (14.3$^{\circ}$)   \\
J1100-7619 & 137               & 1655 $\pm$ 24  & 67 $\pm$ 1.6        & -0.2 - 9.8       & 1.64                & 0.20       & 2.6            & 0.18" x 0.14" (-9.1$^{\circ}$)   \\
J1545-3417 & 88                & 120 $\pm$ 14   & 51 $\pm$ 2.6        & -1.1 - 10.1      & 1.36                & 0.40       & 2.4            & 0.37" x 0.33" (70.1$^{\circ}$)   \\
Sz 69      & 39                & 42 $\pm$ 8.6   & 40 $\pm$ 3.4        & -0.2 - 11.0      & 0.44                & 0.40       & 2.6            & 0.37" x 0.33" (71.0$^{\circ}$)   \\
\hline
\multicolumn{9}{c}{DCN (J=3--2)}                                                                                            \\ \hline
FP Tau     & ---                & $\lesssim$5.7  & $\lesssim$12       & 1.9 - 14.7       & 0.12                & 0.40       & 2.0            & 0.31" x 0.24" (-9.7$^{\circ}$)   \\
J0432+1827 & ---                & $\lesssim$18  & $\lesssim$9.1       & 0.4 - 10.8       & 0.48                & 0.40       & 2.0            & 0.26" x 0.21" (-16.5$^{\circ}$)  \\
J1100-7619 & 62                & 15 $\pm$ 2.7   & 8.3 $\pm$ 1.0       & 2.1 - 7.3        & 0.72                & 0.20       & 1.6            & 0.47" x 0.34" (-3.3$^{\circ}$)   \\
J1545-3417 & ---                & $\lesssim$14  & $\lesssim$8.7       & -1.1 - 10.1      & 0.36                & 0.40       & 2.2            & 0.29" x 0.26" (73.3$^{\circ}$)   \\
Sz 69      & ---                & $\lesssim$8.7  & $\lesssim$8.7       & 1.4 - 9.4        & 0.16                & 0.40       & 2.1            & 0.29" x 0.26" (75.0$^{\circ}$)   \\
\hline
\multicolumn{9}{c}{H$_2$CO 3$_{03}$--2$_{02}$}                                                                                \\ \hline
FP Tau     & ---                & $\lesssim$15   & $\lesssim$15        & 1.9 - 14.7       & 0.36                & 0.40       & 1.8            & 0.30" x 0.24" (-10.0$^{\circ}$)  \\
J0432+1827 & 136               & 80 $\pm$ 8.0   & 6.5 $\pm$ 2.1       & 0.4 - 10.8       & 1.48                & 0.40       & 1.7            & 0.26" x 0.21" (-17.7$^{\circ}$)  \\
J1100-7619 & 219               & 100 $\pm$ 3.7  & 7.7 $\pm$ 1.0       & 2.2 - 7.4        & 2.12                & 0.20       & 1.6            & 0.41" x 0.27" (-2.8$^{\circ}$)   \\
J1545-3417 & ---                & $<$14 & $<$7.2       & -1.1 - 10.1      & 0.40                 & 0.40       & 1.9            & 0.29" x 0.26" (73.3$^{\circ}$)   \\
Sz 69      & ---                & $\lesssim$11  & $\lesssim$6.9       & 1.4 - 9.4        & 0.24                & 0.40       & 1.9            & 0.29" x 0.26" (74.7$^{\circ}$)   \\
\enddata
\tablecomments{\textit{Em.} and \textit{Kep.} are abbreviations of \textit{Emission} and \textit{Keplerian}, respectively.  The emitting radii (column 2) are the radii containing 90\% of the molecular line emission and are presented only for detections.  The integrated and peak fluxes are measured within the Keplerian masks (column 6; Appendix~\ref{sec_appendix_kep}).  Note the difference in unit between the two quantities.  The errors for the integrated and peak fluxes and the channel rms were estimated via bootstrapping of 1000 random samples of non-emission channels.  Samples for the integrated and peak fluxes were extracted from within the Keplerian masks, while the channel rms was extracted from within $2"\times2"$ regions.  Uncertainties do not include $\sim$15\% systematic flux calibration uncertainties.  3$\sigma$ upper limits are given for tentative detections and nondetections and are marked with a $\lesssim$ and $<$, respectively.}
\end{deluxetable*}

We used the following criteria to determine whether or not a molecular line is detected toward a disk:

\begin{enumerate}
	\item Emission is $\geq 3\sigma$ in the velocity-integrated emission map
	\item Peak emission within the Keplerian masks is $\geq 3\sigma$ in at least 3 velocity channels
\end{enumerate}

Lines that satisfy both criteria are \textit{detected}, while lines that satisfy at least one criterion are \textit{tentatively detected}.  Lines that fail both criteria are \textit{nondetected}.

Based on these criteria, we detect $^{12}$CO (J=2--1), $^{13}$CO (J=2--1), C$^{18}$O (J=2--1), HCN (J=3--2), C$_2$H (N=3--2, F=7/2--5/2), and C$_2$H (N=3--2, F=5/2--3/2) toward all 5/5 M4-M5 disks.  We detect DCN (J=3--2) toward one disk (J1100-7619), and we detect H$_2$CO 3$_{03}$--2$_{02}$ toward two disks (J0432+1827 and J1100-7619).  For H$_2$CO 3--2 toward two other disks (FP Tau and Sz 69), and for DCN 3--2 toward four other disks (FP Tau, J0432+1827, J1545-3417, and Sz 69), the emission is tentatively detected and would be worth follow-up observations with deeper integration times.

Channel maps for the detections and tentative detections are presented in Appendix~\ref{sec_appendix_chan}.  Table~\ref{table_contflux} presents the emission measurements for the dust continuum, and Table~\ref{table_emflux} presents the flux measurements for the target molecular lines.  All detections, tentative detections, and nondetections are marked within Table~\ref{table_emflux}.

\subsection{Emission Morphologies}
\label{sec_results_morphologies}

\begin{figure*}
\centering
\resizebox{0.925\hsize}{!}{
    \includegraphics[trim=50pt 25pt 15pt 55pt, clip]{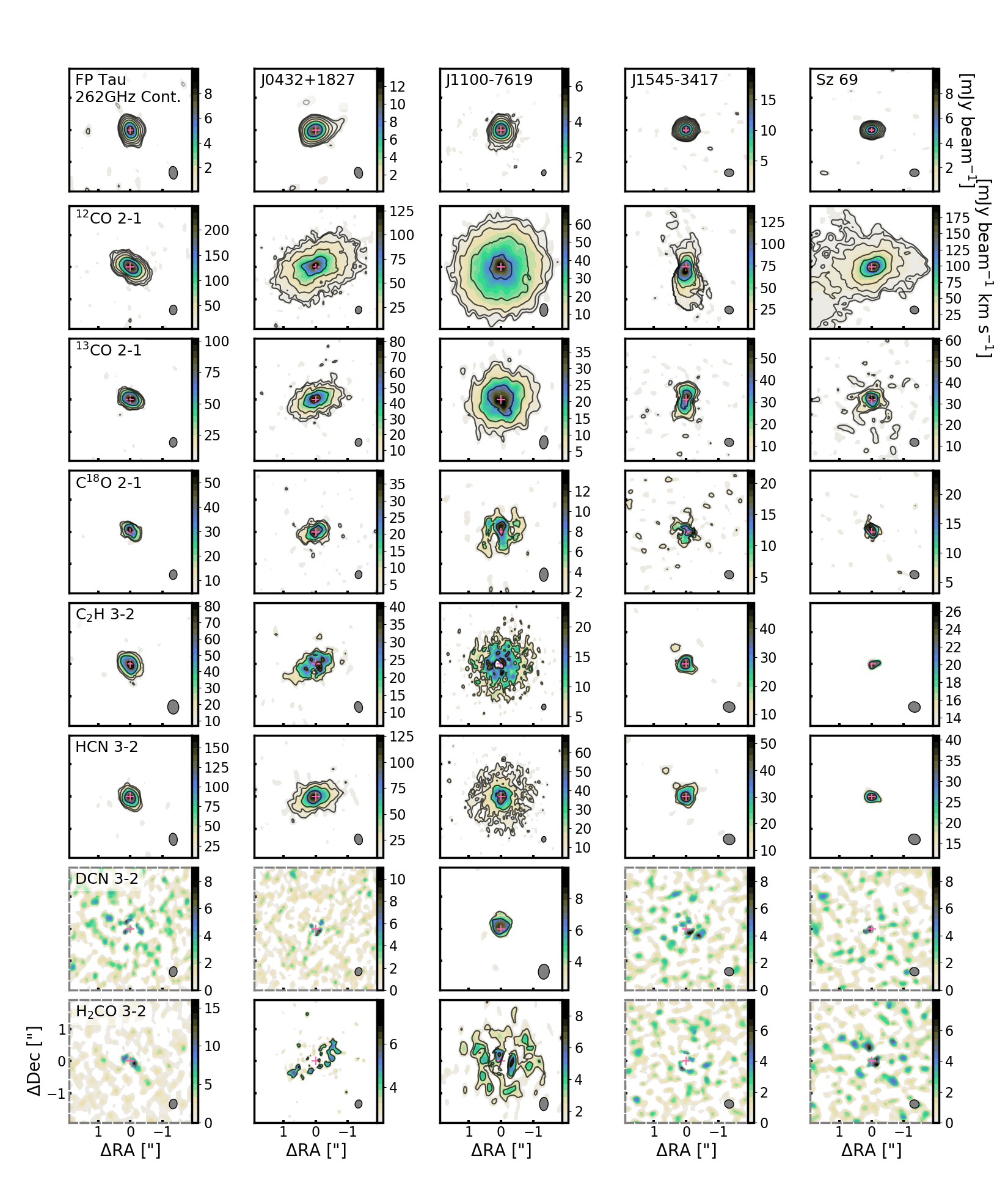}}
\caption{Dust continuum emission and velocity-integrated emission maps for the molecular lines.  Each column corresponds to a different disk.  The top row shows the 1.1mm (262GHz) dust continuum emission for each disk.  All subsequent rows show velocity-integrated molecular line emission.  Contours are shown at the [3$\sigma$, 5$\sigma$, 10$\sigma$, 20$\sigma$...] levels.  $\sigma$ is equal to the rms for the dust continuum emission (Table~\ref{table_contflux}) and to $\sigma_\mathrm{map}$ for the velocity-integrated emission (Section~\ref{sec_analysis_image}).  Disk centers are marked with $+$ signs.  Beams are drawn in the lower right corners.  Subplots for tentative/nondetections are outlined in dashed gray rather than black, and their colorbars start at 0.  Colorbars for detections start at 2$\sigma_\mathrm{map}$.
\label{fig_mom0}}
\end{figure*}

\begin{figure*}
\centering
\resizebox{0.85\hsize}{!}{
    \includegraphics[trim=10pt 35pt 40pt 45pt, clip]{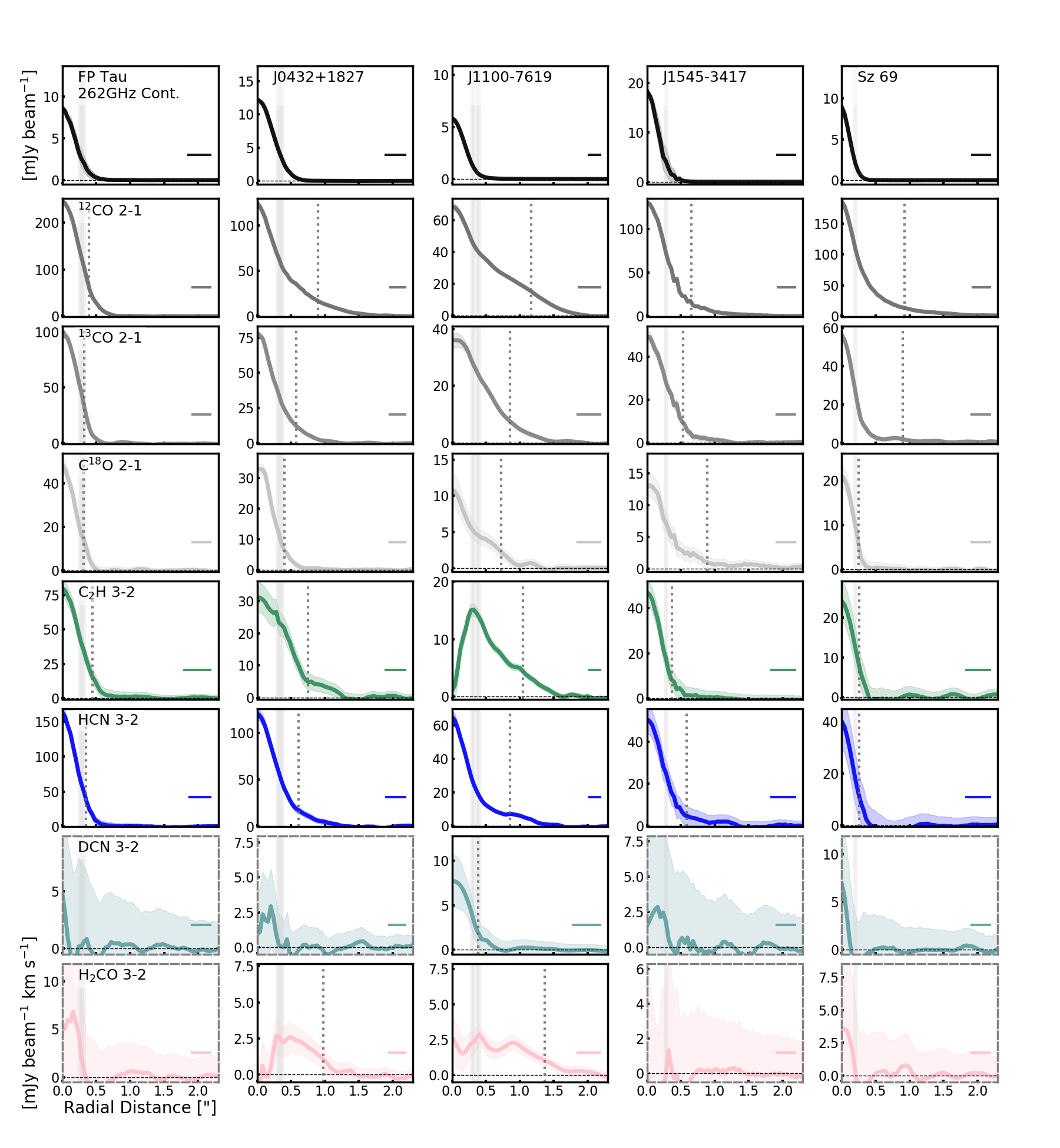}}
\caption{Radial profiles for the 1.1mm (262GHz) dust continuum and the molecular lines in the sample.  Each column corresponds to a different disk.  The top row displays the dust continuum radial profiles (in black).  The subsequent rows display the CO isotopologues (in grays), C$_2$H (green), HCN and DCN (blues), and H$_2$CO (pink) lines.  For the dust continuum radial profiles, the shaded regions are the standard deviation of the emission within each annulus.  For the molecular emission radial profiles, the shaded regions depict the 1$\sigma$ uncertainties of each annulus as described in Section~\ref{sec_analysis_image}.  The vertical light gray shaded regions show the 90\% emitting radii of the 1.1mm and 1.3mm (231GHz) dust continuum.  The vertical dark gray dashed lines mark the 90\% emitting radii for each detected molecular line. Beam sizes are represented by the horizontal bars in the lower right corners.  Subplots for tentative/nondetections are outlined in dashed gray rather than black.
\label{fig_prof}}
\end{figure*}

\begin{figure*}
\centering
\resizebox{0.925\hsize}{!}{
    \includegraphics[trim=15pt 15pt 55pt 55pt, clip]{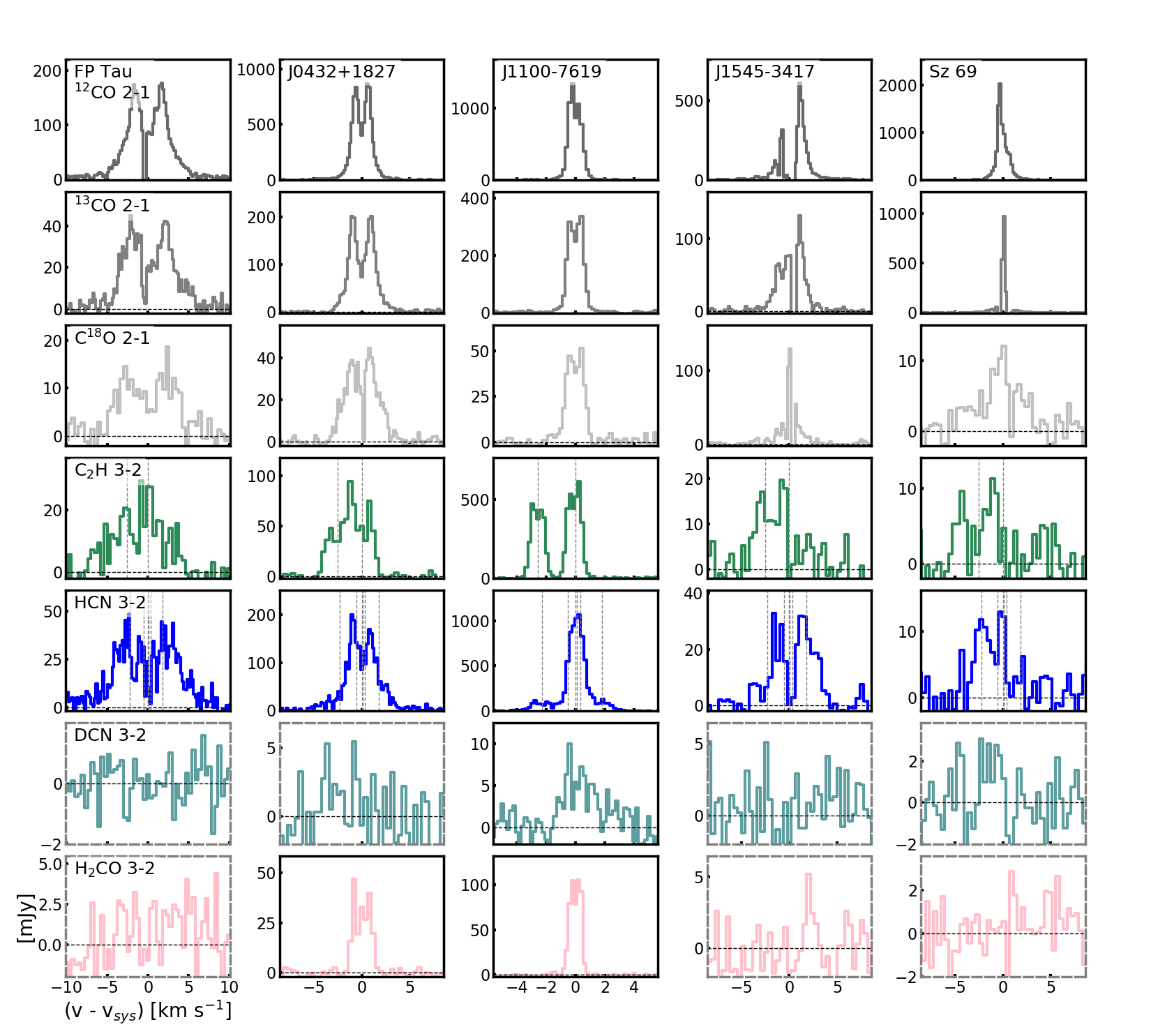}}
\caption{Spectra for the molecular lines in the sample.  Each column corresponds to a different disk.  The rows display the CO isotopologues (in grays), C$_2$H (green), HCN and DCN (blues), and H$_2$CO (pink) lines.  Hyperfine components for C$_2$H 3--2 and HCN 3--2 are marked with vertical dashed gray lines.  The assumed systemic velocities ($v_\mathrm{sys}$) are given in Appendix~\ref{sec_appendix_kep}.  The two hyperfine components for C$_2$H 3--2 toward J1100-7619, centered at $(v - v_\mathrm{sys})$ of $\sim$0km/s and $\sim$-2.5km/s, are clearly separable by eye.  The hyperfine components for HCN 3--2 toward J1100-7619, along with all hyperfine components toward all other disks, are blended.  Subplots for tentative/nondetections are outlined in dashed gray rather than black.
\label{fig_spec}}
\end{figure*}

Figure~\ref{fig_mom0} displays the millimeter dust continuum emission and velocity-integrated molecular line emission maps.  Figures~\ref{fig_prof} and~\ref{fig_spec} display the corresponding radial profiles and spectra, respectively.  All dust continua appear smooth at this resolution (i.e., there are no cavities in the millimeter dust continuum emission).  We use the 90\% emitting radii of the dust continuum to represent the pebble disk edges.  In terms of pebble disk size, J0432+1827 and J1100-7619 are the largest disks, while Sz 69 is the smallest disk.  The 90\% emitting radii for all CO isotopologues, C$_2$H 3--2, and HCN 3--2 emission is either comparable to or exceeds the pebble disk sizes, indicating that the gas of the disks is typically extended relative to the millimeter dust.

We use simple two-dimensional Gaussian fits to the dust continuum emission to estimate the center of each disk, and we qualitatively characterize peaks of the molecular line emission with respect to that disk center.  All five disks in our sample have centrally-peaked CO isotopologue emission distributions.  The $^{13}$CO and C$^{18}$O emission distributions extend beyond the pebble disk edge (i.e., beyond $\sim$30-60AU) for 4/5 and 3/5 disks, respectively.  The disks FP Tau, J1545-3417, and Sz 69 are significantly affected by cloud contamination, which is evident in the CO spectra and the CO channel maps (Appendix~\ref{sec_appendix_chan}).  The CO emission substructure seen beyond the pebble disk edges may be a byproduct of this contamination, but could also be a contribution from an extended disk structure or disk-envelope interaction.  The disk J1100-7619 shows asymmetries in the $^{12}$CO spectrum.  It is not clear what is causing these asymmetries, as the channel maps do not suggest cloud contamination.


2/5 disks have dips or holes in the C$_2$H 3--2 emission.  The inner edge of the hole toward J1100-7619 is aligned with the edge of the pebble disk, between 46-63AU based on the 1.1 and 1.3mm dust continuum.  This means that the majority of the C$_2$H emission is coming from beyond the pebble disk~\citep[e.g.,][]{cite_berginetal2016}.  The hole in C$_2$H emission toward the second disk, J0432+1827, is off-center by $\sim$0.1", appearing in Figure~\ref{fig_mom0} but not in Figure~\ref{fig_prof}.  J0432+1827 additionally shows an emission plateau beyond the pebble disk.  C$_2$H 3--2 emission toward the remaining three disks in our sample appears centrally peaked but is likely barely resolved.  We clearly distinguish each hyperfine emission component (rows 4 and 5 of Table~\ref{table_mol}) in the C$_2$H 3--2 spectrum toward J1100-7619, while the C$_2$H and HCN hyperfine components for all other disks are blended (Figure~\ref{fig_spec}).

All disks have centrally-peaked HCN 3--2 emission morphologies.  The two well-resolved disks, J0432+1827 and J1100-7619, have more compact HCN emission compared to the C$_2$H emission.  Both disks also show plateaus or shelves in the HCN emission beyond the pebble disk edge.  The remaining three disks may also show substructure exterior to the pebble disk at higher spatial resolution.  DCN 3--2 is only detected toward J1100-7619, where the emission is centrally-peaked and extends just beyond the edge of the pebble disk.


When detected, H$_2$CO 3--2 emission shows a hole or depression near the disk center.  J0432+1827 has a central hole in H$_2$CO emission and suggestive rings that extend past the dust continuum.  J1100-7619 has a hole in H$_2$CO emission that is off-center by $<$0.1", which appears in Figure~\ref{fig_mom0} but not Figure~\ref{fig_prof}, with two rings that peak at and beyond the pebble disk edge.


\subsection{Relative Line Fluxes toward Disks around M4-M5 Stars and Solar-Type Stars}
\label{sec_results_relflux}

\begin{figure}
\centering
\resizebox{0.99\hsize}{!}{
    \includegraphics[trim=10pt 10pt 0pt 0pt, clip]{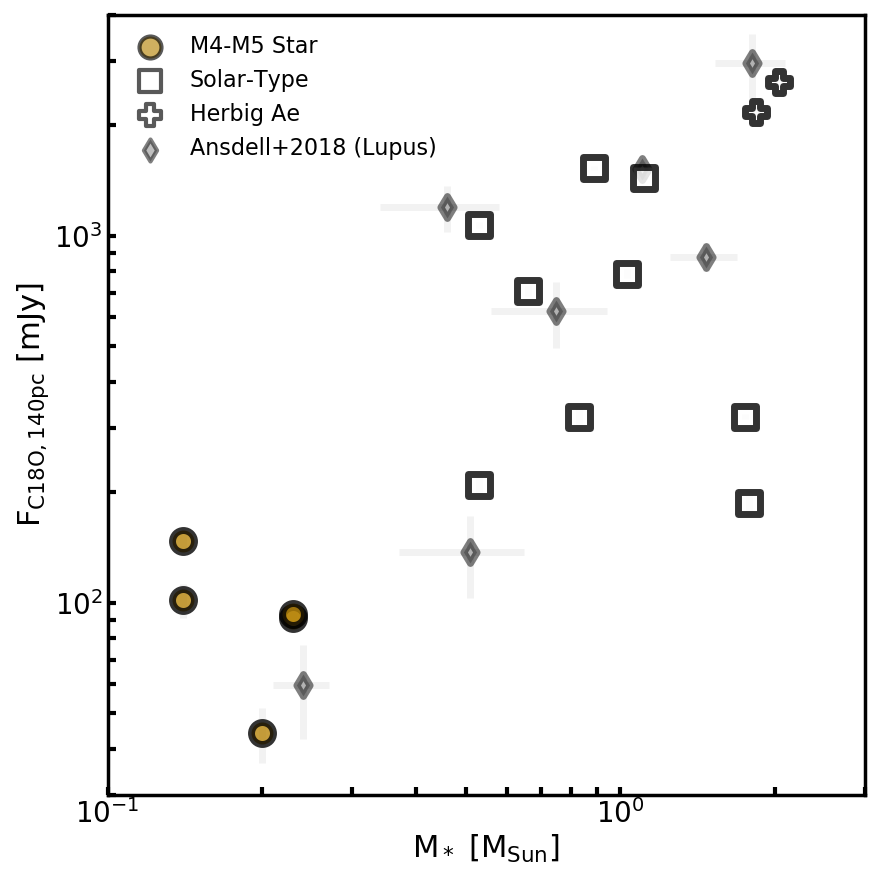}}
\caption{The C$^{18}$O fluxes for the M4-M5 disk sample as a function of stellar mass, compared to disks compiled from the literature.  The dark gold circles are C$^{18}$O fluxes for M4-M5 disk detections from this work.  The white points are ALMA-observed C$^{18}$O (J=2--1) fluxes for solar-type disks (marked with squares) and Herbig Ae disks (marked with crosses).  These disks were compiled from~\cite{cite_huangetal2017},~\cite{cite_bergneretal2019},~\cite{cite_bergneretal2020}, and~\cite{cite_peguesetal2020} and are described in Section~\ref{sec_results_relflux}.  The thin diamonds are disks detected in C$^{18}$O (J=2--1) from~\cite{cite_ansdelletal2018}, which surveyed the Lupus star-forming region.  All fluxes have been scaled to 140pc.
\label{fig_c18osample}}
\end{figure}

\begin{figure*}
\centering
\resizebox{0.925\hsize}{!}{
    \includegraphics[trim=10pt 15pt 10pt 5pt, clip]{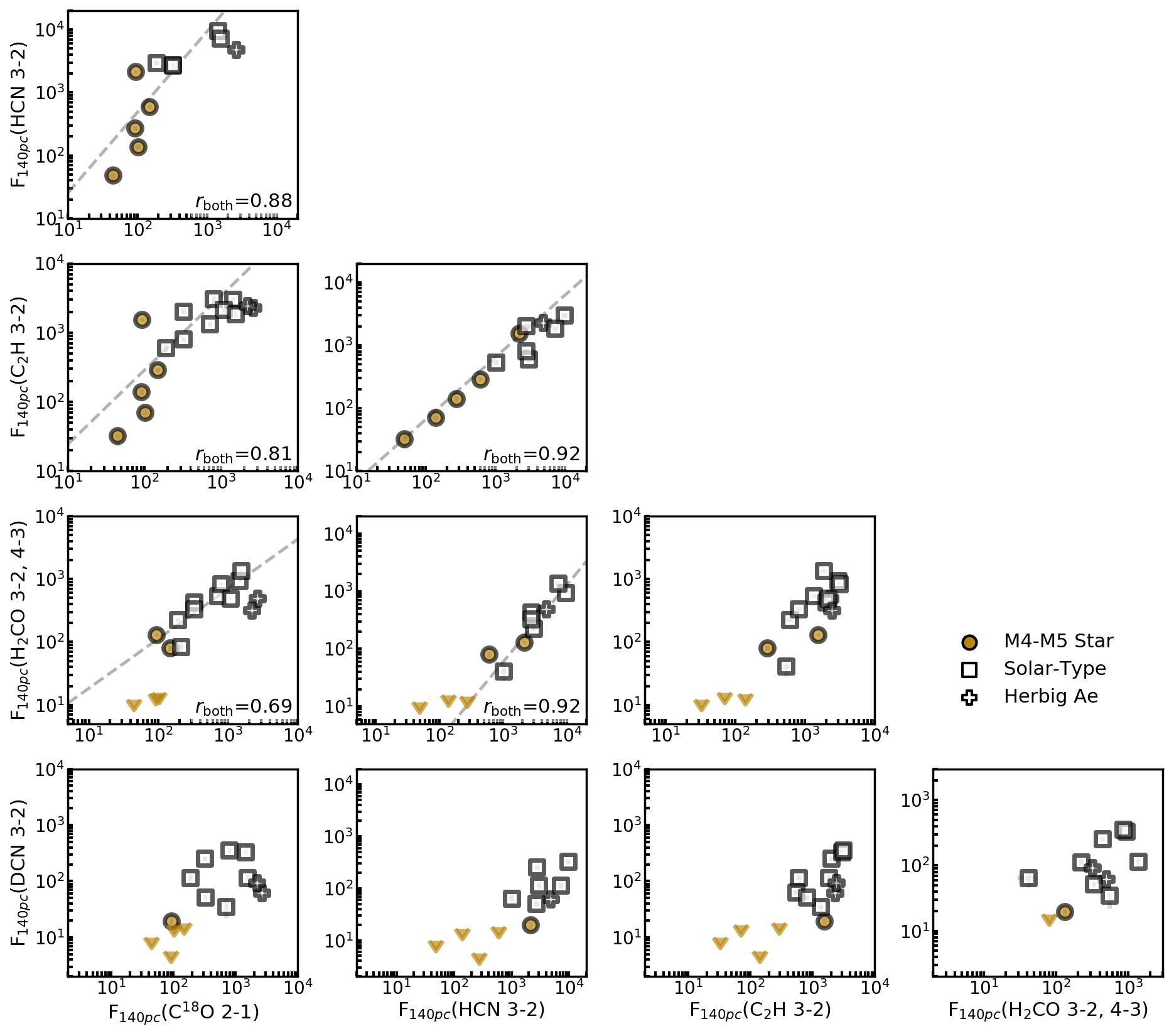}}
\caption{Molecular line fluxes toward the M4-M5 disks (dark gold circles), the solar-type disks (white squares), and the two Herbig Ae disks (white crosses).  The x and y-axes show the integrated fluxes (scaled to 140pc) for a particular molecular line, which are listed along the bottom and left sides, respectively, of the entire figure.  The dark gold triangles are 3$\sigma$ upper limits for M4-M5 disk tentative detections/nondetections and point in the direction of the limit.  Points where both the x and y-axes are upper limits are not shown.  The M4-M5 disks are from this work, while the solar-type and Herbig Ae disks were compiled from ALMA observations of~\cite{cite_huangetal2017, cite_bergneretal2019, cite_bergneretal2020, cite_peguesetal2020}.  Spearman correlation coefficients across all detections are written in the bottom-right of each plot whenever they are statistically significant.  The dashed gray lines are linear fits to the log data with statistically significant Spearman correlations (excluding the upper limits).  These lines indicate how well the data would adhere to a power law distribution in linear space.
\label{fig_fluxvsflux}}
\end{figure*}

\begin{figure*}
\centering
\resizebox{0.9\hsize}{!}{
    \includegraphics[trim=0pt 10pt 0pt 0pt, clip]{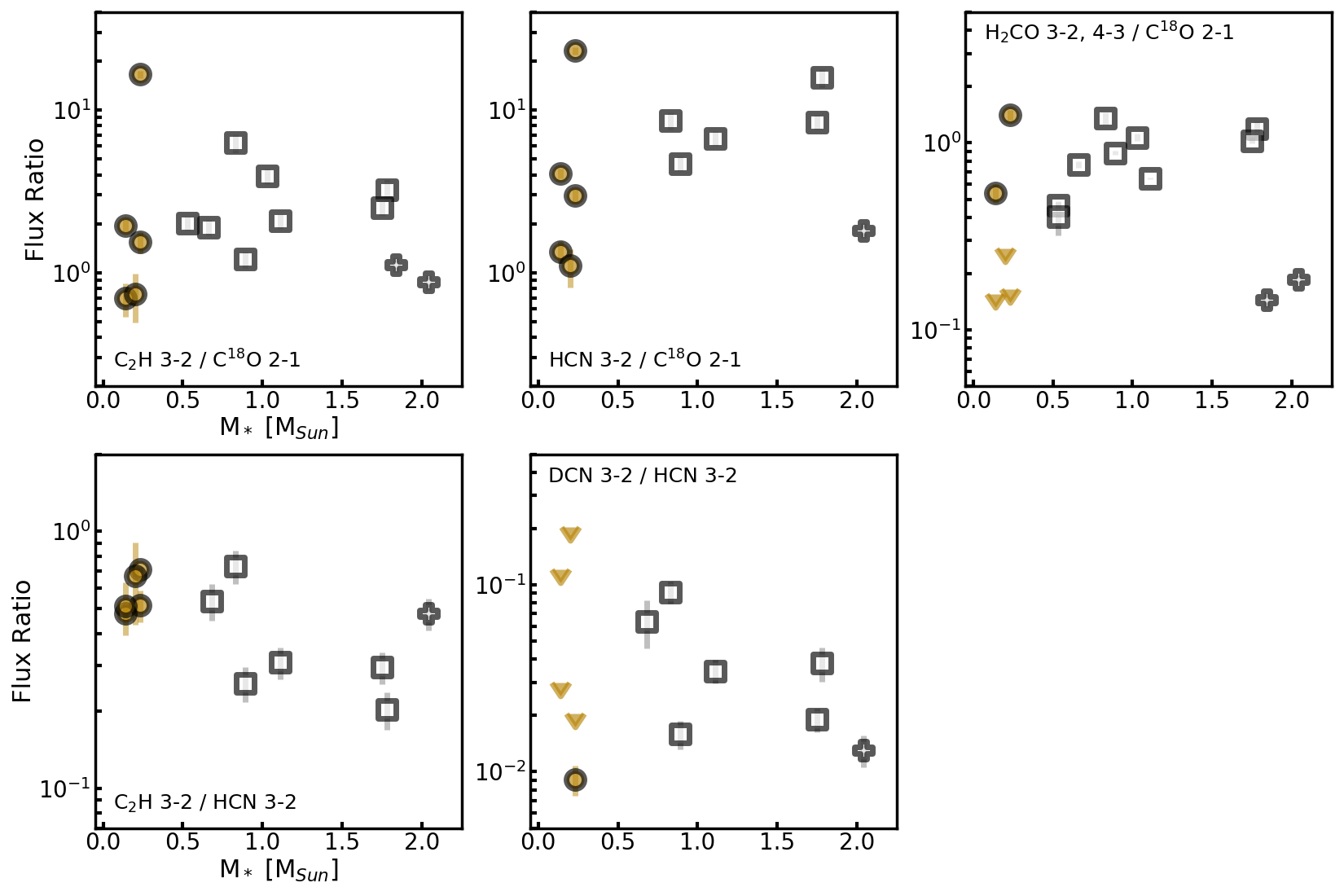}}
\caption{Relative flux ratios for different molecular lines toward the M4-M5 disks (dark gold circles and dark gold triangles for detections and 3$\sigma$ upper limits, respectively), the solar-type disks (white squares), and the two Herbig Ae disks (white crosses), plotted as a function of stellar mass.  The M4-M5 disks are from this work, while the solar-type and Herbig Ae disks were compiled from ALMA observations of~\cite{cite_huangetal2017, cite_bergneretal2019, cite_bergneretal2020, cite_peguesetal2020}.
\label{fig_fluxvsmstar}}
\end{figure*}

\begin{figure*}
\centering
\resizebox{0.85\hsize}{!}{
    \includegraphics[trim=0pt 0pt 0pt 0pt, clip]{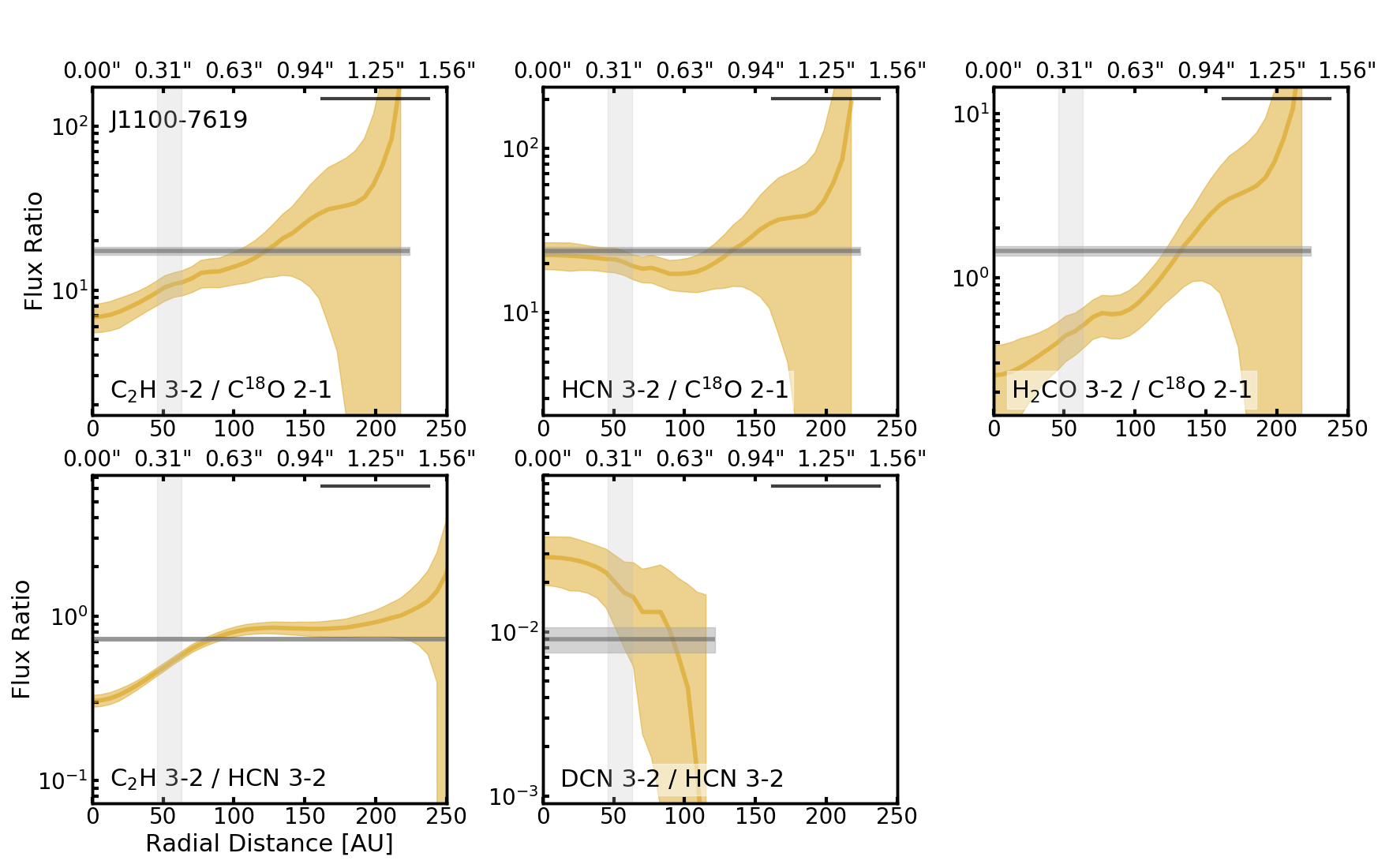}}
\caption{Radially resolved flux ratios for different molecular lines observed toward J1100-7619.  The error per annulus is given by $|F_\mathrm{t}/F_\mathrm{b}| \times \sqrt{(\sigma_\mathrm{t}/F_\mathrm{t})^2 + (\sigma_\mathrm{b}/F_\mathrm{b})^2}$, where $F$ is the flux in that annulus, $\sigma$ is the radial profile error (described in Section~\ref{sec_analysis_image}), and the subscripts $t$ and $b$ refer to the molecular lines at the top and bottom, respectively, of the ratio.  The beam sizes are represented by the horizontal bars at the top right of each plot.  The beams of the line images were smoothed and circularized to the same size before the ratios were computed.  The horizontal gray lines and shaded regions depict the disk-averaged flux ratios and error, respectively.  These regions extend horizontally to the minimum boundary within which the disk-averaged fluxes were measured (i.e., the minimum Keplerian mask extent for the molecular lines; Appendix~\ref{sec_appendix_kep}).  The vertical gray regions are from the edges of the 1.1mm (262GHz) to the 1.3mm (231GHz) dust continuum (Table~\ref{table_contflux}).
\label{fig_j1100_radflux}}
\end{figure*}

We now compare relative line fluxes (i.e., the flux for one molecular line with respect to the flux of a different molecular line) for the M4-M5 disks in this work to relative line fluxes measured for solar-type and Herbig Ae disks in the literature.  The literature disk sample was compiled from the chemistry surveys of~\cite{cite_huangetal2017},~\cite{cite_bergneretal2019},~\cite{cite_bergneretal2020}, and~\cite{cite_peguesetal2020}.  We use only the molecular line+disk pairs in these surveys that were detected with ALMA, leading to thirteen unique disks in all with at least two detections of C$^{18}$O, C$_2$H, DCN, HCN, and H$_2$CO line emission.  C$^{18}$O 2--1 was detected with ALMA toward 11/13 disks, C$_2$H 3--2 toward 10/13 disks, DCN 3--2 toward 10/13 disks, HCN 3--2 toward 7/13 disks, and either H$_2$CO 3$_{03}$-2$_{02}$ or 4$_{04}$-3$_{03}$ toward 13/13 disks.
We consider both H$_2$CO 3$_{03}$--2$_{02}$ and H$_2$CO 4$_{04}$--3$_{03}$ line fluxes from the literature, because both lines have similar flux behaviors~\citep{cite_peguesetal2020}.
Stellar masses and spectral types for the T Tauri disks in this combined literature sample range from $\sim$0.4-1.8M$_\Sun$ and M1-G7, respectively.  Two Herbig Ae disks (HD 163296 and MWC 480, A-star disks with stellar masses of $\sim$1.8-2.0M$_\Sun$) are included in the literature disk sample.  We include the two Herbig Ae disks in relevant figures and tables for completeness, but since there are only two of them, we focus mainly on comparisons with the solar-type disk sample in later discussion.

Figure~\ref{fig_c18osample} plots distance-normalized C$^{18}$O fluxes as a function of stellar mass across the combined sample of disks.  The combined sample appears to follow the same trend in C$^{18}$O vs. stellar mass.  The addition of the M4-M5 disks suggests a positive dependence of C$^{18}$O flux on stellar mass across the sample, which was not distinguishable from the solar-type and Herbig Ae disk samples alone.

Figure~\ref{fig_fluxvsflux} compares distance-normalized line fluxes between the M4-M5 and literature disk samples.  The Spearman correlation coefficients ($r$), which show how well the data can be described by a monotonic function, are shown when the associated p-value is $\leq$0.01.  All correlation coefficients and measures of statistical significance are presented and explained in Appendix~\ref{sec_appendix_linear}.  The line fluxes are often lower for the M4-M5 disks than for the solar-type and Herbig Ae disks.  J1100-7619, which is particularly bright in C$_2$H and HCN emission, is a notable exception.  Despite differences in individual fluxes, trends found between fluxes for solar-type and Herbig Ae disks seem to apply to M4-M5 disks.  The strongest correlations across the M4-M5, solar-type, and Herbig Ae disk samples are found between C$_2$H 3--2 and HCN 3--2 ($r$=0.92, p-value$<$0.001) and H$_2$CO and HCN 3--2 ($r$=0.92, p-value$<$0.001).  Notably C$_2$H 3--2 and HCN 3--2 are perfectly correlated for the M4-M5 disks alone ($r_\mathrm{M4-M5}$=1.00, p-value$<$0.001).  For the combined sample, the trends between the C$_2$H 3--2 vs. C$^{18}$O 2--1 line fluxes and the HCN 3--2 vs. C$^{18}$O 2--1 line fluxes are similar to each other ($r$=0.81, p-value$<$0.001 and $r$=0.88, p-value$<$0.001, respectively).  We also find a weak correlation between H$_2$CO line fluxes and C$^{18}$O 2--1 ($r$=0.69, p-value=0.01).  All other pairs of line fluxes do not appear to be significantly correlated.

We note that while we expect H$_2$CO and DCN line emission to be optically thin, studies have shown that C$_2$H and HCN line emission is often optically thick~\citep[e.g.,][]{cite_bergneretal2019}.  It is thus possible that the C$_2$H 3--2 and HCN 3--2 molecular line fluxes trace the distribution size rather than the underlying molecular abundance.  We investigate this possibility in Appendix~\ref{sec_appendix_13CO}, and we find evidence that the strong correlation between the C$_2$H 3--2 and HCN 3--2 fluxes likely cannot be explained by optical depth effects alone.

We plot a subset of the disk flux ratios for the M4-M5, solar-type, and Herbig Ae disks against stellar mass in Figure~\ref{fig_fluxvsmstar}.  The flux ratio subset includes molecular line emission relative to C$^{18}$O 2--1 emission, which tells us how the molecular line emission relates to the disk gas.  We also include the strongly correlated C$_2$H 3--2 vs. HCN 3--2 and the deuterated fraction DCN 3--2 vs. HCN 3--2.
We note that the C$_2$H 3--2 / HCN 3--2 flux ratios for the M4-M5 disks are similar in value to each other.  The C$_2$H 3--2 / HCN 3--2 flux ratios for the M4-M5 disks are high relative to the typical C$_2$H 3--2 / HCN 3--2 flux ratios for the solar-type disks.  Otherwise, we find no clear trend for any of the disk molecular line flux ratios in either the individual or combined disk samples.  The ratios across all samples appear flat relative to stellar mass and also to stellar luminosity (not shown).  The lack of any trends between molecular line flux ratios and stellar mass, despite the clear trends between the millimeter dust continuum fluxes, C$^{18}$O fluxes, and stellar masses in Figures~\ref{fig_disksample} and~\ref{fig_c18osample}, suggests that the underlying disk chemistry is similar for the M4-M5 and solar-type stars in the outer disk regions probed by our disk-integrated fluxes.


\subsection{Case Study: J1100-7619}
\label{sec_results_j1100}

J1100-7619 presents the brightest molecular emission in our sample and was observed with the highest spatial resolution.  We therefore use J1100-7619 to explore how the chemistry changes with disk radius around an M4-M5 star.


\subsubsection{Radial Fluxes}
\label{sec_results_j1100_fluxes}

Figure~\ref{fig_j1100_radflux} shows disk-averaged and azimuthally-averaged (radial) molecular flux ratios toward J1100-7619 for the same molecule pairs as in Figure~\ref{fig_fluxvsmstar}.  The (C$_2$H 3--2 / C$^{18}$O 2--1) and (H$_2$CO 3--2 / C$^{18}$O 2--1) ratios both increase across the disk.  There is a notable bump in the (H$_2$CO 3--2 / C$^{18}$O 2--1) ratio that appears at $\sim$80AU, just beyond the edge of the pebble disk.  The (DCN 3--2 / HCN 3--2) and (HCN 3--2 / C$^{18}$O 2--1) ratios are both roughly constant across the disk where significant signal-to-noise exists, with values of $\sim$(1.5-3.0) $\times 10^{-2}$ and $\sim$20-30, respectively.  The (C$_2$H 3--2 / HCN 3--2) ratio increases across the pebble disk, and then flattens out beyond the pebble disk edge.  Past $\sim$100 AU, the (C$_2$H 3--2 / HCN 3--2) ratio is notably constant, with values of $\sim$0.8-0.9.


\subsubsection{Excitation Temperatures, Column Densities, and Optical Depths}
\label{sec_results_j1100_fits}

\begin{figure*}
\centering
\resizebox{0.99\hsize}{!}{
    \includegraphics[trim=12.5pt 12.5pt 9pt 3.5pt, clip]{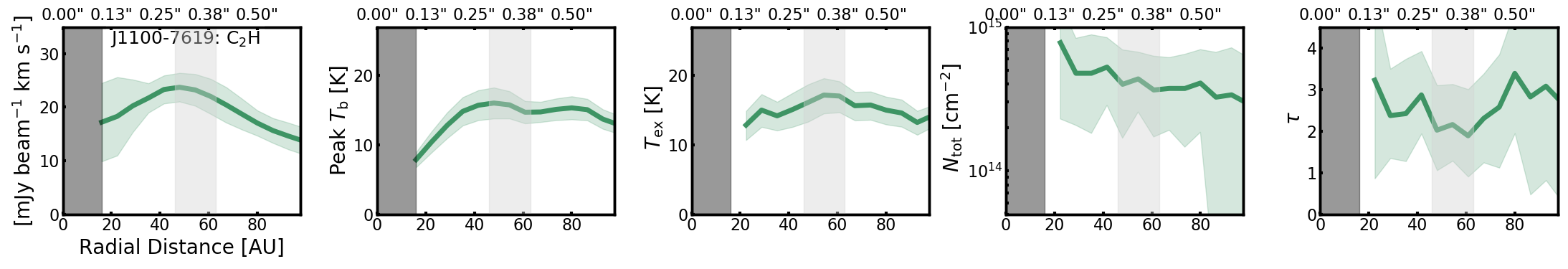}}
\resizebox{0.99\hsize}{!}{
    \includegraphics[trim=12.5pt 12.5pt 9pt 3.5pt, clip]{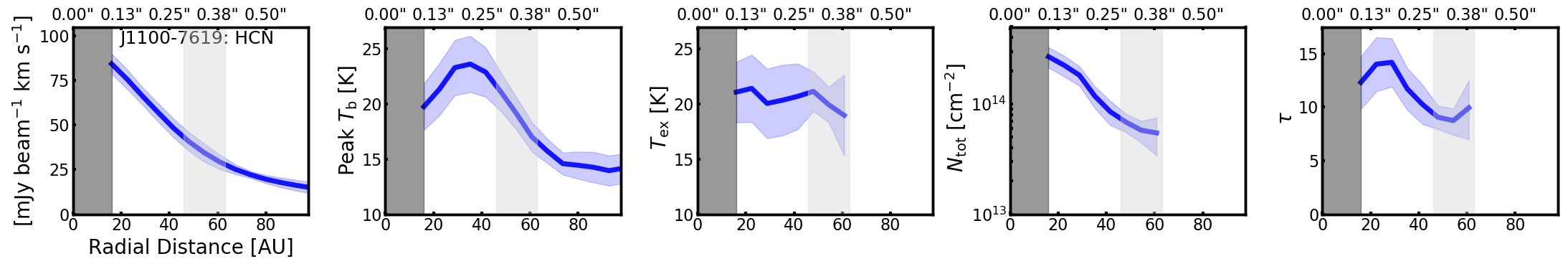}}
\caption{Radial flux profiles (column 1), peak brightness temperatures (column 2), and excitation temperatures, column densities, and optical depths (columns 3, 4, and 5, respectively) estimated from the pixel-by-pixel hyperfine fits toward J1100-7619.  The top and bottom rows display the results for C$_2$H (green) and HCN (blue), respectively.  For columns 1 and 2, the shaded uncertainties are the standard deviations of the emission/temperature within each 0.04" annulus.  The profiles shown in columns 3 through 5 are the weighted azimuthal averages of the values within each 0.04" annulus (Appendix~\ref{sec_appendix_hyperfine}).  The vertical dark gray region to the left of each plot represents the beam size.  We exclude annuli within the beam size to avoid unresolved emission within that region.  The beams for both molecules were circularized prior to the hyperfine fits.  The vertical pale gray regions are from the edge of the 1.1mm (262GHz) dust continuum to the edge of the 1.3mm (231GHz) dust continuum (Table~\ref{table_contflux}).
\label{fig_j1100_hypprof}}
\end{figure*}

\begin{figure*}
\centering
\resizebox{0.495\hsize}{!}{
    \includegraphics[trim=10pt 28.5pt 9.5pt 11pt, clip]{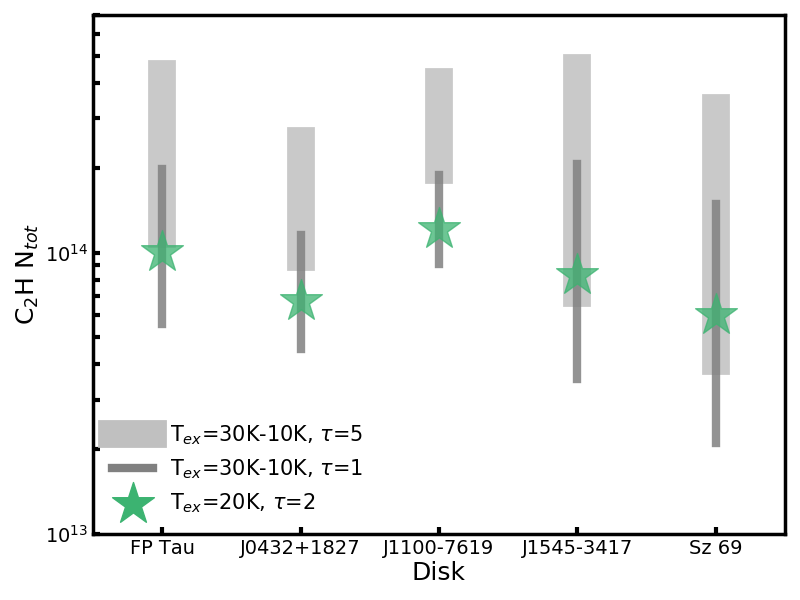}}
\resizebox{0.495\hsize}{!}{
    \includegraphics[trim=10pt 28.5pt 9.5pt 11pt, clip]{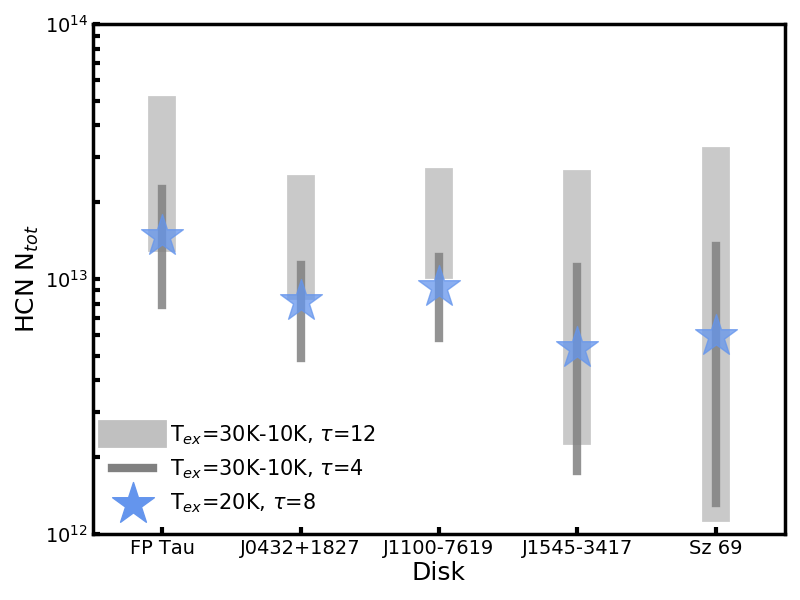}}
    %
\resizebox{0.495\hsize}{!}{
    \includegraphics[trim=10pt 10.5pt 9.5pt 8.5pt, clip]{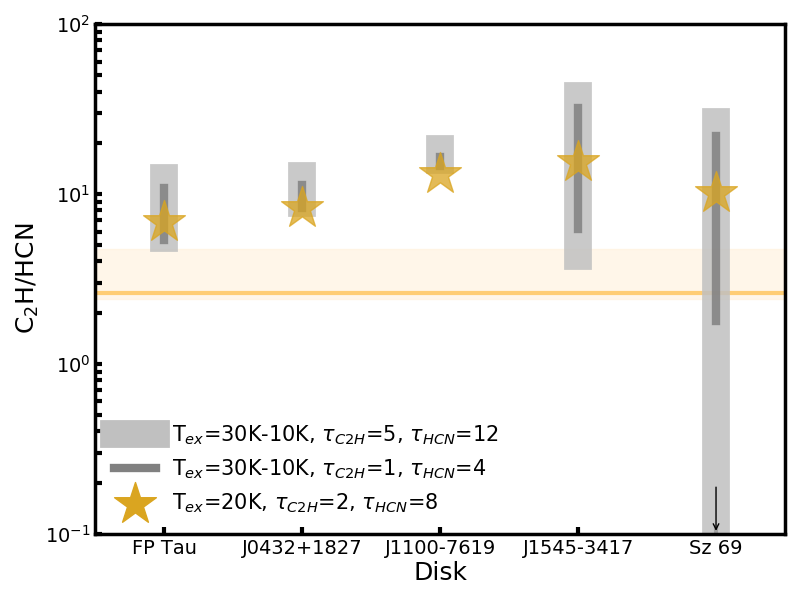}}
\resizebox{0.495\hsize}{!}{
    \includegraphics[trim=10pt 10.5pt 9.5pt 8.5pt, clip]{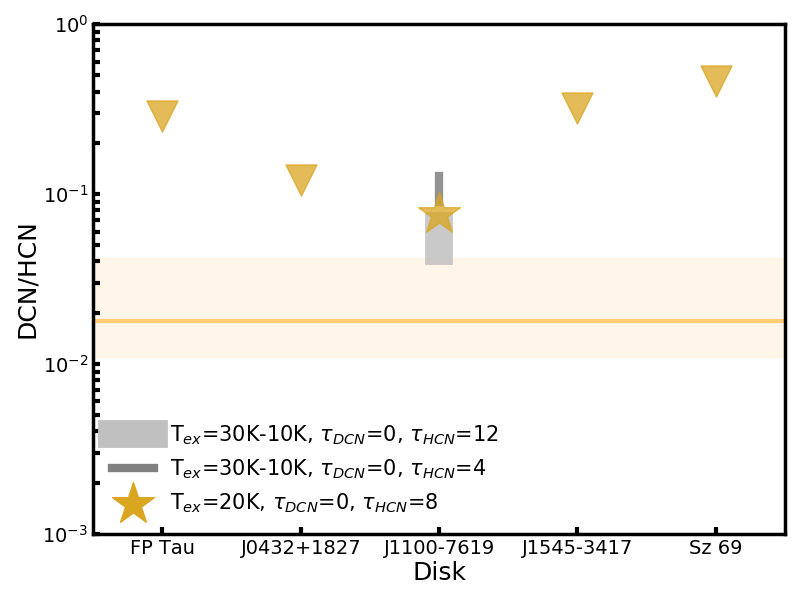}}
\caption{Disk-averaged column density estimates for C$_2$H (top left) and HCN (top right) for the entire M4-M5 disk sample, as well as estimates of the disk-averaged C$_2$H/HCN (bottom left) and DCN/HCN (bottom right) ratios.  The estimates were computed using the methodology described in Section~\ref{sec_analysis_coldens} and Appendix~\ref{sec_appendix_Ntot}.  For detections, the emission and disk areas were measured interior to where the emission radial profile reached 1$\sigma$.  For upper limits, the 3$\sigma$ errors and disk areas were measured within the bounds of the Keplerian masks (Appendix~\ref{sec_appendix_kep}).  Each estimate assumes an excitation temperature and optical depth (listed at the bottom left of each panel).  The stars and triangles represent median estimates and 3$\sigma$ upper limits, respectively.  Each gray bar gives a range of estimates assuming excitation temperatures from 10-30K and include error.  For the ratio plots, the horizontal dark lines and shaded regions correspond to the median and 16$^\mathrm{th}$-84$^\mathrm{th}$ percentile range, respectively, for a subset of ALMA-observed solar-type and Herbig Ae disks from the literature~\citep{cite_huangetal2017, cite_bergneretal2020} (Appendix~\ref{sec_appendix_Ntot}).
\label{fig_diskavgNtot}}
\end{figure*}

Where sufficient signal-to-noise exists for J1100-7619, we use the pixel-by-pixel hyperfine fitting procedure described in Section~\ref{sec_analysis_hyperfine} and Appendix~\ref{sec_appendix_hyperfine}.  This procedure allows us to estimate excitation temperatures, column densities, and optical depths for C$_2$H and HCN within each pixel.  Although we proceed with this procedure, we note that these hyperfine fits are subject to significant intrinsic uncertainties (see Appendix~\ref{sec_appendix_hyperfine}).

To increase the velocity resolution of the hyperfine fits, we reimaged the C$_2$H 3--2, C$_2$H (N=3--2, J=5/2-3/2), and HCN 3--2 line emission at channel widths of 0.14km s$^{-1}$ prior to performing the fits.  After the fitting procedure, we performed weighted averages of pixel values within each annulus around the host star.  The weighting process is described in Appendix~\ref{sec_appendix_hyperfine}.  We used this procedure to create weighted radial profiles of the excitation temperatures, column densities, and optical depths, which are plotted in Figure~\ref{fig_j1100_hypprof}.

C$_2$H excitation temperatures are $\sim$10-20K and appear roughly constant across the disk.  These low temperatures suggest either thermal emission from the disk midplane or subthermal emission from an upper disk layer.  C$_2$H optical depths range from $\sim$1-4 and are also roughly constant with radius.  C$_2$H column densities are $\sim$(10-1)$ \times 10^{14}$ cm$^{-2}$ over $\sim$20-100AU and appear roughly flat with radius.

The signal-to-noise for HCN is sufficient only for probing interior to the pebble disk edge (from $\sim$20-60AU).  In this region, HCN excitation temperatures are roughly constant.  These temperatures are slightly warmer but still consistent with the C$_2$H excitation temperatures.
HCN is optically thick in this region, with optical depths decreasing from $\sim$15-7.  HCN column densities are from $\sim$(30-3)$\times 10^{13}$ cm$^{-2}$.  The HCN column density profile decreases steeply in comparison to the C$_2$H column density profile over the same portion of the disk.

\cite{cite_bergneretal2019} used their pixel-by-pixel hyperfine procedure to derive C$_2$H and HCN excitation temperature and column density constraints for their sample of solar-type and Herbig Ae disks.  When we compare the J1100-7619 results with their solar-type and Herbig Ae disk sample, we find that our estimated excitation temperatures and optical depths across the pebble disk for J1100-7619 are consistent with estimates for the disks around more massive stars.


\subsection{Disk-Averaged Column Densities}
\label{sec_results_Ntot}

We now use the methodology described in Section~\ref{sec_analysis_coldens} and Appendix~\ref{sec_appendix_Ntot} to estimate disk-averaged C$_2$H, DCN, and HCN column densities for the entire M4-M5 disk sample.  For each molecular line and disk, we take the extent of the emission to be where the emission radial profile decreases to 1$\sigma$.  We extract emission only within that radius, and we use that radius to define the angular size $\Omega$ of the disk (see Appendix~\ref{sec_appendix_Ntot}).

We assume that the disk-averaged excitation temperatures range from 10-30K for both molecules, with a median value of 20K.  For disk-averaged optical depths, we assume 1-5 with a median of 2 for C$_2$H and 4-12 with a median of 8 for HCN.  We also estimate disk-averaged column densities for DCN, assuming that DCN is optically thin and has the same excitation temperature and conditions as C$_2$H and HCN.

Figure~\ref{fig_diskavgNtot} displays the estimated disk-averaged C$_2$H and HCN column densities for the M4-M5 disk sample using the disk-averaged median, minimum, and maximum excitation temperature and optical depth constraints.  The median C$_2$H column density estimates are (5-10)$ \times 10^{13}$ cm$^{-2}$ across the sample.  Changing the excitation temperature and opacity result in factors of 2-4 higher or lower estimates of the column density.  These values, which were estimated using the C$_2$H (N=3--2, J=7/2--5/2) line, are consistent with those estimated using the C$_2$H (N=3--2, J=5/2--3/2) line (not shown).  The median HCN column density estimates span a factor of 3, from (4-12)$ \times 10^{12}$ cm$^{-2}$.  Changes in optical depth from 4 to 12 lead to factors $\lesssim$3 in difference between estimates, while changes in excitation temperature from 10K to 30K lead to factors of $\sim$3-10 (excluding Sz 69).  For the bright and well-resolved disk J1100-7619, DCN is detected, and the estimated column densities (not shown) span $\sim$(4-10)$ \times 10^{11}$ cm$^{-2}$ from 10-30K with a median of $\sim$7$ \times 10^{11}$ cm$^{-2}$.

Figure~\ref{fig_diskavgNtot} also displays the estimated disk-averaged C$_2$H/HCN and DCN/HCN column density ratios.  These are plotted in comparison to the 16$^\mathrm{th}$, median, and 84$^\mathrm{th}$ percentiles derived from a sample of disk-averaged ratios from the literature for solar-type and Herbig Ae disks~\citep{cite_huangetal2017, cite_bergneretal2020}.  We used only the disk-averaged ratios from these references that were observed with ALMA and had comparable excitation temperature and optical depth assumptions.  Detailed descriptions of the excitation temperatures and optical depths assumed in the literature are given in Appendix~\ref{sec_appendix_Ntot}.

For all M4-M5 disks, estimates of the disk-averaged C$_2$H/HCN ratio do not change significantly with optical depth.  Median estimates of the ratio range from $\sim$7 for FP Tau to $\sim$11 for J1545-3417.  Notably all C$_2$H/HCN estimates for the M4-M5 disks exceed the 84$^\mathrm{th}$ percentile of the solar-type and Herbig Ae disk C$_2$H/HCN estimates.  The DCN/HCN estimates are relatively well constrained for J1100-7619.  This disk has a median DCN/HCN value of $\sim$0.072 and ranges from $\sim$0.04-0.1 across the extremes.  This single value exceeds the 84$^\mathrm{th}$ percentile disk-averaged DCN/HCN ratio found for the solar-type and Herbig Ae disks, although we would need more DCN detections toward M4-M5 disks to determine if this finding is representative of a larger sample.


\section{Discussion}
\label{sec_discussion}

\subsection{Chemical Origins of Observed Emission Patterns}
\label{sec_discussion_origins}

\begin{figure}
\centering
\resizebox{0.99\hsize}{!}{
    \includegraphics[trim=10pt 10pt 55pt 55pt, clip]{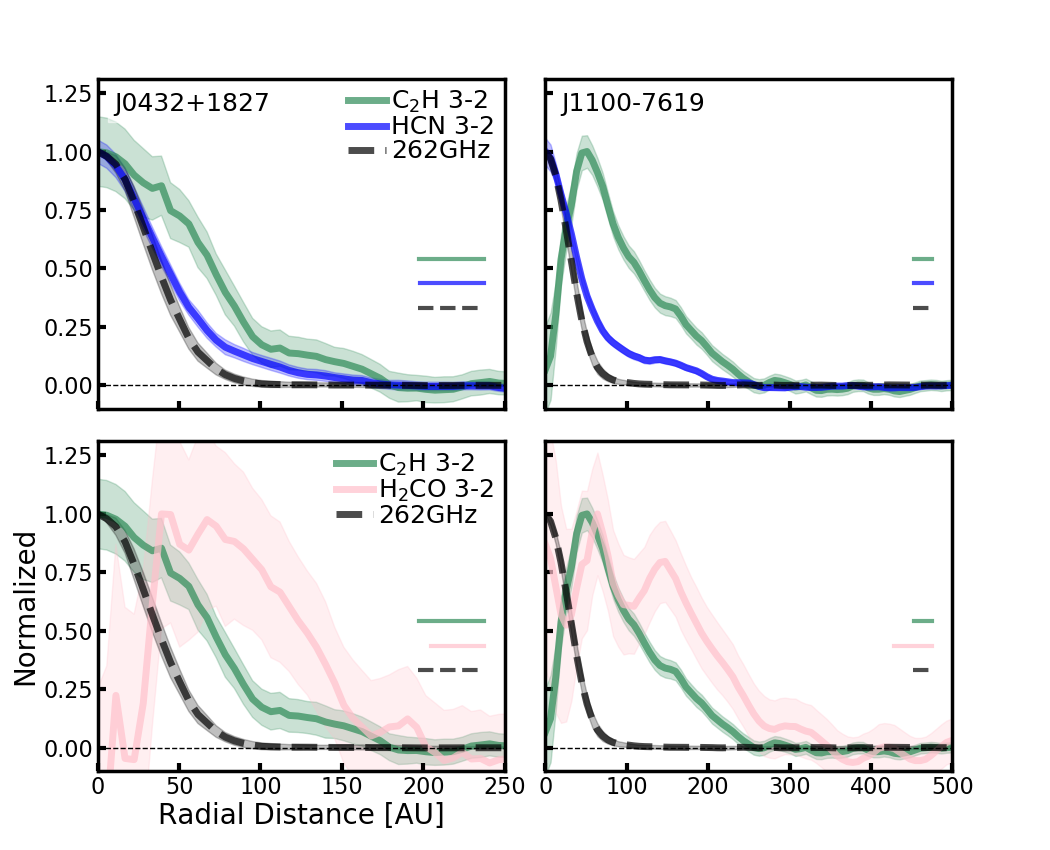}}
\caption{Normalized radial profiles of molecular line emission (adapted from Figure~\ref{fig_prof}) toward J0432+1827 (left column) and J1100-7619 (right column).  The profiles depicted are of C$_2$H 3--2 (green), HCN 3--2 (blue), and H$_2$CO 3--2 (pink), with the 1.1mm dust continuum emission profiles (dashed black) also shown for comparison.  The beam sizes are represented by the horizontal bars in the lower right corners.
\label{fig_profnorm}}
\end{figure}

The molecular emission morphologies we observe are products of the local chemistry and environment.  Here we discuss the observed emission morphologies across our M4-M5 disk sample in the context of chemical pathways and environmental conditions that may form them.  For recent, in-depth discussion of formation pathways and conditions for each molecule, refer to~\cite{cite_huangetal2017} (HCN and DCN),~\cite{cite_bergneretal2019} (C$_2$H and HCN), and~\cite{cite_peguesetal2020} (H$_2$CO).

\subsubsection{C$_2$H and HCN}

Observational studies have found significant positive correlations between HCN and C$_2$H line fluxes for solar-type disks~\citep[stellar masses of $\sim$0.5-1.8M$_\Sun$][]{cite_bergneretal2019}, along with significant positive correlations between CN and C$_2$H line fluxes for both low-mass M-star (stellar masses $<$0.5M$_\Sun$) and solar-type disks~\citep{cite_miotelloetal2019}.  These studies concluded that production of these molecules can be traced to the same or similar chemical and/or physical driver(s).  Theoretical studies of T Tauri disks~\citep{cite_duetal2015, cite_berginetal2016} have found that depletions of oxygen relative to carbon and heightened UV irradiation are likely major drivers, as they can cause increased production of C$_2$H, other hydrocarbons, and cyanides in the outer T Tauri disk regions.  Simulated observations of C$_2$H have shown that this increased production can be observed as rings in the hydrocarbon emission~\citep{cite_berginetal2016}.

It is difficult to isolate physical/chemical driver(s) and how they are tied to emission morphologies without detailed modeling.  That being said, we note two intruiging similarities between C$_2$H and HCN beyond the pebble disk edges of our sample: morphologies and relative fluxes and column densities, which together point toward a related chemistry beyond the pebble disk.  First, the two well-resolved disks in the sample (J0432+1827 and J1100-7619; Figure~\ref{fig_profnorm}) show slope changes and suggestive plateaus in both C$_2$H and HCN emission beyond the pebble disk.  Second, the disk (C$_2$H 3--2 / HCN 3--2) flux ratios (Figures~\ref{fig_fluxvsflux} and~\ref{fig_fluxvsmstar}) appear perfectly correlated across the M4-M5 disk sample.  The radially resolved (C$_2$H 3--2 / HCN 3--2) flux ratio toward J1100-7619 helps us understand this correlation, because the ratio is virtually constant beyond the pebble disk edge (Figure~\ref{fig_j1100_radflux}).
The estimated disk-averaged C$_2$H/HCN column density ratios (Figure~\ref{fig_diskavgNtot}), which are influenced most heavily by contributions from the outer disk, are similar across the M4-M5 disk sample.

It is important to note how the C$_2$H and HCN column density profiles measured across the pebble disk for J1100-7619 (Figure~\ref{fig_j1100_hypprof}) are different in shape, with the HCN column densities decreasing more steeply across the pebble disk than the C$_2$H column densities.  This difference may indicate that the C$_2$H and HCN chemistry is decoupled and/or affected by different physical/chemical processes interior to the pebble disk edge.
This further suggests that the similarities and correlations between C$_2$H and HCN morphologies, fluxes, and disk-averaged column densities discussed here are driven by the emission \textit{beyond} the pebble disk.

\subsubsection{H$_2$CO, CO, and C$_2$H}
\label{sec_discussion_h2covsco}

H$_2$CO can form efficiently through either gas-phase pathways, which are expected to be most efficient in warm and dense disk regions, or through the grain-surface hydrogenation of CO ice, which is expected to be prominent beyond the CO snowline~\citep[e.g.,][]{cite_hiraokaetal1994, cite_fockenbergetal2002, cite_hiraokaetal2002, cite_watanabeetal2002, cite_hidakaetal2004, cite_watanabeetal2004, cite_atkinsonetal2006, cite_fuchsetal2009}.
Toy disk models, detailed chemical modeling, and observations of H$_2$CO in multiple T Tauri disks have shown that both pathways can contribute significantly to H$_2$CO emission in disks~\cite[e.g.,][]{cite_loomisetal2015, cite_carneyetal2017, cite_obergetal2017, cite_peguesetal2020}, producing central peaks and/or outer rings in the emission.

H$_2$CO is clearly detected toward two disks (J0432+1827 and J1100-7619; Figure~\ref{fig_profnorm}) in the M4-M5 disk sample.  For both disks, there are peaks in the H$_2$CO emission that are at/beyond the edge of the pebble disk.
The hole in C$_2$H emission toward J0432+1827 is off-center relative to the dust continuum estimate by $\sim$0.1" and difficult to interpret, but we do note intriguing alignments in the H$_2$CO and C$_2$H emission for J1100-7619 (Figure~\ref{fig_profnorm}).  The first ring in H$_2$CO emission toward this disk is roughly aligned with the peak in the C$_2$H emission at $\sim$60AU, and both peaks are located near the edge of the pebble disk.  The second ring in H$_2$CO emission is roughly aligned with a small bump in the C$_2$H emission at $\sim$150AU.  The H$_2$CO emission peaks do not appear to be direct consequences of the CO snowline: assuming a simple radial temperature profile with a power law exponent of 0.55, a temperature of 142K at 1 AU~\citep[the averages for M-star disks from][]{cite_andrewsetal2007}, and CO midplane freeze-out temperatures of 18-26K~\citep[e.g.,][]{cite_obergetal2011b}, we estimate upper limits on the CO snowline of $\sim$22-43 AU, which are interior to both H$_2$CO peaks.


These alignments suggest a link in H$_2$CO and C$_2$H formation in these disks.  It is possible that UV photodesorption of CO ice explains both the H$_2$CO emission and the apparent link in H$_2$CO and C$_2$H formation beyond the pebble disk.  On grain surfaces, this non-thermal mechanism may ultimately release emission from H$_2$CO produced via CO ice hydrogenation~\citep[e.g., discussion by][]{cite_walshetal2014, cite_loomisetal2015, cite_obergetal2017, cite_fraudetal2019}.  Simultaneously in the gas phase, this same mechanism is hypothesized to cause increased gas-phase atomic carbon abundances and lead to enhanced hydrocarbon production, including of C$_2$H~\citep{cite_duetal2015}.  Those enhanced hydrocarbons may produce H$_2$CO through its gas-phase pathways as well~\citep[e.g., through the neutral-neutral reaction of CH$_3$ and O;][]{cite_fockenbergetal2002, cite_atkinsonetal2006}.


\subsection{M4-M5 vs. Solar-Type Outer Disk Chemistry}
\label{sec_discussion_solar}

\subsubsection{Hydrocarbons and Hydrocyanides}

C$_2$H and HCN are precursor molecules for more complex hydrocarbon and hydrocyanide chemistry.
We note two findings that suggest similar connections between C$_2$H and HCN chemistry for our M4-M5 disks and for solar-type disks from the literature.  First, radially resolved C$_2$H and HCN excitation temperatures and optical depths toward the bright and well-resolved disk J1100-7619, measured interior to the pebble disk edge (Figure~\ref{fig_j1100_hypprof}), are roughly consistent with C$_2$H and HCN measurements for solar-type disks presented in~\cite{cite_bergneretal2019}.  Second, the C$_2$H 3--2 and HCN 3--2 fluxes for the M4-M5 disks appear to follow the same trend as the solar-type disks (Figure~\ref{fig_fluxvsflux}).  These fluxes do not show any trends across stellar type with stellar mass (Figure~\ref{fig_fluxvsmstar}).
These results imply similar underlying physical/chemical processes for the M4-M5 disks and the solar-type disks, consistent with the predictions for brown dwarf disks relative to T Tauri disks from thermochemical modeling~\citep{cite_greenwoodetal2017}.

At the same time, we find evidence suggesting that C$_2$H production is enhanced relative to HCN production in the disks around our cooler M4-M5 stars, in comparison to disks around the warmer solar-type stars.
Assuming that the C$_2$H and HCN emission are originating from similar layers beyond the pebble disk, we find that M4-M5 disk-averaged C$_2$H/HCN column density ratios are higher than the 84$^\mathrm{th}$ percentile C$_2$H/HCN ratio for solar-type disks estimated by~\cite{cite_bergneretal2019} (Figure~\ref{fig_diskavgNtot}).  The C$_2$H 3--2 / HCN 3--2 flux ratios are also higher for the M4-M5 disks compared to the solar-type disks.
These enhancements imply greater hydrocarbon production in the M4-M5 disks relative to the solar-type disks.  This finding is consistent with previous infrared chemistry observations and models of the inner regions ($<$10AU) of low-mass M-star disks, brown dwarf disks, and more massive T Tauri disks.
These previous studies found relatively high C$_2$H$_2$/HCN flux and column density ratios and more carbon-rich atmospheres in the disks around low-mass M-stars and brown dwarfs.  They inferred that hydrocarbon chemistry in the innermost regions of disks around these low-mass stars is enhanced relative to the more massive T Tauri stars~\citep[][]{cite_pascuccietal2009, cite_walshetal2015, cite_pascuccietal2013}.

We can only draw preliminary conclusions from our work, as our sample size is small and biased toward gas-rich M4-M5 disks.
As an initial view into disk chemistry around M4-M5 stars, our results suggest that similar patterns of hydrocarbon and hydrocyanide chemistry may exist beyond the pebble disk edges around M4-M5 and solar-type stars.  However, there is some mechanism(s) or process(es), such as the difference in radiation field, that appears to be producing more hydrocarbons in disks around the cooler M4-M5 stars.  It is not clear if this mechanism(s) is the same as that driving the hydrocarbon enhancements previously discovered in the inner $<$10AU of disks around low-mass M-stars and brown dwarfs.  More observations and models of low-mass M-star disk chemistry, especially those that explore chemistry beyond the pebble disk, will help us investigate these tentative conclusions.

\subsubsection{Formaldehyde}

H$_2$CO emission morphologies for the two M4-M5 disks J0432-1827 and J1100-7619 show central/off-center holes in H$_2$CO emission, as well as substantial rings in H$_2$CO emission at/beyond the pebble disk edge (Figure~\ref{fig_profnorm}).  In contrast,~\cite{cite_peguesetal2020} found mostly centrally peaked H$_2$CO emission morphologies for their sample dominated by large solar-type disks (observed with a mean resolution of 0.59" $\pm$ 0.22"), with H$_2$CO emission substructure that was significant but of lesser magnitude than the central emission peaks.
It is possible that this apparently relatively enhanced production of H$_2$CO emission beyond the pebble disk edge for the M4-M5 disks is due to (1) a relatively large reservoir of CO ice for the M4-M5 disks, which are much colder than their solar-type disk counterparts, and/or (2) a relatively small region of the M4-M5 disks interior to the CO snowline, where H$_2$CO would be produced mainly through warm, efficient gas-phase pathways.  Notably we only have two disks firmly detected in H$_2$CO, and so we need a larger sample of low-mass M-star disks to verify this tentative conclusion.


\section{Summary}
\label{sec_summary}

We have conducted an ALMA survey of CO isotopologues and small organic molecules toward a sample of five M4-M5 disks.  We summarize our key findings below:

\begin{enumerate}
    \item We detect $^{12}$CO 2--1, $^{13}$CO 2--1, C$^{18}$O 2--1, C$_2$H 3--2, and HCN 3--2 toward 5/5 disks.  We detect H$_2$CO 3--2 toward 2/5 disks and tentatively detect it toward two other disks.  We detect DCN 3--2 toward 1/5 disks and tentatively detect it toward all other disks (see Section~\ref{sec_results_detections}).
    \item The dust continuum emission toward all five disks in the sample is smooth at this resolution (0.12-0.29").  The $^{12}$CO 2--1, $^{13}$CO 2--1, and C$^{18}$O 2--1 emission is centrally peaked across the sample, with substructure that may be due to significant cloud contamination, extended disk structure, and/or disk-envelope interactions for 3/5 disks.  For the two well-resolved disks, HCN 3--2 emission is centrally peaked with suggestive substructure beyond the pebble disk edge, C$_2$H 3--2 and H$_2$CO 3--2 have central/off-center depressions or holes in emission and substructure/rings beyond the pebble disk, and DCN 3--2 is centrally peaked and compact (see Section~\ref{sec_results_morphologies}).
    \item The C$_2$H and HCN line fluxes are correlated for the M4-M5 disk sample, as has been seen previously for a sample consisting mostly of solar-type disks from the literature.  There are no clear relationships between the molecular line flux ratios and stellar masses/luminosities for either the M4-M5 or solar-type disk samples (see Section~\ref{sec_results_relflux}).
    \item Radially resolved flux ratios of (C$_2$H 3--2 / C$^{18}$O 2--1) and (H$_2$CO 3--2 / C$^{18}$O 2--1) toward the well-resolved disk J1100-7619 increase monotonically with distance.  (DCN 3--2 / HCN 3--2) and (HCN 3--2 / C$^{18}$O 2--1) are roughly constant across the disk where sufficient signal-to-noise exists.  (C$_2$H 3--2 / HCN 3--2) increases with radius across the pebble disk, and then is constant beyond the pebble disk edge (see Section~\ref{sec_results_j1100_fluxes}).
    \item Pixel-by-pixel fits of the C$_2$H and HCN hyperfine structure toward J1100-7619 reveal that C$_2$H excitation temperatures, optical depths, and column densities are $\sim$10-20K, $\sim$1-4, and $\sim$(1-10)$\times 10^{14}$ cm$^{-2}$, respectively, across the disk.  HCN excitation temperatures and optical depths are from $\sim$15-25K and $\sim$7-15, respectively, interior to the pebble disk edge where sufficient signal-to-noise exists.  HCN column densities in the same disk region range from $\sim$(3-30)$\times 10^{13}$ cm$^{-2}$ (see Section~\ref{sec_results_j1100_fits}).
    \item For typical assumptions on excitation and assuming similar disk layers of origin for the C$_2$H and HCN emission, disk-averaged column density values for the M4-M5 disk sample are (5-10)$\times 10^{13}$ cm$^{-2}$ for C$_2$H and (4-12)$\times 10^{12}$ cm$^{-2}$ for HCN.  Disk-averaged C$_2$H/HCN column density ratios for the M4-M5 disk sample all exceed the 84$^\mathrm{th}$ percentile estimated for a sample of solar-type disks from the literature.  This is similar, and perhaps related to, the hydrocarbon enhancements observed in the inner ($<$10AU) regions of M-star disks by infrared surveys.  The single disk-averaged DCN/HCN column density estimate toward J1100-7619 is higher than the solar-type disk estimates from the literature (see Section~\ref{sec_results_Ntot}).
    \item We find evidence that C$_2$H and HCN share similar physical/chemical driver(s) beyond the pebble disks of our M4-M5 disk sample, as has been theorized for solar-type disks, while their relative chemistry appears more distinct interior to the pebble disk edge.  Similar radial substructure of the H$_2$CO and C$_2$H emission may point to a linked dependence on the UV photodesorption of CO ice beyond the pebble disk (see Sections~\ref{sec_discussion_origins} and~\ref{sec_discussion_solar}).
\end{enumerate}

We stress that our sample is biased toward bright, gas-rich M4-M5 disks, and that these findings serve as an \textit{initial} window into $>$10AU disk chemistry for low-mass M-stars.  Spatially resolved molecular line observations toward a larger sample of low-mass M-star disks would allow us to explore these patterns of disk chemistry over a more representative population of young, cool stars.

\acknowledgments

Jamila Pegues gratefully acknowledges the support of the National Science Foundation (NSF) Graduate Research Fellowship through Grant Numbers DGE1144152 and DGE1745303, and Karin \"Oberg gratefully acknowledges the support of the Simons Foundation through a Simons Collaboration on the Origins of Life (SCOL) PI Grant (Number 321183).  L. Ilsedore Cleeves gratefully acknowledges support from the Davide and Lucille Packard Foundation and the Virginia Space Grant Consortium.  Viviana V. Guzm\'{a}n gratefully acknowledges support from FONDECYT Iniciación 11180904.  Feng Long gratefully acknowledges support from the Smithsonian Institution as an SMA Fellow.  Support for this work was also provided by NASA through the NASA Hubble Fellowship grants \#HST-HF2-51429.001-A and \#HST-HF2-51460.001-A awarded by the Space Telescope Science Institute, which is operated by the Association of Universities for Research in Astronomy, Inc., for NASA, under contract NAS5-26555.

This paper makes use of the following ALMA data: ADS/JAO.ALMA\#2017.1.01107.S.  ALMA is a partnership of ESO (representing its member states), NSF (USA) and NINS (Japan), together with NRC (Canada), MOST and ASIAA (Taiwan), and KASI (Republic of Korea), in cooperation with the Republic of Chile. The Joint ALMA Observatory is operated by ESO, AUI/NRAO and NAOJ.  The National Radio Astronomy Observatory is a facility of the National Science Foundation operated under cooperative agreement by Associated Universities, Inc.
 
This work has made use of data from the European Space Agency (ESA) mission
{\it Gaia} (\url{https://www.cosmos.esa.int/gaia}), processed by the {\it Gaia}
Data Processing and Analysis Consortium (DPAC,
\url{https://www.cosmos.esa.int/web/gaia/dpac/consortium}). Funding for the DPAC
has been provided by national institutions, in particular the institutions
participating in the {\it Gaia} Multilateral Agreement.

All data reduction scripts and computer code used for this research were written in Python (version 2.7).  All plots were generated using Python's Matplotlib package~\citep{cite_matplotlib}.  This research also made use of Astropy (\url{http://www.astropy.org}), a community-developed core Python package for Astronomy~\citep{cite_astropy2013, cite_astropy2018}, as well as the NumPy~\citep{cite_numpy} and SciPy~\citep{cite_scipy} Python packages.



\begin{appendix}

\section{Keplerian Mask Parameters}
\label{sec_appendix_kep}

Table~\ref{table_kepmask} gives the Keplerian mask parameters used for the M4-M5 disk sample.

\begin{deluxetable*}{lccccccccccccc}
\tablewidth{0.99\textwidth}
\tablecaption{Keplerian Mask Parameters. \label{table_kepmask}}
\tablehead{
Disk       & $^{12}$CO $\theta_\alpha$  & $^{12}$CO $\theta_\beta$  & $V_\mathrm{LSR}$ & $V_0$  & $R_0$  & $q$ & \multicolumn{7}{c}{Mask Extent (")}                \\
           & ($^\circ$)          & ($^\circ$)          &            {(}km s$^{-1}${)} & {(}km s$^{-1}${)} & {(}AU{)}  &    & $^{12}$CO & $^{13}$CO & C$^{18}$O & C$_2$H & DCN  & HCN  & H$_2$CO
}
\colnumbers \startdata
\hline
FP Tau     & 213.3               & 46.7                & 8.32      & 0.25       & 100      & 0.3     & 1.24                      & 0.68                      & 0.56      & 1.44   & 0.12 & 1.20 & 0.36    \\
J0432+1827 & 157.5               & 42.9                & 5.60      & 0.25       & 100      & 0.3     & 2.16                      & 1.32                      & 1.32      & 1.32   & 0.48 & 1.36 & 1.48    \\
J1100-7619 & 108.8               & 20.9                & 4.75      & 0.22       & 100      & 0.3     & 2.04                      & 1.92                      & 1.36      & 1.48   & 0.72 & 1.64 & 2.12    \\
J1545-3417 & 264.2               & 53.1                & 4.50$^{[1]}$      & 0.25       & 100      & 0.3     & 2.00                      & 1.32                      & 1.96      & 1.16   & 0.36 & 1.36 & 0.40    \\
Sz 69      & 167.2               & 20.5                & 5.30$^{[1]}$      & 0.25       & 100      & 0.3     & 2.68                      & 2.68                      & 0.96      & 0.4    & 0.16 & 0.44 & 0.24   
\enddata
\tablecomments{All Keplerian masks were generated using~\cite{cite_kepmask}.  The masks assume that the combined thermal and turbulent line width is defined by $\Delta V \sim V_0(R_0/100)^q$~\citep{cite_yenetal2016}.  We estimated the geometric angle parameters, denoted here as $\theta_\alpha$ and $\theta_\beta$, using the masks, fixed broadening parameters, fixed systemic velocity ($V_\mathrm{LSR}$), and a grid-search algorithm.  $\theta_\alpha$ and $\theta_\beta$ represent the position and inclination angles, respectively.  We stress that these are \textit{parameters}, not measurements, which were chosen purely to maximize the fit of the Keplerian masks.  We use the unconventional $\alpha$ and $\beta$ denotation here to avoid confusion with angles measured in other studies from dust or line emission.  For J1545-3417 and Sz 69, systemic velocities are based on values and uncertainties available in the literature~\citep{cite_yenetal2018}.  For FP Tau, J0432+1827, and J1100-7619, the systemic velocities were selected based on inspection of the CO isotopologue emission.  The Keplerian mask radii are the distances at which the radial emission profiles first reached zero. \textit{References: [1]~\cite{cite_yenetal2018}.}}
\end{deluxetable*}


\section{Hyperfine Constraints}
\label{sec_appendix_hyperfine}

\begin{figure*}
\centering
\resizebox{0.925\hsize}{!}{
    \includegraphics[]{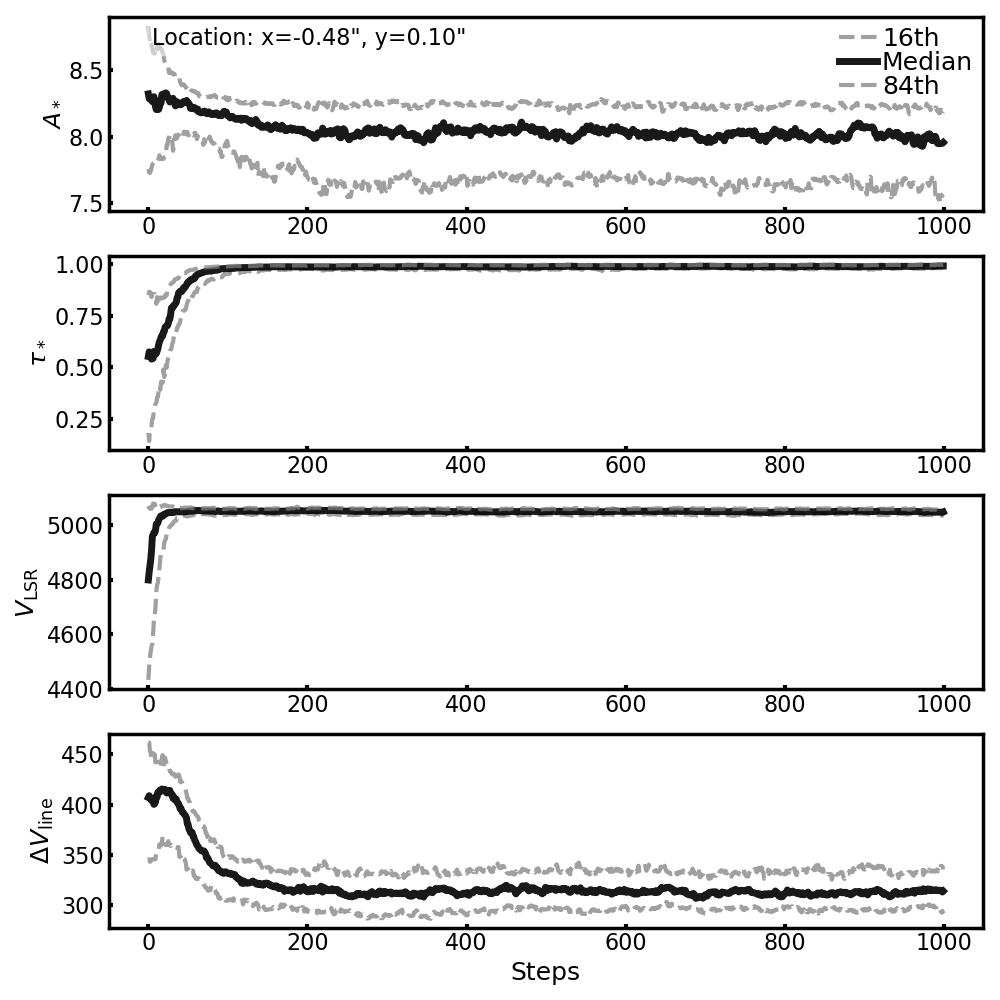}
    \includegraphics[]{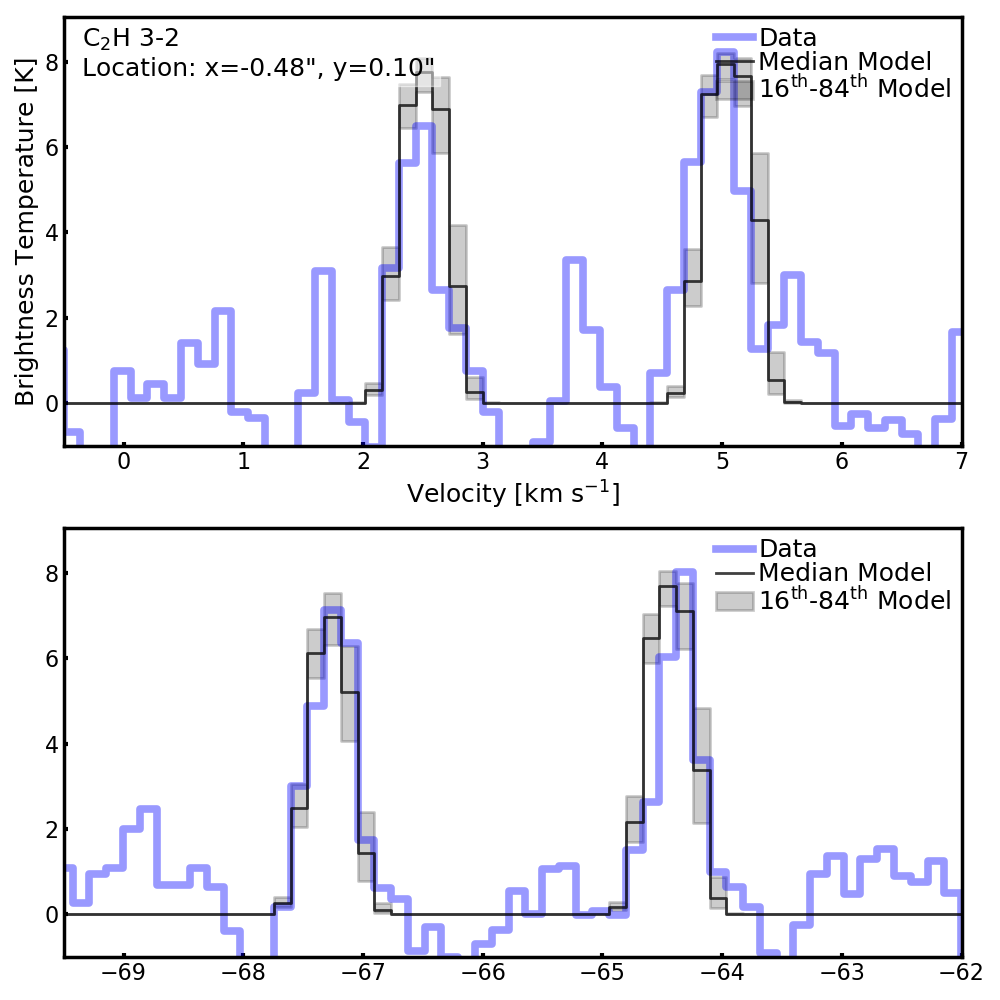}
    }
\resizebox{0.925\hsize}{!}{
    \includegraphics[trim=0pt 0pt 0pt -70pt, clip]{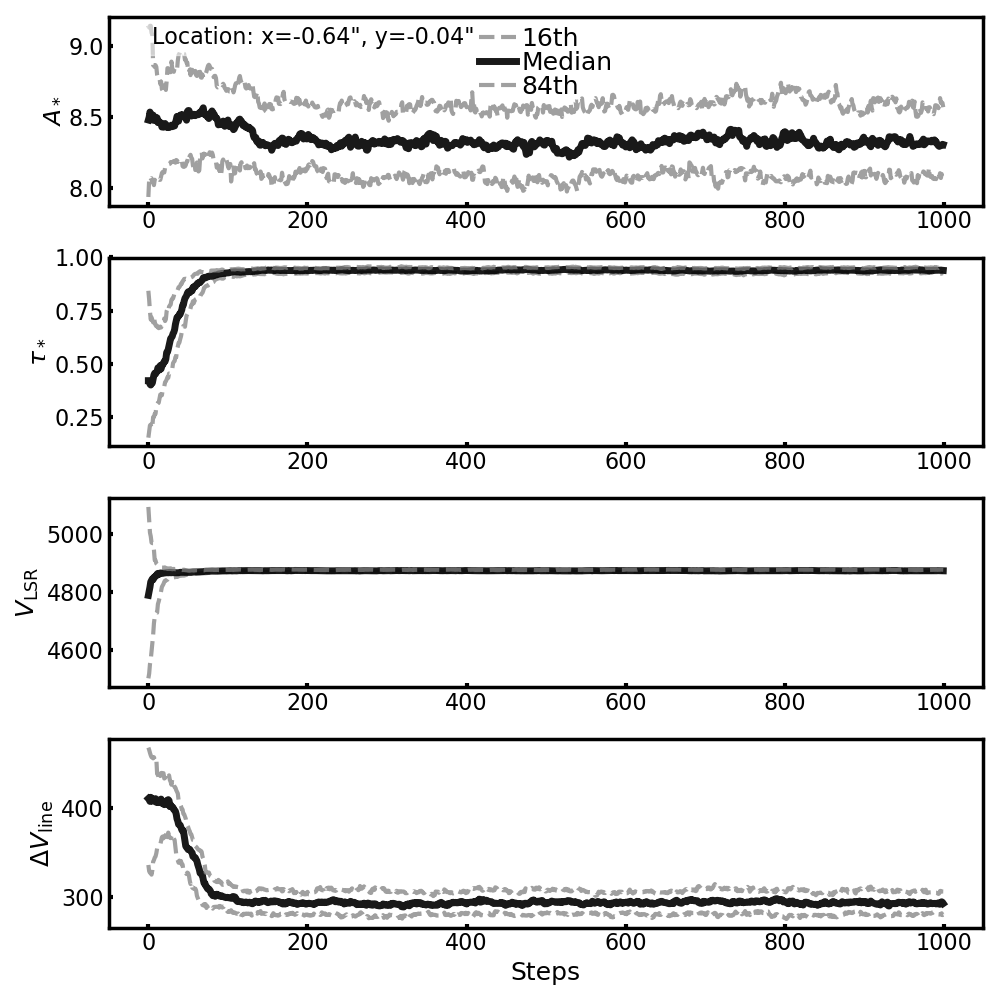}
    \includegraphics[trim=0pt 0pt 0pt -70pt, clip]{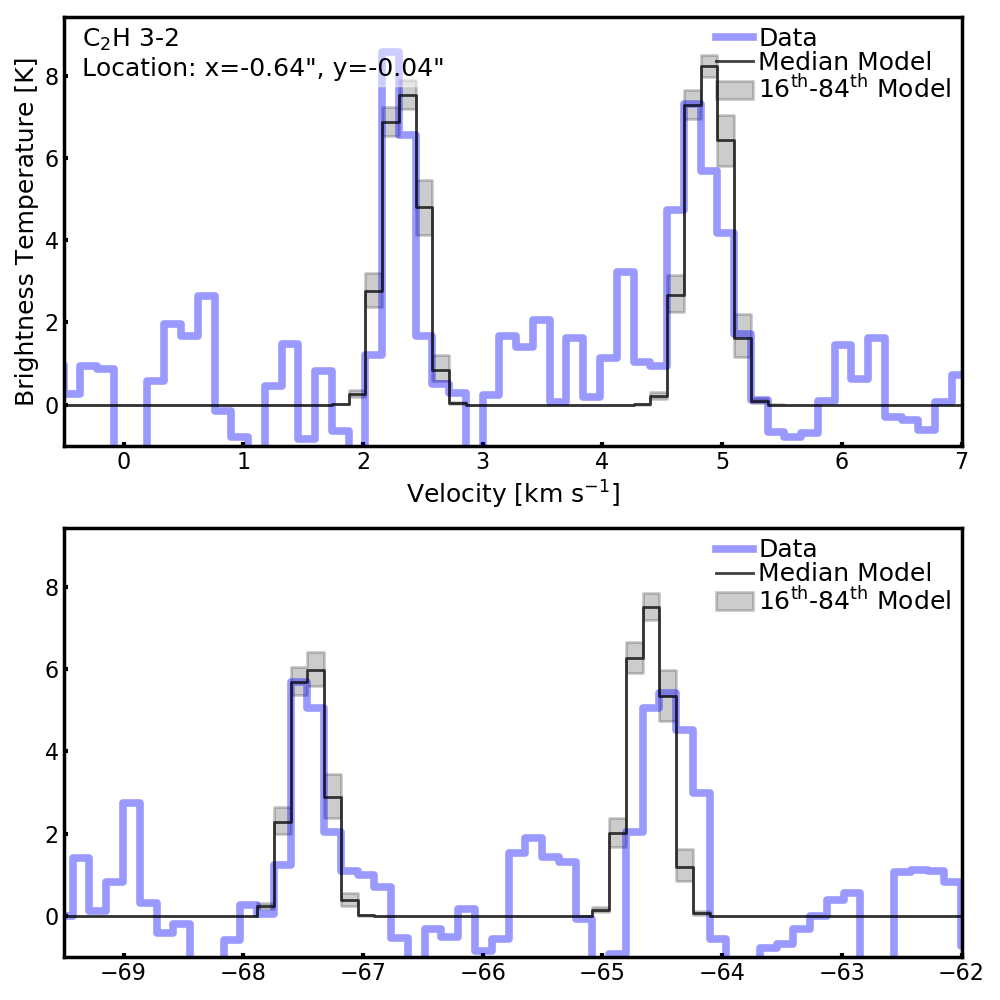}
    }
\caption{Example C$_2$H 3--2 and C$_2$H (N=3--2, J=5/2-3/2) hyperfine fits toward J1100-7619 within a specific pixel.  The location of the pixel is given in arcsec along the top left of each panel.  \textit{Left:} The 16$^\mathrm{th}$ percentile, median, and 84$^\mathrm{th}$ percentile \texttt{emcee} chains for each of the four model parameters (from top row to bottom row, they are: $\hat{A}^*$, $\hat{\tau}^*$, $\hat{V}_\mathrm{LSR}$, and $\hat{\Delta V}_\mathrm{line}$).  \textit{Right:} The median and percentile hyperfine spectrum fits for C$_2$H 3--2 (top of each panel) and C$_2$H (N=3--2, J=5/2-3/2) (bottom of each panel).
\label{fig_exhyperfits_C2H}}
\end{figure*}

\begin{figure*}
\centering
\resizebox{0.925\hsize}{!}{
    \includegraphics[]{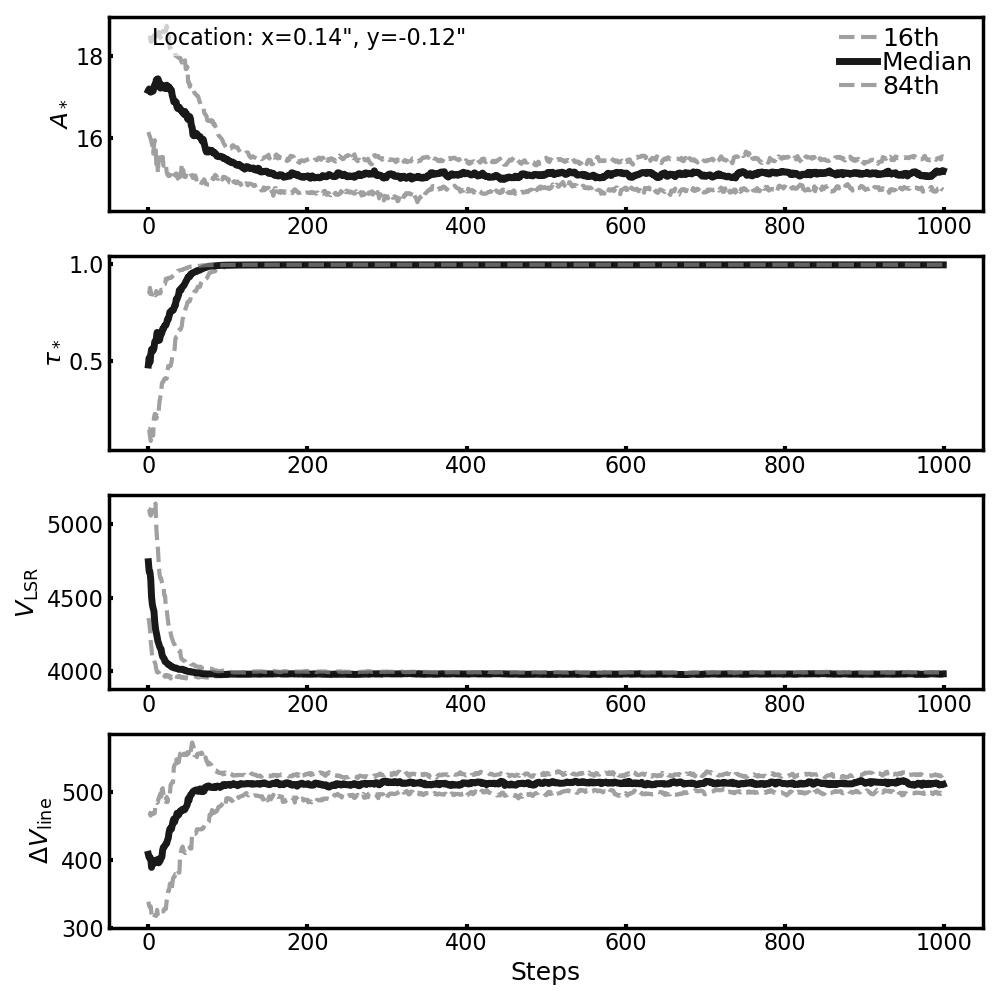}
    \includegraphics[]{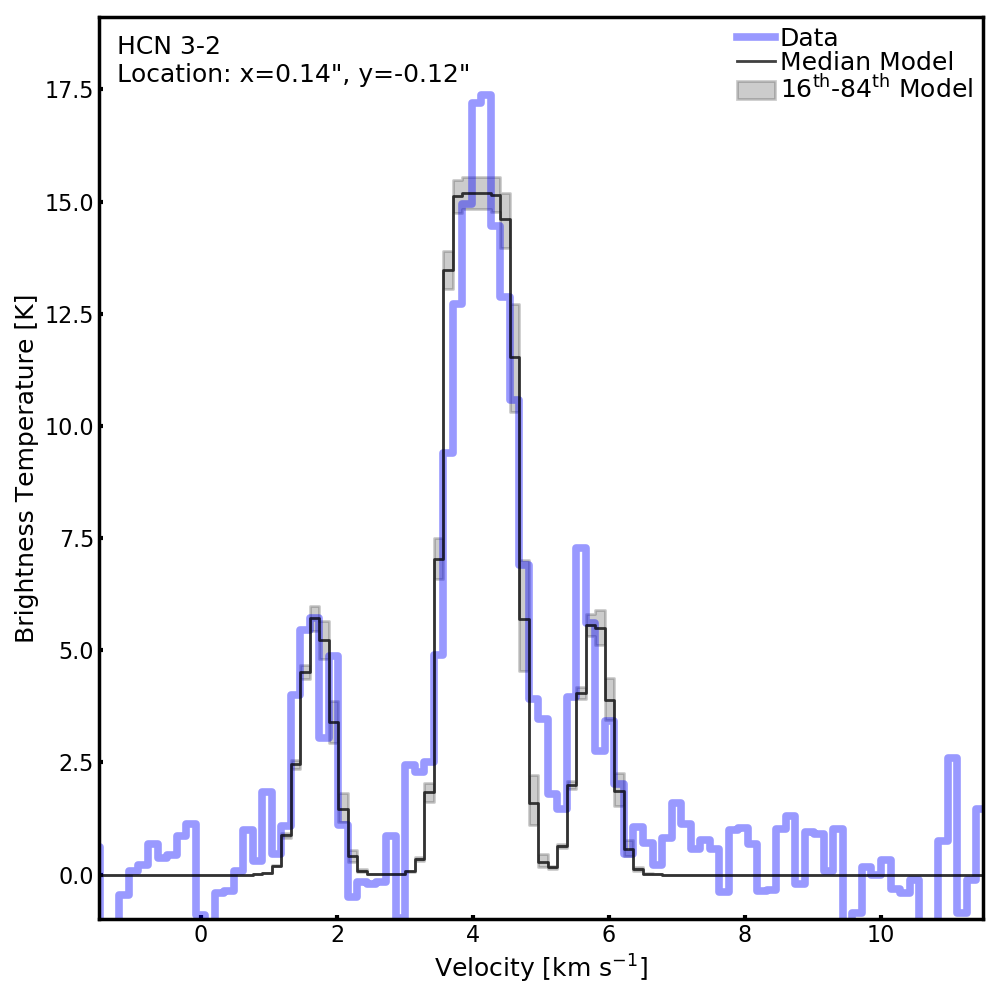}
    }
\resizebox{0.925\hsize}{!}{
    \includegraphics[trim=0pt 0pt 0pt -50pt, clip]{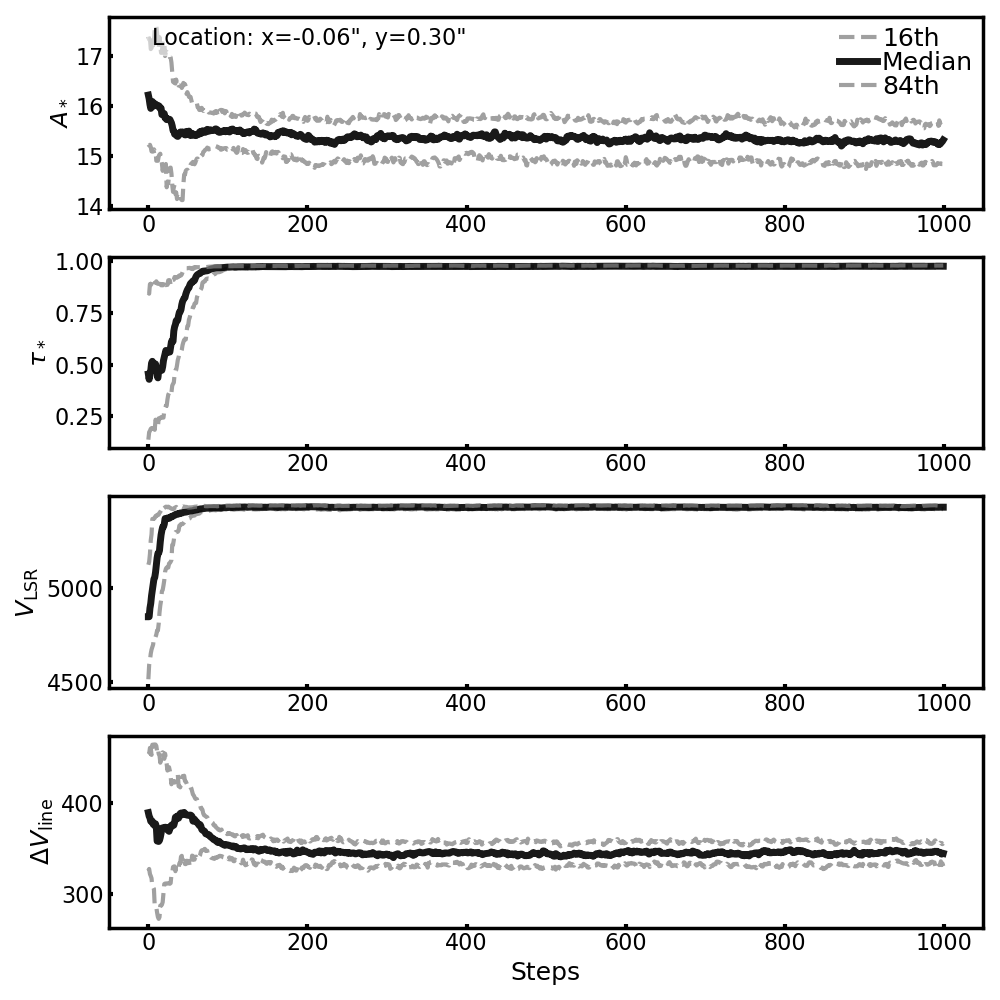}
    \includegraphics[trim=0pt 0pt 0pt -50pt, clip]{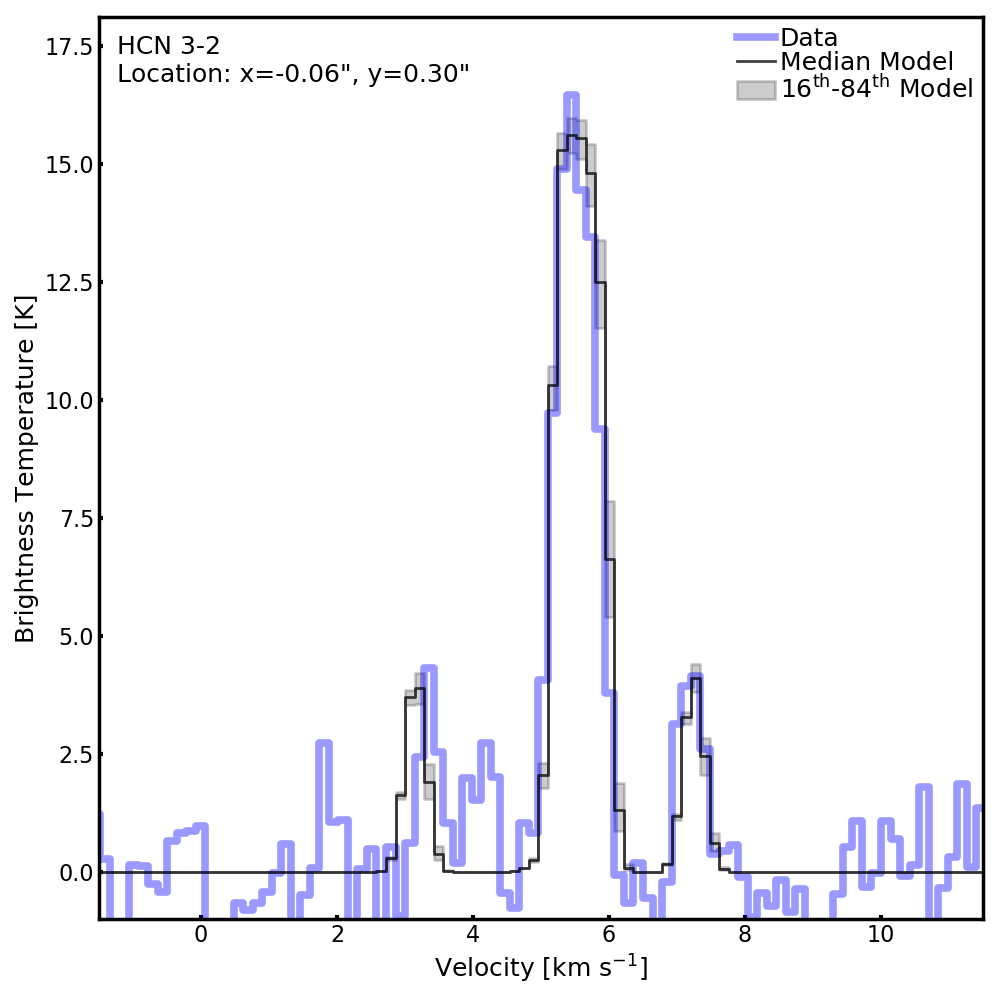}
    }
\caption{Example HCN 3--2 hyperfine fits toward J1100-7619 within a specific pixel.  The location of the pixel is given in arcsec along the top left of each panel.  \textit{Left:} The 16$^\mathrm{th}$ percentile, median, and 84$^\mathrm{th}$ percentile \texttt{emcee} chains for each of the four model parameters (from top row to bottom row, they are: $\hat{A}^*$, $\hat{\tau}^*$, $\hat{V}_\mathrm{LSR}$, and $\hat{\Delta V}_\mathrm{line}$).  \textit{Right:} The median and percentile hyperfine spectrum fits.
\label{fig_exhyperfits_HCN}}
\end{figure*}

\subsection{Pixel-by-Pixel Hyperfine Fits}

We estimate excitation temperatures, column densities, and optical depths for C$_2$H and HCN in the disk J1100-7619 using a pixel-by-pixel hyperfine fitting procedure (Section~\ref{sec_analysis_hyperfine}, Section~\ref{sec_results_j1100_fits}).  This procedure is based on the work of~\cite{cite_bergneretal2019}, which in turn was adapted from the fitting procedure of~\cite{cite_estallelaetal2017} and the calculations of~\cite{cite_mangumetal2015}.  By fitting to the spectrum through each pixel, we reduce line blending of the hyperfine components of the spectra, which allows for tighter constraints on the fit.  We also avoid systematic errors due to changes in line width, line center, and line peak as a function of disk location.

The model assumes that the molecule is in local thermodynamic equilibrium (LTE), that the line width is the same for each hyperfine component, and that the excitation conditions are homogeneous along the line of sight within the given region.  We fit the same model across both C$_2$H 3--2 and C$_2$H (N=3--2, J=5/2-3/2), because the additional emission from C$_2$H (N=3--2, J=5/2-3/2) further improves our model constraints for C$_2$H.  Note that more hyperfine components exist for C$_2$H 3--2 beyond those listed in Table~\ref{table_mol}; however, those unlisted hyperfine components have very weak line intensities, and likely would not significantly affect our spectrum fits if we included them.

The hyperfine model contains four parameters: an amplitude ($\hat{A}^*$), a transformed optical depth for the main hyperfine component ($\hat{\tau}^*_\mathrm{m}$), a line width ($\hat{\Delta V}_\mathrm{line}$), and a central velocity reference point ($\hat{V}_\mathrm{LSR}$).  Note that we use the $^\wedge$ symbol to differentiate these model parameters from constants and other quantities in the equations.  The amplitude and transformed optical depth parameters are directly related to the excitation temperature, column density, and total optical depth of the hyperfine line at that pixel.

We now write down the final equations used for this procedure.  First, the spectrum model is in brightness temperature units and is a function of velocity.  We denote this model as $\hat{H}\{V_\mathrm{k}\}$, where we again use the $^\wedge$ symbol to differentiate between the model spectrum $\hat{H}$ and the actual spectrum $H$ later on.

\begin{align}
    \hat{H}\{V_\mathrm{k}\} &= \frac{\hat{A}^*}{\hat{\tau}_\mathrm{m}^*} \bigg(1 - \exp(-\tau_\mathrm{k}) \bigg),
\label{eq_Hk}
\end{align}

\noindent where $V_\mathrm{k}$ is the central velocity of channel $k$.  $\hat{A}^*$ is the amplitude parameter (in brightness temperature units) and $\hat{\tau}_\mathrm{m}^*$ is the transformed optical depth parameter (bounded such that $0 < \hat{\tau}_\mathrm{m}^* < 1$).  $\tau_\mathrm{k}$ is the optical depth of this $k^\mathrm{th}$ channel, summed over all $L$ hyperfine components:

\begin{align}
    \tau_\mathrm{k} &= \sum_\mathrm{i}^\mathrm{L} \frac{\tau_\mathrm{m} (S_\mathrm{i}\mu^2/S_\mathrm{m}\mu^2) G_\mathrm{i}\{V_\mathrm{k}\}}{\Delta V_\mathrm{chan}},
\label{eq_tk}
\end{align}

\noindent where $\tau_\mathrm{m}$ is the optical depth of the main hyperfine component, calculated as:

\begin{equation}
    \tau_\mathrm{m} = 
    \begin{cases}
        -\ln (1 - \hat{\tau}_\mathrm{m}^*)   &\text{if } \hat{\tau}_\mathrm{m}^* > 0.01 \\
        \hat{\tau}_\mathrm{m}^* + \frac{(\hat{\tau}_\mathrm{m}^*)^2}{2} + \frac{(\hat{\tau}_\mathrm{m}^*)^3}{3}  &\text{if }\hat{\tau}_\mathrm{m}^* \leq 0.01,
    \end{cases}
\end{equation}

\noindent and $(S_\mathrm{i}\mu^2/S_\mathrm{m}\mu^2)$ is the strength of the $i^\mathrm{th}$ hyperfine component relative to the main hyperfine component (Table~\ref{table_mol}).  $\Delta V_\mathrm{chan}$ is the width of each velocity channel (identical across all channels).  $G_\mathrm{i}\{V_\mathrm{k}\}$ is the value of the line profile for the $i^\mathrm{th}$ hyperfine component at velocity channel $k$, assumed to be:

\begin{equation}
    G_\mathrm{i}\{V_\mathrm{k}\} = 
        \hat{\Delta V}_\mathrm{line} \sqrt{\frac{\pi}{16\ln(2)}} \bigg(  \text{erf}(x^+_\mathrm{i,k}) - \text{erf}(x^-_\mathrm{i,k}) \bigg),
\end{equation}

\noindent where $\hat{\Delta V}_\mathrm{line}$ is the line width parameter of the model, assumed to be the same for each hyperfine component, and erf($z$) represents the error function for input $z$.  If $|x^+_\mathrm{i,k} - x^-_\mathrm{i,k}| < 10^{-4}$, then $G_\mathrm{i}\{V_\mathrm{k}\}$ is instead approximated as:

\begin{align}
    G_\mathrm{i}\{V_\mathrm{k}\} \approx & \frac{\hat{\Delta V}_\mathrm{line}}{2 \sqrt{\ln(2)} }
    \bigg(x^+_\mathrm{i,k} - x^-_\mathrm{i,k}\bigg) \nonumber \\
    &\times \exp \bigg(-\bigg(\frac{x^+_\mathrm{i,k} + x^-_\mathrm{i,k}}{2}\bigg)^2 \bigg).
\end{align}

\noindent $x^+_\mathrm{i,k}$ and $x^-_\mathrm{i,k}$ are calculated as:

\begin{align}
    x^\pm_\mathrm{i,k} &= 2\sqrt{\ln(2)} \bigg( \frac{V_\mathrm{k} \pm (\Delta V_\mathrm{chan}/2) - \hat{V}_\mathrm{LSR} - \Delta V_\mathrm{i}}{\hat{\Delta V}_\mathrm{line}} \bigg),
\end{align}

\noindent where $\hat{V}_\mathrm{LSR}$ is the reference point parameter of the model.  $\Delta V_\mathrm{i}$ is the shift in velocity between the $i^\mathrm{th}$ and main hyperfine components and can be calculated from the frequencies (Table~\ref{table_mol}).

We estimated the four parameters of the model (the amplitude parameter $\hat{A}^*$, the transformed optical depth parameter $\hat{\tau}_\mathrm{m}^*$, the line width parameter $\hat{\Delta V}_\mathrm{line}$, and the velocity reference parameter $\hat{V}_\mathrm{LSR}$) per pixel in the image using the \texttt{EnsembleSampler} function from the \texttt{emcee} package~\citep{cite_emcee}.  The log-likelihood function used for \texttt{emcee} included error in the resolved peaks (i.e., 3 peaks for HCN 3--2 and 4 peaks across C$_2$H 3--2 and C$_2$H 3--2) and error across all model values:

\begin{equation}
    \ln \mathcal{L} = 
        -\sum_\mathrm{k} \bigg( \frac{H\{V_\mathrm{k}\}-\hat{H}\{V_\mathrm{k}\}}{\epsilon_\mathrm{chan}} \bigg)^2 - (\epsilon_\mathrm{peak} \times 100),
\end{equation}

\noindent where $H\{V_\mathrm{k}\}$ and $\hat{H}\{V_\mathrm{k}\}$ are the actual and model spectra, respectively, at velocity channel $k$.  $\epsilon_\mathrm{chan}$ is the channel rms converted to brightness temperature.  $\epsilon_\mathrm{peak}$ is the summed relative error in the peaks of the resolved hyperfine components (where 'relative error' is the absolute difference in the actual value and model value, divided by the actual value) and applies an additional penalty when the model and actual spectrum peaks are misaligned.  The actual spectrum is converted from flux density units (i.e., units of [power per distance$^2$ per frequency]) to brightness temperature units using~\citep[e.g.,][]{cite_textbooknrao}:

\begin{align}
T &= \bigg(\frac{\mathcal{F} c_0^2}{2 k_\mathrm{B} \nu_0^2 \Omega_\mathrm{area}}\bigg),
\label{eq_fluxtoT}
\end{align}

\noindent where $\mathcal{F}$ is the flux density, $c_0$ is the speed of light, $k_B$ is the Boltzmann constant, and $\nu_0$ is the frequency of the main hyperfine component.  $\Omega_\mathrm{area}$ is the solid angle of the emitting area (in this case, the solid angle of the pixel).  Note that in practice, we calculate $\mathcal{F}$ by converting our observed fluxes from units of Jy/beam, to Jy/pixel, and finally to (W m$^{-2}$ Hz$^{-1}$/pixel), before applying Equation~\ref{eq_fluxtoT} in S.I. units.

We used 100 \texttt{emcee} walkers per pixel and walked them for 1000 steps.  Initial estimates of each walker for $\hat{A}^*$, $\hat{\tau}_\mathrm{m}^*$, $\hat{\Delta V}_\mathrm{line}$, and $\hat{V}_\mathrm{LSR}$ were selected from Uniform distributions in the ranges of [peak of the emission in brightness temperature units $\pm$ 10\%], (0, 1), [the disk's systemic velocity $\pm$ 10\%], and [300m s$^{-1}$, 500m s$^{-1}$], respectively.  We disregarded the first 950 steps for each walker to account for the burn-in phase, creating a sampling distribution of $100 \times 50$ for each parameter.  We took the median parameter of each sampling distribution, $\hat{A}^{*(50)}$, $\hat{\tau}^{*(50)}$, $\hat{\Delta V}^{(50)}_\mathrm{line}$, and $\hat{V}^{(50)}_\mathrm{LSR}$, to be the final estimate of each pixel's four parameters.  We used the difference in the 16$^\mathrm{th}$ and 84$^\mathrm{th}$ percentile values relative to the median values as estimates of the lower and upper errors, respectively.

Figures~\ref{fig_exhyperfits_C2H} and~\ref{fig_exhyperfits_HCN} show example hyperfine \texttt{emcee} chains and fits for the C$_2$H and HCN molecular lines, respectively.  The bottom panel of Figure~\ref{fig_exhyperfits_HCN} illustrates the importance of the HCN 3--2 satellite components discussed previously in anchoring the fit.  This same panel also illustrates the intrinsic limitations of the underlying model at high optical depths.  We discuss these limitations later in this section.

Adapting from~\cite{cite_mangumetal2015},~\cite{cite_estallelaetal2017} and~\cite{cite_bergneretal2019}, we estimate the excitation temperature $T_\mathrm{ex}$ from the following series of calculations:

\begin{align}
\mathcal{T}_\mathrm{ex} &= \frac{\hat{A}^{*(50)}}{\hat{\tau}^{(50)} f} + \mathcal{B}_\mathrm{cont} \label{eq_Texset_Tex1} \\
T_\mathrm{ex} &= \frac{h \nu_0}{k_\mathrm{B} \ln(h \nu_0 / (k_\mathrm{B} \mathcal{T}_\mathrm{ex}) + 1)},  \label{eq_Texset_Tex2}
\end{align}

\noindent where Equation~\ref{eq_Texset_Tex1} determines the Planck-corrected excitation temperature from the model parameters, and Equation~\ref{eq_Texset_Tex2} calculates the excitation temperature in brightness temperature units from the Planck-corrected excitation temperature.  We calculate $\mathcal{B}_\mathrm{cont}$ from Equation~\ref{eq_fluxtoT}, using the dust continuum flux at that pixel for $\mathcal{F}$ (set to be a very small number if the continuum at that pixel is 0) and the solid angle of the pixel for $\Omega_\mathrm{area}$.  $h$ is the Planck constant, $k_\mathrm{B}$ is the Boltzmann constant, and $f$ is the filling factor (assumed to be $f=1$ for J1100-7619).  Note that unlike the previous studies, we do not calculate the Planck-corrected temperature of the background continuum prior to Equation~\ref{eq_Texset_Tex1}, as this is an over-correction of the background brightness temperature.  The effect of this change is negligible at these wavelengths, continuum fluxes, and level of uncertainty.

We then estimate the total optical depth $\tau$ as:

\begin{align}
\tau &= -\frac{\ln(1 - \hat{\tau}^{*(50)})}{R_\mathrm{m}},
\end{align}

\noindent where $R_\mathrm{m}$ here is the strength of the \textit{main} hyperfine transition relative to other same-level J transitions.  $R_\mathrm{i}$ is generally calculated for any hyperfine transition as $R_\mathrm{i} = \frac{(S_\mathrm{i}\mu^2)}{\sum_\mathrm{Jn=Ji} (S_\mathrm{n}\mu^2)}$.  As an example for a main hyperfine line, $R_\mathrm{m}=0.527$ for C$_2$H (N=3--2, J=7/2-5/2) is calculated from $\frac{2.2870}{2.2870 + 1.7110}$, which is the line intensities of the main line (N=3--2, J=7/2-5/2) divided by the sum of the line intensities for all other J=7/2-5/2 lines.

Finally, we calculate $N_\mathrm{tot}$ as~\citep{cite_mangumetal2015}:

\begin{align}
\overline{N}_\mathrm{tot} =& \bigg( \frac{3 h Q\{T_\mathrm{ex}\}}{8 \pi^3 (S_\mathrm{m}\mu^2)} \bigg) \bigg( \frac{\exp(E_\mathrm{u}/T_\mathrm{ex})}{\exp(h\nu_\mathrm{0}/(k_\mathrm{B} T_\mathrm{ex})) - 1} \bigg) \nonumber \\
&\times \bigg(\frac{\int_\mathrm{k} \hat{H}_\mathrm{m}dV}{\mathcal{T}_\mathrm{ex} - \mathcal{B}_\mathrm{cont}} \bigg)
\bigg(\frac{\tau R_\mathrm{m}}{(1 - \exp(-\tau R_\mathrm{m}))}\bigg),
\label{eq_Ntot}
\end{align}

\noindent where $Q\{T_\mathrm{ex}\}$ is the value of the partition function at temperature $T_\mathrm{ex}$~\citep[taken from the CDMS database,][]{cite_cdms2016}.  $S_\mathrm{m}\mu^2$ and $E_\mathrm{u}$ are the absolute line intensity and upper level energy, respectively, of the main hyperfine component (Table~\ref{table_mol}).  Note that the column density equation in~\citep{cite_mangumetal2015} explicitly includes the upper degeneracy level $g_\mathrm{u}$.  For our calculations, that factor of degeneracy is already implicitly included in the CDMS values for $S\mu^2$ (Table~\ref{table_mol}).  Finally, $\int_\mathrm{k} \hat{H}_\mathrm{m}dV$ is the integral of \textit{only} the main hyperfine component of the model across all velocity channels.  $\hat{H}_\mathrm{m}$ can be calculated by using Equation~\ref{eq_tk} for only $i=m$, and then plugging that result into Equation~\ref{eq_Hk}.  Note that any i$^\mathrm{th}$ hyperfine component can be used in this way to determine $N_\mathrm{tot}$; we use the main hyperfine component here because it has the strongest emission.

\subsection{Weighted Radial Profiles}

To generate the weighted radial profiles of Figure~\ref{fig_j1100_hypprof}, we first define $\sigma_\mathrm{pix,Q}^+ = \sqrt{(Q_{84} - Q_{50})^2 + (Q_{50} \times f_{50})^2}$ and $\sigma_\mathrm{pix,Q}^- = \sqrt{(Q_{16} - Q_{50})^2 + (Q_{50} \times f_{50})^2}$ as the upper and lower errors, respectively, for each pixel and each measurement.  Here $Q$ refers to the measured quantity (e.g., the excitation temperature) and the subscripts 16, 50, and 84 refer to the 16$^\mathrm{th}$, median, and 84$^\mathrm{th}$ percentiles, respectively.  $f_{50}$ is the difference between the integrated fluxes of the median and actual spectrum, relative to the integrated flux of the actual spectrum.

we next extract all valid pixels within each deprojected annulus around the disk center.  ``Valid" pixels are those that fulfill all of the following three criteria:  (1) The signal-to-noise of the faintest resolved hyperfine component is at least three (there are three total resolved components for HCN and four for C$_2$H).  (2) The relative errors for all four parameters for this pixel ($\hat{A}^{*(50)}$, $\hat{\tau}^{*(50)}$, $\hat{\Delta V}^{(50)}_\mathrm{line}$, and $\hat{V}^{(50)}_\mathrm{LSR}$) are $\leq0.20$.  Here the relative error is the maximum over $[|P_{16} - P_{50}|/P_{50}, |P_{84} - P_{50}|/P_{50}|]$, where $P$ refers to each parameter.  (3) $f_{50} \leq 0.20.$

We then perform a weighted average on the valid pixels within each annulus to produce the profiles of Figure~\ref{fig_j1100_hypprof}.  The weight per pixel and per measurement is $(1/\sigma^2)$, where $\sigma = \sqrt{(\sigma_\mathrm{pix,Q}^+)^2 + (\sigma_\mathrm{pix,Q}^-)^2}$.  The upper and lower shaded error ranges for the measurements in Figure~\ref{fig_j1100_hypprof} are the rms across all $\sigma_\mathrm{pix,Q}^+$ and $\sigma_\mathrm{pix,Q}^-$, respectively, for each annulus.

\subsection{Intrinsic Sources of Error}

We note that the pixel-by-pixel hyperfine fits, and thus the weighted radial profiles, have significant intrinsic uncertainties.  First, the C$_2$H and HCN pixel-by-pixel spectra toward J1100-7619 have relatively small line widths ($\sim$0.2-0.5km s$^{-1}$) because of the disk's low inclination angle.  This fact, coupled with the relatively coarse velocity resolution of the data (0.14km s$^{-1}$), means that the hyperfine procedure has a limited number of data points to fit for each hyperfine component.  Second, the pixel-by-pixel fitting procedure does not account for uncertainties due to correlations between neighboring pixels.  Third, the error in the fits is intrinsically large.  This is particularly true for HCN, where the procedure is dependent on the fainter satellite hyperfine components as an 'anchor' for the overall fit.  Finally, the underlying column density equation (Equation~\ref{eq_Ntot}, Equation~\ref{eq_Ntotb}) makes Gaussian assumptions about the molecular line shape that break down as the optical depth increases.  At the optical depths we are encountering for C$_2$H and HCN, uncertainties intrinsic to the model may become as large as $\sim$40\%.  We thus stress that Figure~\ref{fig_j1100_hypprof} should be interpreted as rough constraints, rather than precise measurements, of the values, as should the column density estimates determined in Section~\ref{sec_results_Ntot}.



\section{Column Density Estimates}
\label{sec_appendix_Ntot}

\subsection{Methodology}

Assuming an excitation temperature $T_\mathrm{ex}$ and total optical depth $\tau$ for a hyperfine line, we can estimate the column density $\overline{N}_\mathrm{tot}$, which is averaged over the disk region, using the total emission across all hyperfine components within that disk region.  We use the column density equation adapted from~\cite{cite_mangumetal2015}, which presents $\overline{N}_\mathrm{tot}$ as measured from a specific $i^\mathrm{th}$ hyperfine emission component:

\begin{align}
\overline{N}_\mathrm{tot} =& \bigg( \frac{3 h Q\{T_\mathrm{ex}\}}{8 \pi^3 (S_\mathrm{i}\mu^2)} \bigg) \bigg( \frac{\exp(E_\mathrm{u}/T_\mathrm{ex})}{\exp(h\nu_\mathrm{i}/(k_\mathrm{B} T_\mathrm{ex})) - 1} \bigg) \nonumber \\
&\times \bigg(\frac{\int_\mathrm{k} H_\mathrm{i}dV}{\mathcal{T}_\mathrm{ex} - \mathcal{B}_\mathrm{cont}} \bigg)
\bigg(\frac{\tau R_\mathrm{i}}{(1 - \exp(-\tau R_\mathrm{i}))}\bigg),
\label{eq_Ntotb}
\end{align}

\noindent where $h$ is the Planck constant, $k_\mathrm{B}$ is the Boltzmann constant, and $Q\{T_\mathrm{ex}\}$ is the value of the partition function at temperature $T_\mathrm{ex}$~\citep[taken from the CDMS database,][]{cite_cdms2016}. $\nu_\mathrm{i}$, $S_\mathrm{i}\mu^2$, and $E_\mathrm{u}$ are the frequency, absolute line intensity, and upper level energy, respectively, of the $i^\mathrm{th}$ hyperfine component.  $R_\mathrm{i}$ is the line intensity of the $i^\mathrm{th}$ component relative to the summed line intensities of all other same J-level transitions (the equation and an example calculation are in Appendix~\ref{sec_appendix_hyperfine}).  $\mathcal{T}_\mathrm{ex}$ and $\mathcal{B}_\mathrm{cont}$ are the Planck excitation temperature and continuum brightness temperature, respectively, calculated using the conversion procedures in Appendix~\ref{sec_appendix_hyperfine}.  Finally, $\int_\mathrm{k} H_\mathrm{i}dV$ is the integral across $k$ velocity channels of \textit{only} the $i^\mathrm{th}$ hyperfine component of emission in brightness temperature units.  This quantity is converted from flux density units using the conversion procedure in Appendix~\ref{sec_appendix_hyperfine}.  Note that the column density equation in~\cite{cite_mangumetal2015} explicitly includes a factor of the upper degeneracy level $g_\mathrm{u}$.  For our calculations, that factor of degeneracy is already included implicitly within the CDMS values for $S_\mathrm{i}\mu^2$.

Since we are assuming an excitation temperature $T_\mathrm{ex}$ and total optical depth $\tau$, Equation~\ref{eq_Ntotb} reduces to the following:

\begin{align}
\longrightarrow \overline{N}_\mathrm{tot} =& D_\mathrm{i} \{T_\mathrm{ex}, \tau\} \times \int_\mathrm{k} H_\mathrm{i}dV,
\label{eq_Ntotb2}
\end{align}

\noindent where $D_\mathrm{i} \{T_\mathrm{ex}, \tau\}$ is a term dependent on $T_\mathrm{ex}$, $\tau$, and the molecular line characteristics of \textit{only} the $i^\mathrm{th}$ hyperfine component.

When the hyperfine components are blended, the quantity $(\int_\mathrm{k} H_\mathrm{i}dV)$ is not known.  We only know the total flux density integrated across all hyperfine components, which we convert to brightness temperature units $(\int_\mathrm{k} H_\mathrm{tot}dV)$ using the brightness temperature conversion in Appendix~\ref{sec_appendix_hyperfine}.  Even so, we have enough information to set up a system of ($L$+1) equations with ($L$+1) unknowns:

\begin{align}
\int_\mathrm{k} H_\mathrm{tot}dV =& \sum_{i=1}^L \int_\mathrm{k} H_\mathrm{i}dV, \nonumber \\
\overline{N}_\mathrm{tot} =& D_1\{T_\mathrm{ex}, \tau\} \times \int_\mathrm{k} H_\mathrm{1}dV, \nonumber \\
\overline{N}_\mathrm{tot} =& D_2\{T_\mathrm{ex}, \tau\} \times \int_\mathrm{k} H_\mathrm{2}dV, \nonumber \\
    &\vdots \nonumber \\
\overline{N}_\mathrm{tot} =& D_\mathrm{L}\{T_\mathrm{ex}, \tau\} \times \int_\mathrm{k} H_\mathrm{L}dV,
\end{align}

\noindent where $L$ is the total number of hyperfine components for this hyperfine line (i.e., $L$=2 for C$_2$H (N=3--2, J=7/2-5/2), $L$=2 for C$_2$H (N=3--2, J=5/2-3/2), and $L$=6 for HCN 3--2).  Note that while we combined the C$_2$H (N=3--2, J=7/2-5/2) and C$_2$H (N=3--2, J=5/2-3/2) hyperfine spectra in Section~\ref{sec_analysis_hyperfine} and Appendix~\ref{sec_appendix_hyperfine}, we treat the lines separately for this general analysis and use the C$_2$H (N=3--2, J=5/2-3/2) line to check consistency with the C$_2$H (N=3--2, J=7/2-5/2) results.

The unknowns of this system of equations are $(\int_\mathrm{k} H_\mathrm{i}dV)$ for each $i^\mathrm{th}$ hyperfine component and $\overline{N}_\mathrm{tot}$.  The latter equations can be subtracted from each other to eliminate $\overline{N}_\mathrm{tot}$; e.g. $(D_\mathrm{i}\{T_\mathrm{ex}, \tau\} \times \int_\mathrm{k} H_\mathrm{i}dV) - (D_\mathrm{j}\{T_\mathrm{ex}, \tau\} \times \int_\mathrm{k} H_\mathrm{j}dV) = 0$ for hyperfine components $i$ and $j$.  These equations can then be written in the form of \textbf{Mx} = \textbf{b}, where \textbf{M} is the matrix (dimensions $L \times L$) of known coefficients:

\[
	\mathbf{M} =
	\left[ {\begin{array}{ccccccc}
		 1 &    1 &     1 &   1 &  \dots & 1 &   1 \\
		 D_1 &     -D_2 &     0 &  0 &   \dots &   0 &  0 \\
		 0 &     D_2 &     -D_3 &  0 &   \dots &   0 &  0 \\
		 0 &     0 &     D_3 &     -D_4 &  \dots & 0 & 0 \\
		 \dots &     \dots &     \dots &     \dots &  \ddots & \vdots & \vdots \\
		 0 &     0 & 0 & \dots & 0 &  D_{L-1} &     -D_L
	\end{array} } \right],
\]

\noindent (where we have temporarily dropped the explicit $\{T_\mathrm{ex}, \tau\}$ notation simply to reduce the width of the matrix text).  \textbf{x} is the vector (dimensions $L \times 1$) of unknown flux components:

\[
	\mathbf{x} =
	\left[ {\begin{array}{c}
		 \int_\mathrm{k} H_\mathrm{1}dV \\
		 \int_\mathrm{k} H_\mathrm{2}dV \\
		 \vdots \\
		 \int_\mathrm{k} H_\mathrm{L}dV
	\end{array} } \right],
\]

\noindent and \textbf{b} is the vector (dimensions $L \times 1$) of equation solutions (which are mostly 0):

\[
	\mathbf{b} =
	\left[ {\begin{array}{c}
		 \int_\mathrm{k} H_\mathrm{tot}dV \\
		 0 \\
		 \vdots \\
		 0
	\end{array} } \right].
\]

\noindent We calculate \textbf{M} and \textbf{b}, and then we use standard linear algebra practices to solve for \textbf{x} (e.g., \texttt{numpy}'s \texttt{linalg.solve} function in Python).  We then plug one of the solved hyperfine flux components from \textbf{x} (e.g., the main hyperfine flux component) back into Equation~\ref{eq_Ntotb2} to recover $\overline{N}_\mathrm{tot}$.

\subsection{Literature Sample}

We now describe the literature samples used to determine the 16$^\mathrm{th}$, median, and 84$^\mathrm{th}$ percentiles of C$_2$H/HCN and DCN/HCN~\citep{cite_huangetal2017, cite_bergneretal2020} in Figure~\ref{fig_diskavgNtot}.

There were a total of seven disks with ALMA detections of both C$_2$H 3--2 and HCN 3--2 emission presented in~\cite{cite_bergneretal2020}.  When estimating the disk-averaged column densities,~\cite{cite_bergneretal2020} assumed optical depths for C$_2$H and HCN of $\sim$1.1 and $\sim$5.8, respectively, and an excitation temperature of 30K.  The 16$^\mathrm{th}$, median, and 84$^\mathrm{th}$ percentiles of these estimates (2.4, 2.6, 4.7) are used in the C$_2$H/HCN panel of Figure~\ref{fig_diskavgNtot}.

There were a total of five disks with ALMA observations of the optically thin lines DCN 3--2 and H$^{13}$CN (J=3--2), and one disk with ALMA observations of DCN 3--2 and HCN 3--2 (assumed to be optically thin within the emitting area), presented in~\cite{cite_huangetal2017}.  These observations were used to estimate the DCN/HCN abundance ratios for all six disks as proxies for the overall D/H ratios.  There were also DCN/HCN column density ratios for three additional disks presented in~\cite{cite_bergneretal2020}.  \cite{cite_huangetal2017} considered excitation temperatures of 15K and 75K, while \cite{cite_bergneretal2020} assumed 30K.  Here we use the 16$^\mathrm{th}$, median, and 84$^\mathrm{th}$ percentiles across the combined 15K and 30K estimates (0.011, 0.018, 0.042) in the DCN/HCN panel of Figure~\ref{fig_diskavgNtot}.  The column density ratios are insensitive to temperature changes at these excitation temperatures~\citep{cite_huangetal2017, cite_bergneretal2020}.


\section{Channel Maps}
\label{sec_appendix_chan}

Channel maps of detected/tentatively detected emission toward FP Tau, J0432+1827, J1100-7619, J1545-3417, and Sz 69 are provided in the following sections.

\subsection{Channel Maps for FP Tau}

\begin{figure*}
\centering
\resizebox{0.99\hsize}{!}{
    \includegraphics{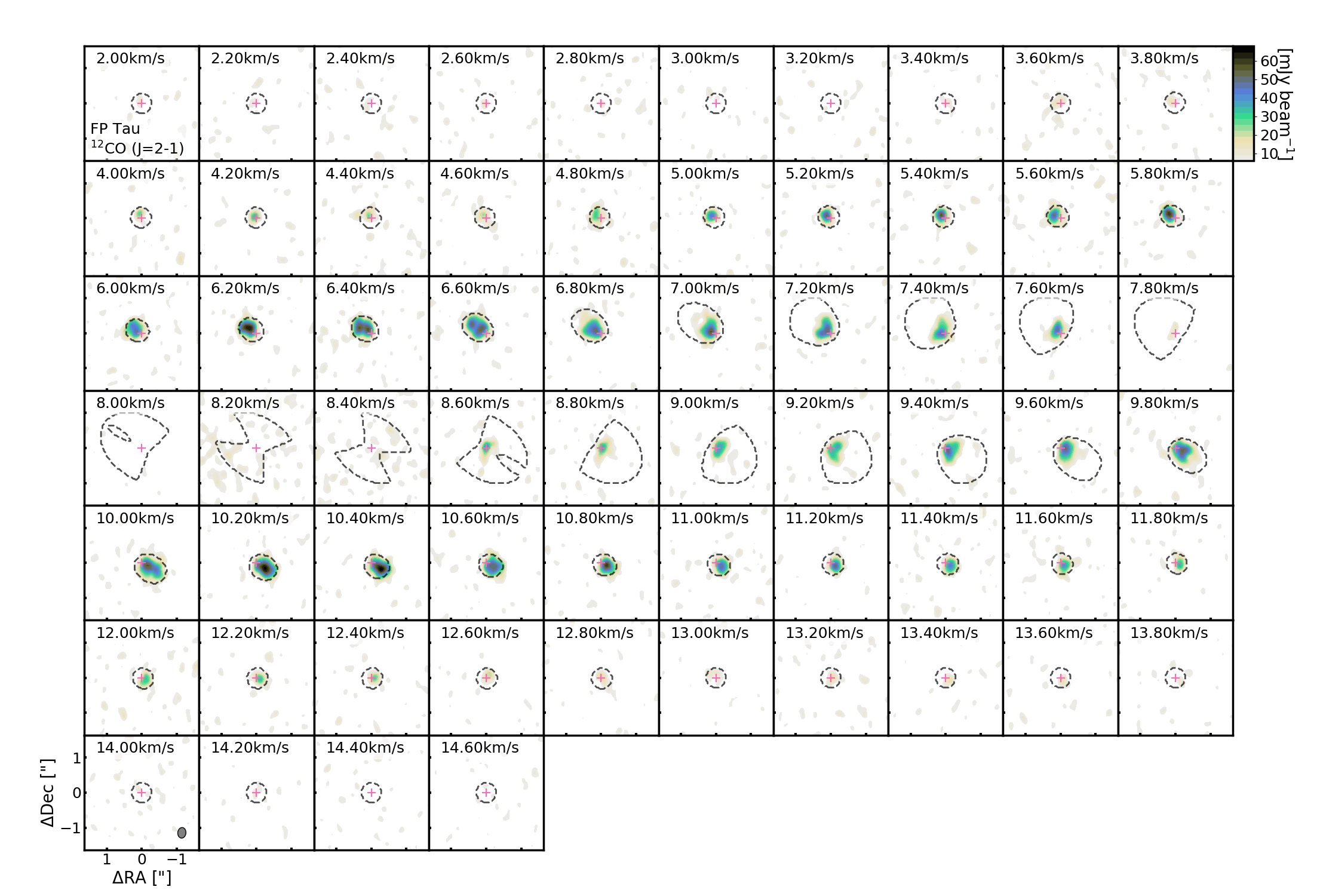}}
\caption{$^{12}$CO toward FP Tau above 2$\sigma$.
\label{figset_fptau_12CO}}
\end{figure*}

\begin{figure*}
\centering
\resizebox{0.99\hsize}{!}{
    \includegraphics{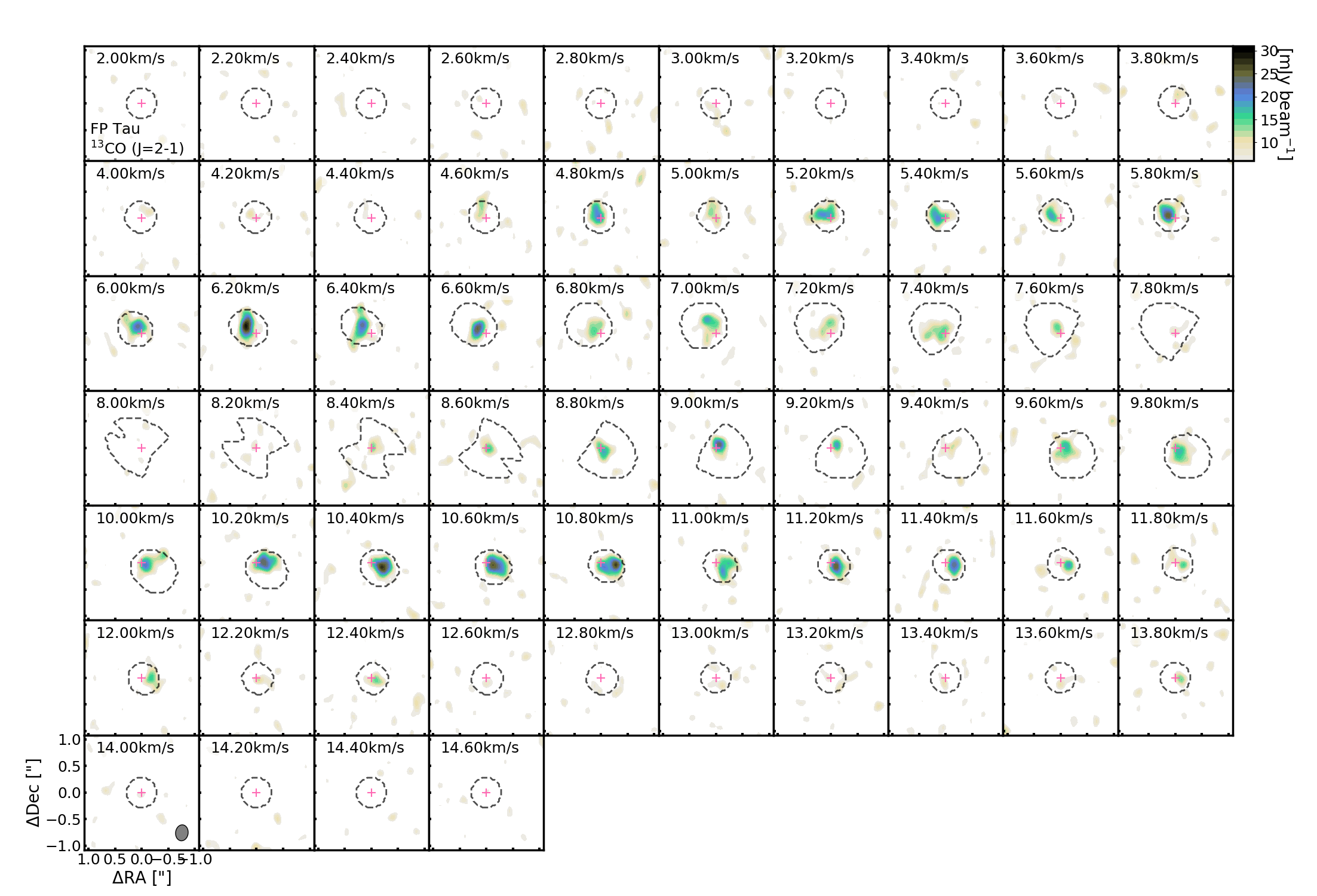}}
\caption{$^{13}$CO toward FP Tau above 2$\sigma$.
\label{figset_fptau_13CO}}
\end{figure*}

\begin{figure*}
\centering
\resizebox{0.99\hsize}{!}{
    \includegraphics{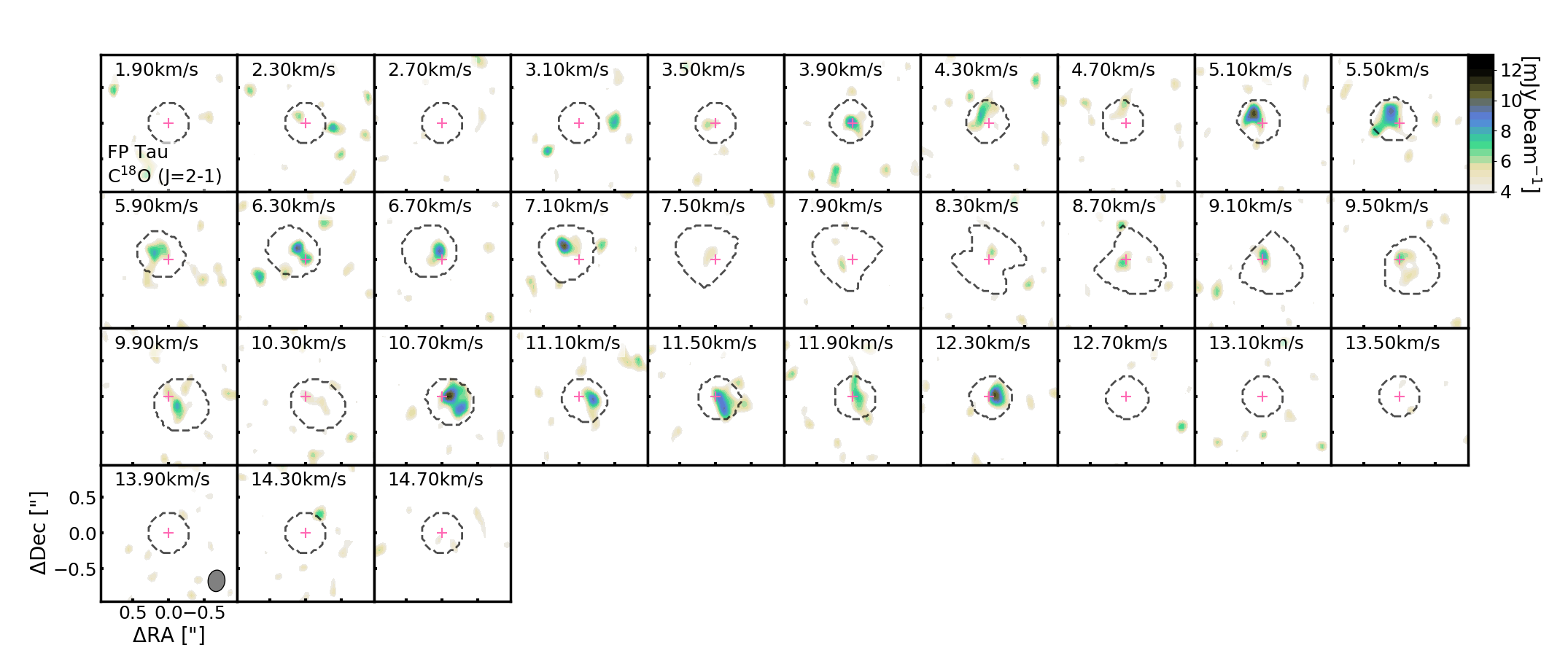}}
\caption{C$^{18}$O toward FP Tau above 2$\sigma$.
\label{figset_fptau_C18O}}
\end{figure*}

\begin{figure*}
\centering
\resizebox{0.99\hsize}{!}{
    \includegraphics{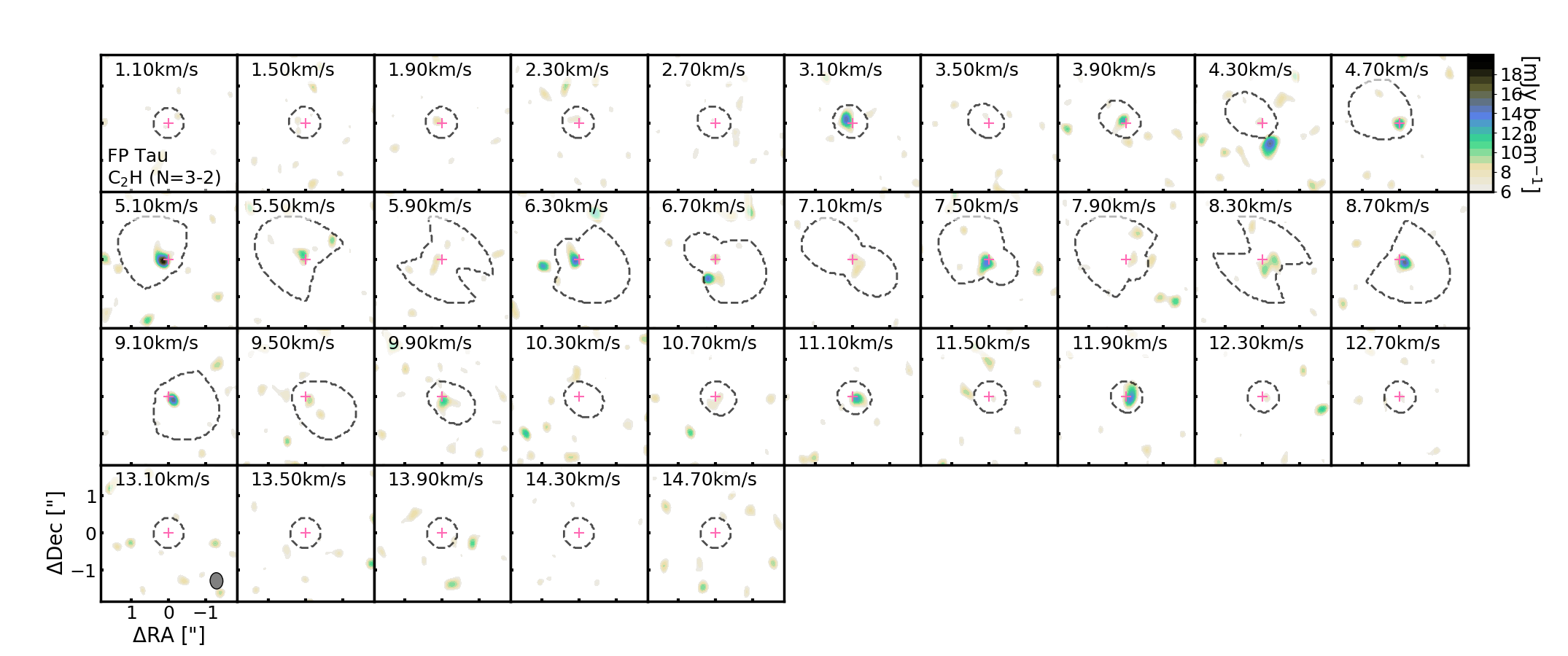}}
\caption{C$_2$H 3--2 toward FP Tau above 2$\sigma$.
\label{figset_fptau_C2H}}
\end{figure*}

\begin{figure*}
\centering
\resizebox{0.99\hsize}{!}{
    \includegraphics{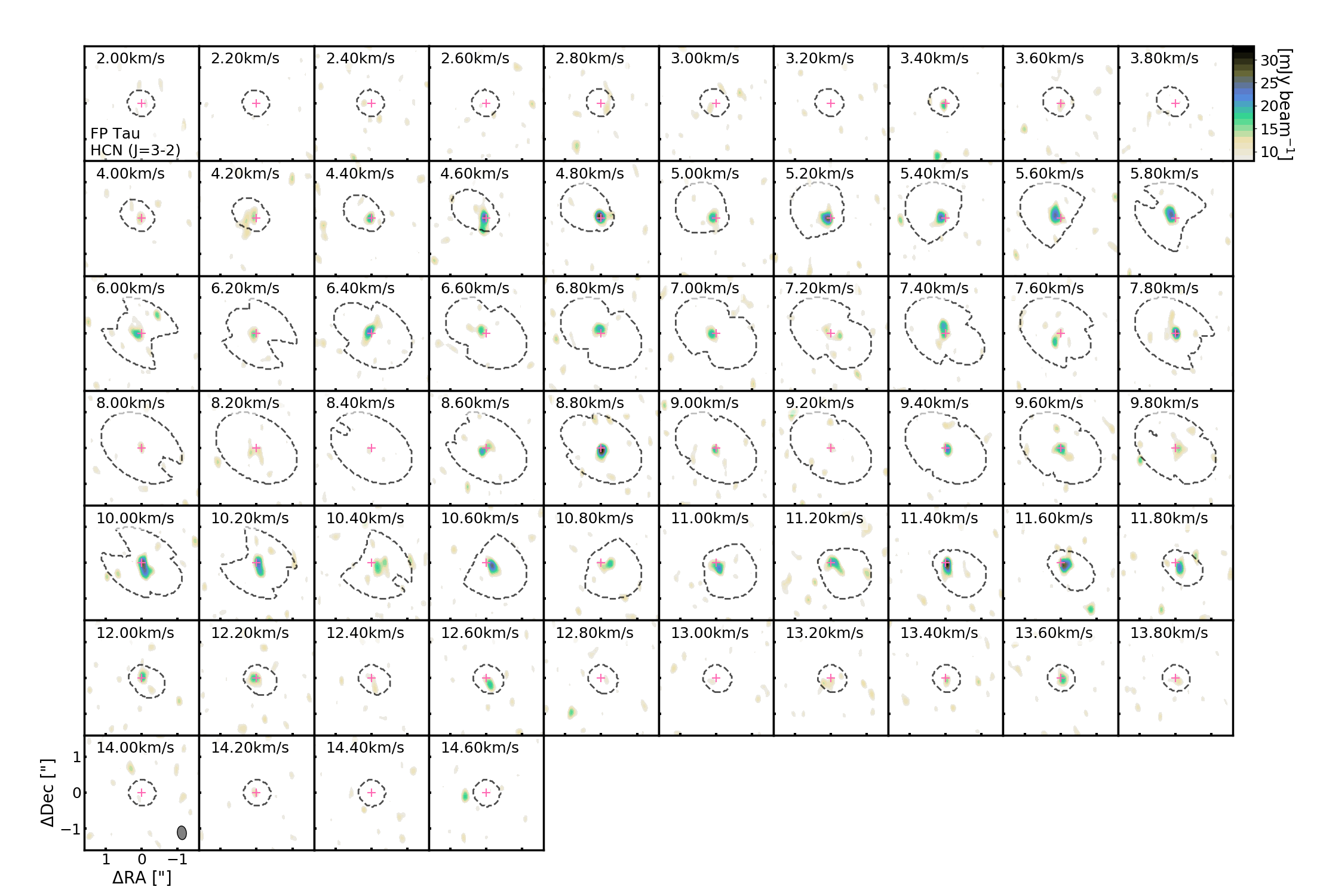}}
\caption{HCN 3--2 toward FP Tau above 2$\sigma$.
\label{figset_fptau_HCN}}
\end{figure*}

\begin{figure*}
\centering
\resizebox{0.99\hsize}{!}{
    \includegraphics{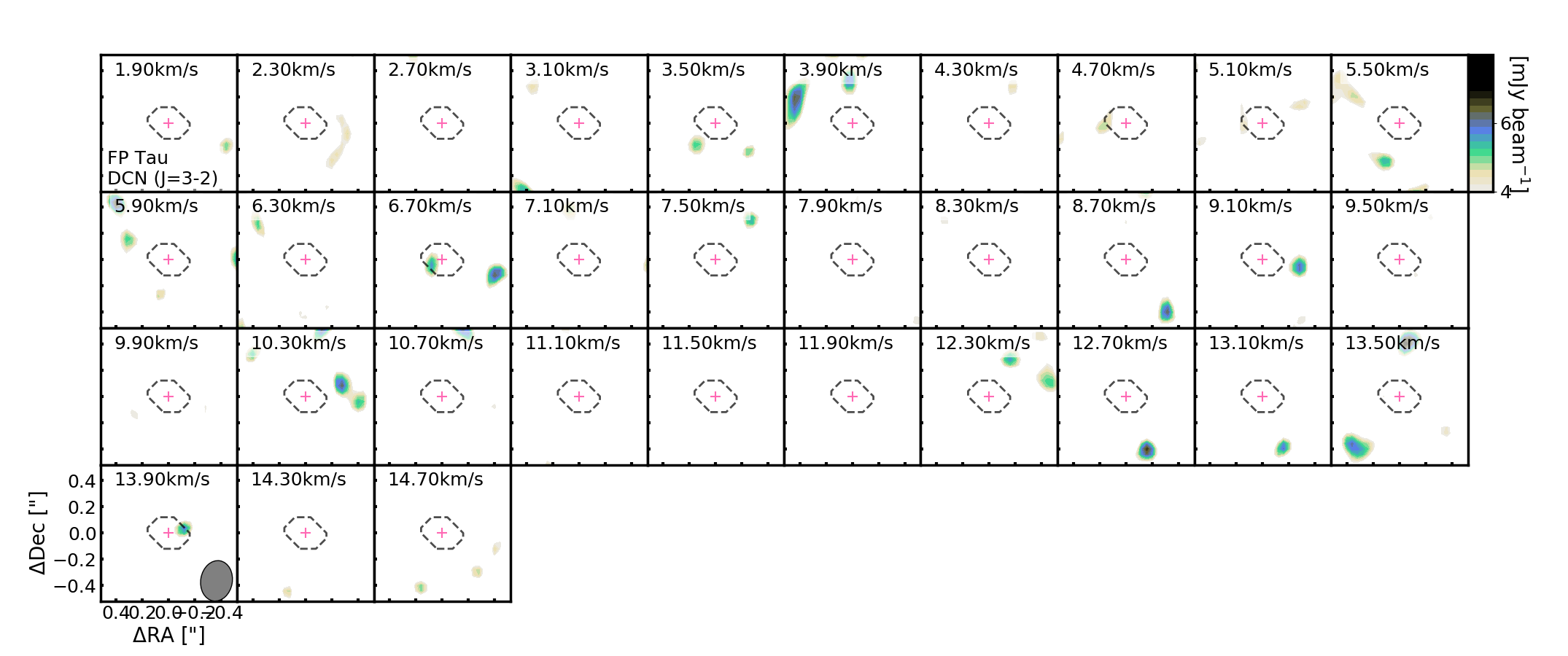}}
\caption{DCN 3--2 toward FP Tau above 2$\sigma$.
\label{figset_fptau_DCN}}
\end{figure*}

\begin{figure*}
\centering
\resizebox{0.99\hsize}{!}{
    \includegraphics{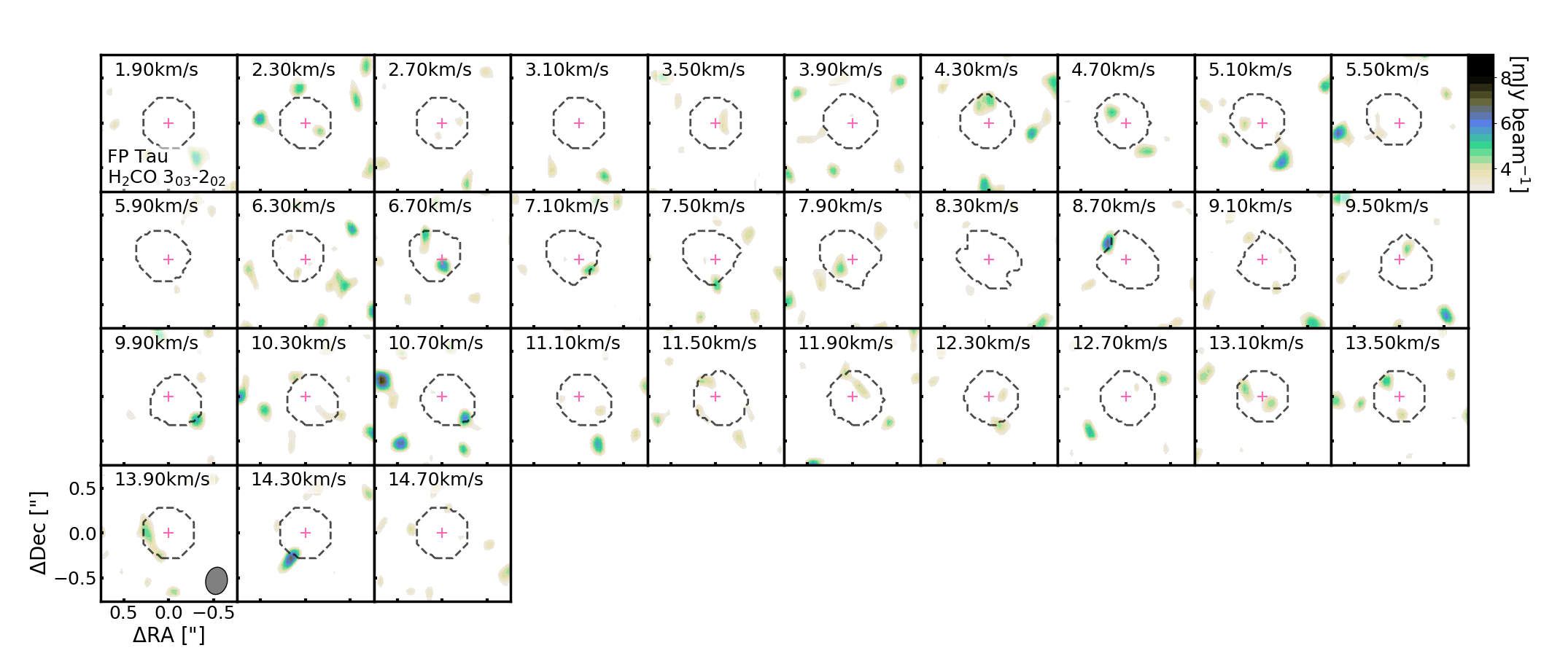}}
\caption{H$_2$CO 3--2 toward FP Tau above 2$\sigma$.
\label{figset_fptau_H2CO}}
\end{figure*}

Figures~\ref{figset_fptau_12CO} through~\ref{figset_fptau_H2CO} display channel maps of detected/tentatively detected emission toward FP Tau.

\subsection{Channel Maps for J0432+1827}

\begin{figure*}
\centering
\resizebox{0.99\hsize}{!}{
    \includegraphics{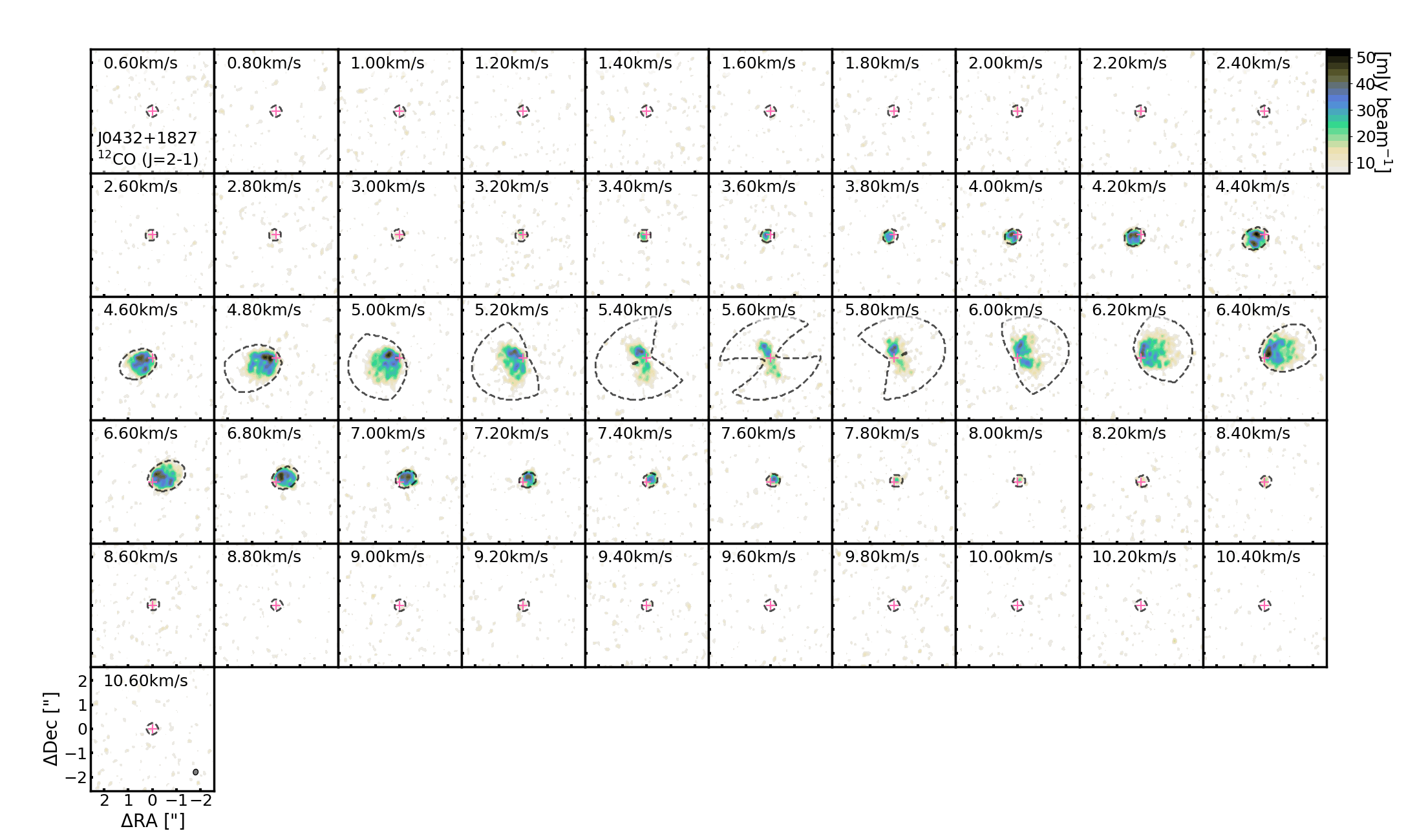}}
\caption{$^{12}$CO toward J0432+1827 above 2$\sigma$.
\label{figset_j0432_12CO}}
\end{figure*}

\begin{figure*}
\centering
\resizebox{0.99\hsize}{!}{
    \includegraphics{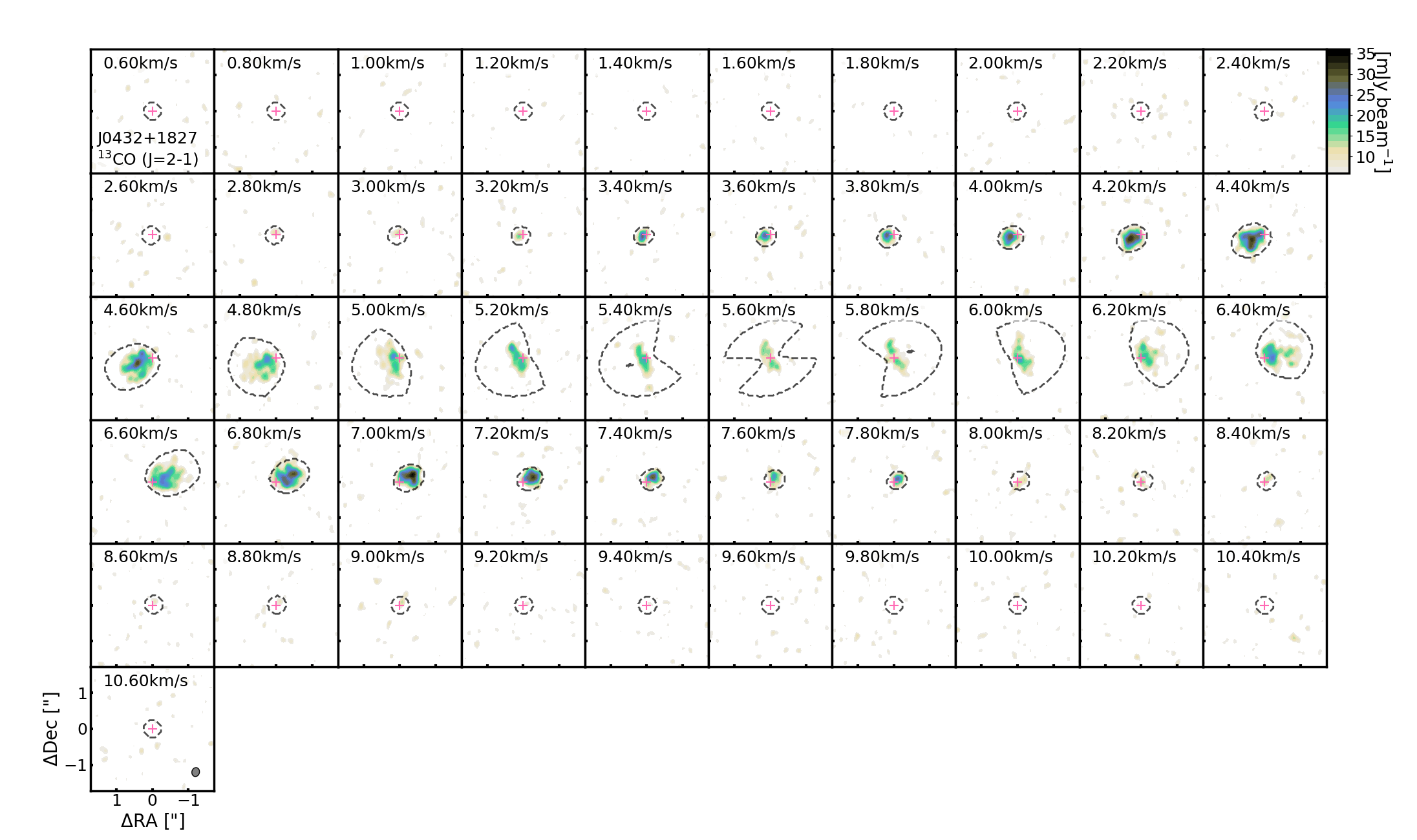}}
\caption{$^{13}$CO toward J0432+1827 above 2$\sigma$.
\label{figset_j0432_13CO}}
\end{figure*}

\begin{figure*}
\centering
\resizebox{0.99\hsize}{!}{
    \includegraphics{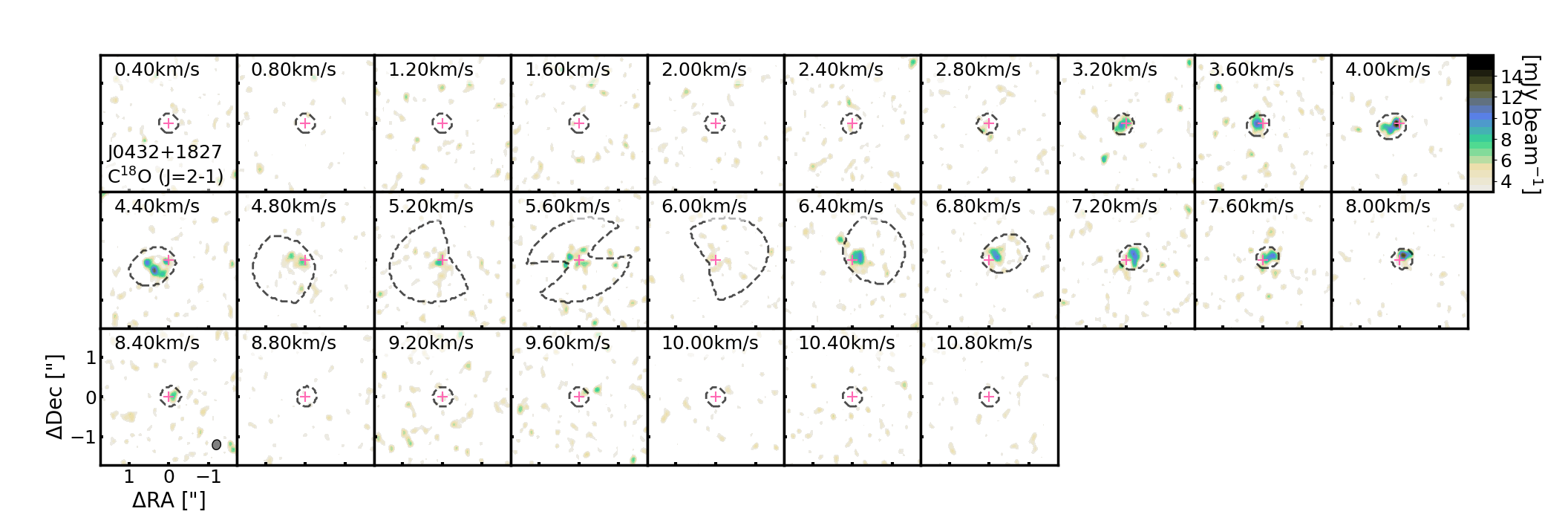}}
\caption{C$^{18}$O toward J0432+1827 above 2$\sigma$.
\label{figset_j0432_C18O}}
\end{figure*}

\begin{figure*}
\centering
\resizebox{0.99\hsize}{!}{
    \includegraphics{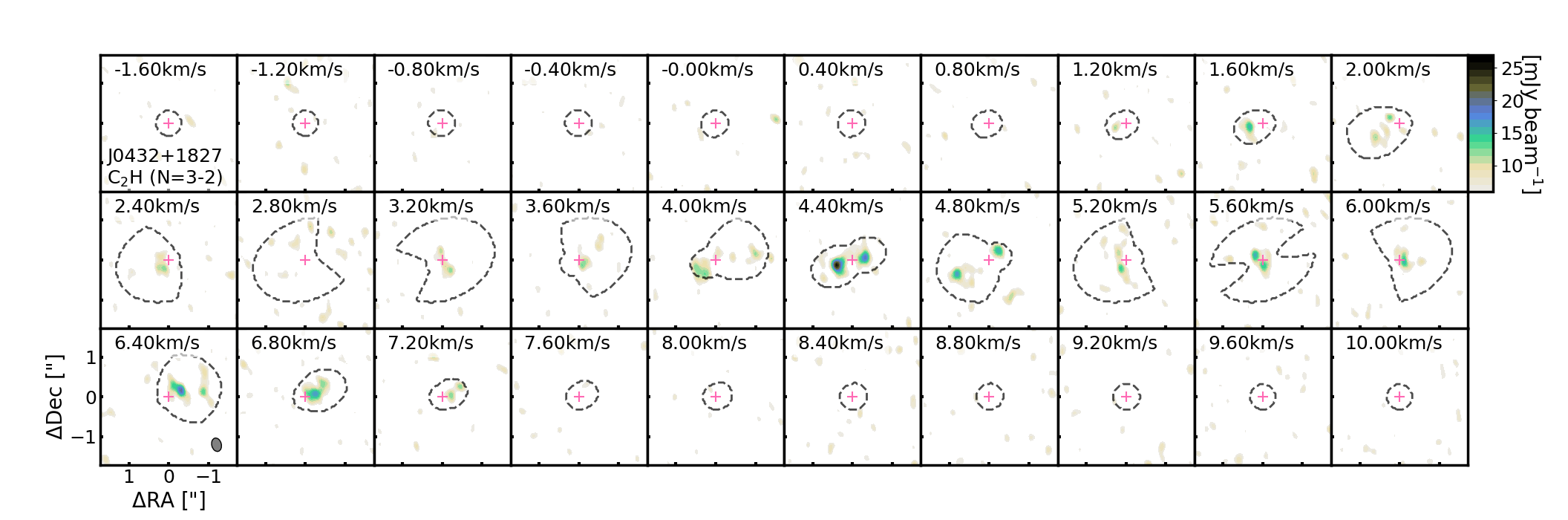}}
\caption{C$_2$H 3--2 toward J0432+1827 above 2$\sigma$.
\label{figset_j0432_C2H}}
\end{figure*}

\begin{figure*}
\centering
\resizebox{0.99\hsize}{!}{
    \includegraphics{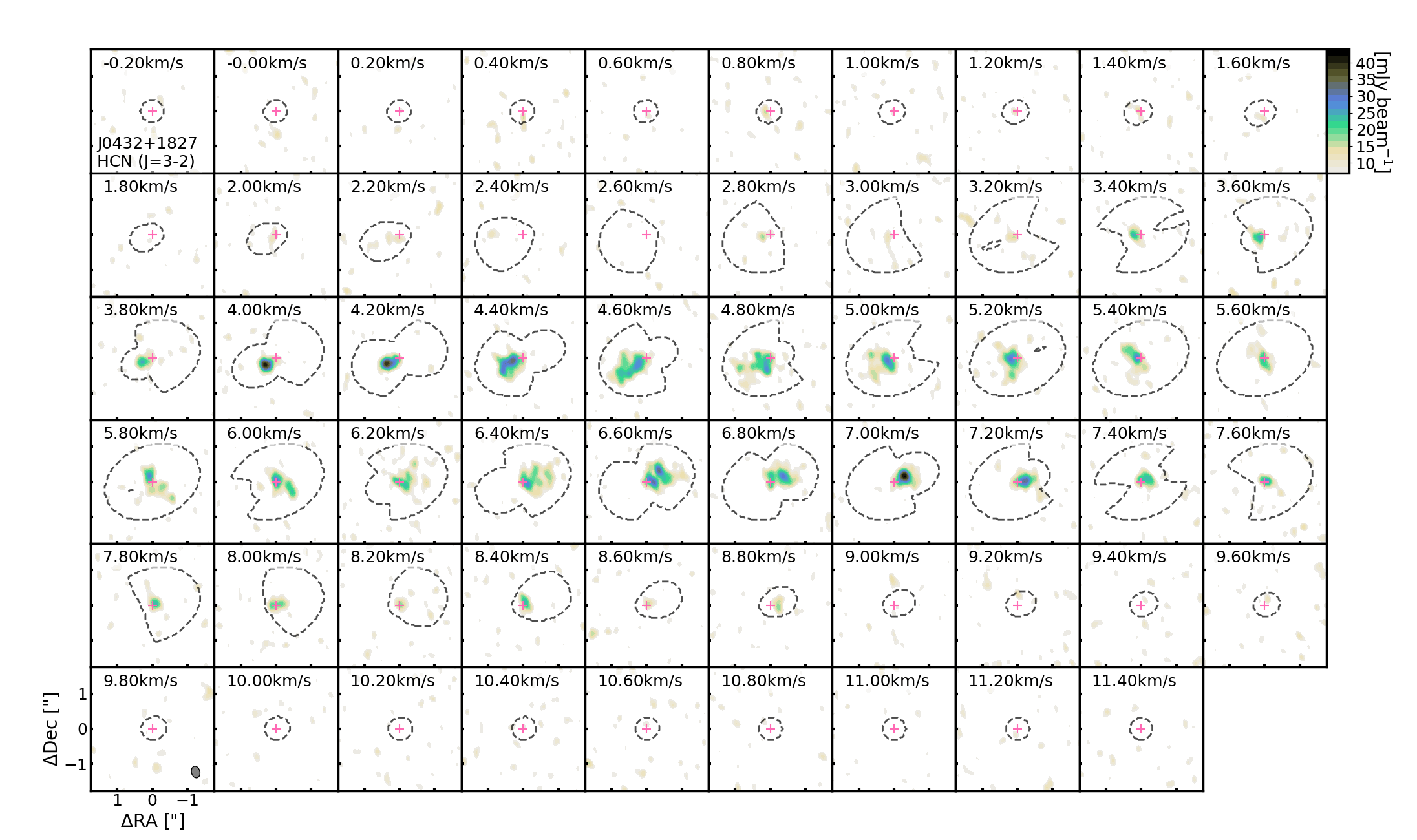}}
\caption{HCN 3--2 toward J0432+1827 above 2$\sigma$.
\label{figset_j0432_HCN}}
\end{figure*}

\begin{figure*}
\centering
\resizebox{0.99\hsize}{!}{
    \includegraphics{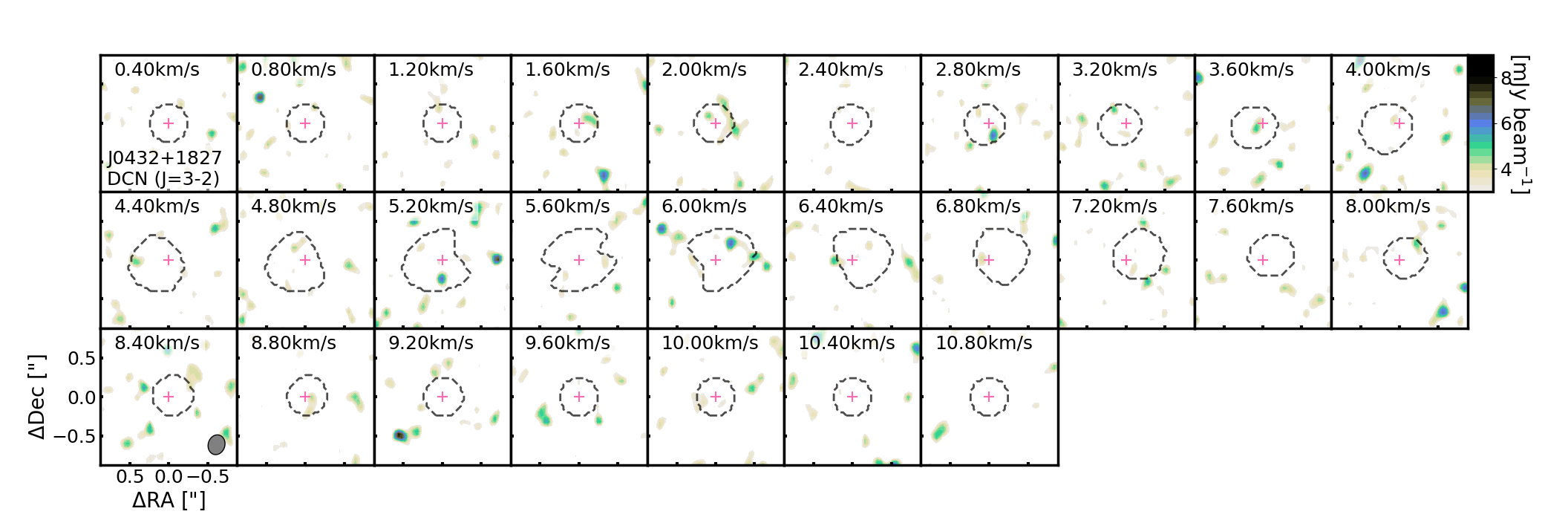}}
\caption{DCN 3--2 toward J0432+1827 above 2$\sigma$.
\label{figset_j0432_DCN}}
\end{figure*}

\begin{figure*}
\centering
\resizebox{0.99\hsize}{!}{
    \includegraphics{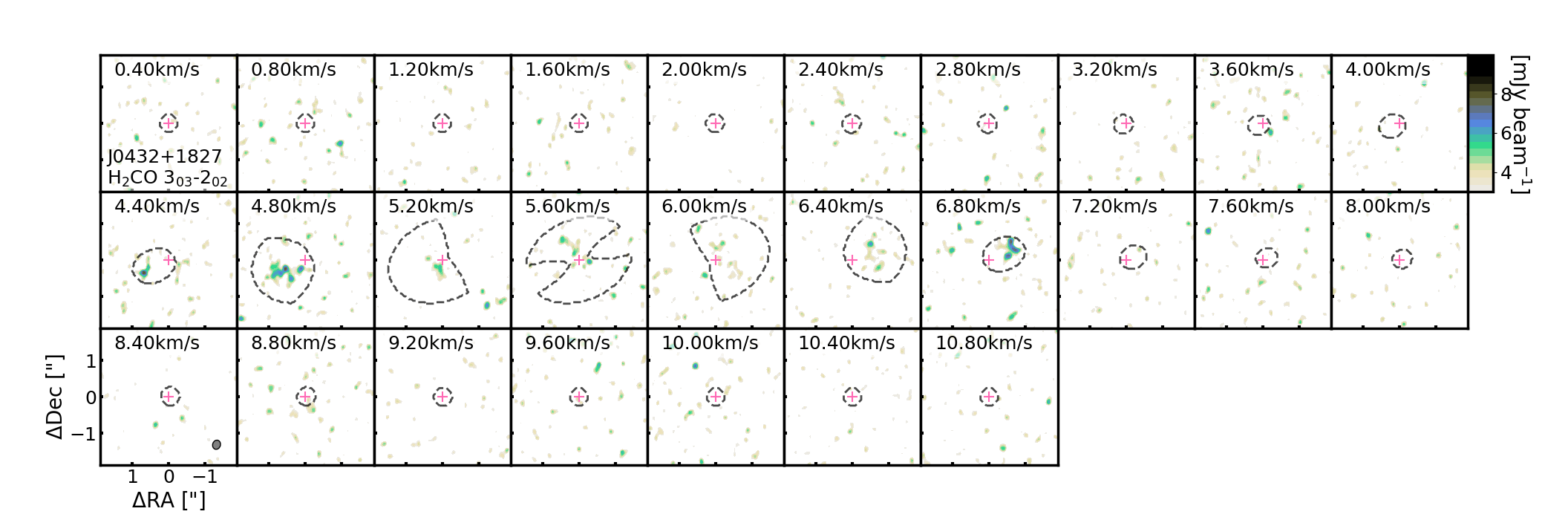}}
\caption{H$_2$CO 3--2 toward J0432+1827 above 2$\sigma$.
\label{figset_j0432_H2CO}}
\end{figure*}

Figures~\ref{figset_j0432_12CO} through~\ref{figset_j0432_H2CO} display channel maps of detected/tentatively detected emission toward J0432+1827.

\subsection{Channel Maps for J1100-7619}

\begin{figure*}
\centering
\resizebox{0.99\hsize}{!}{
    \includegraphics{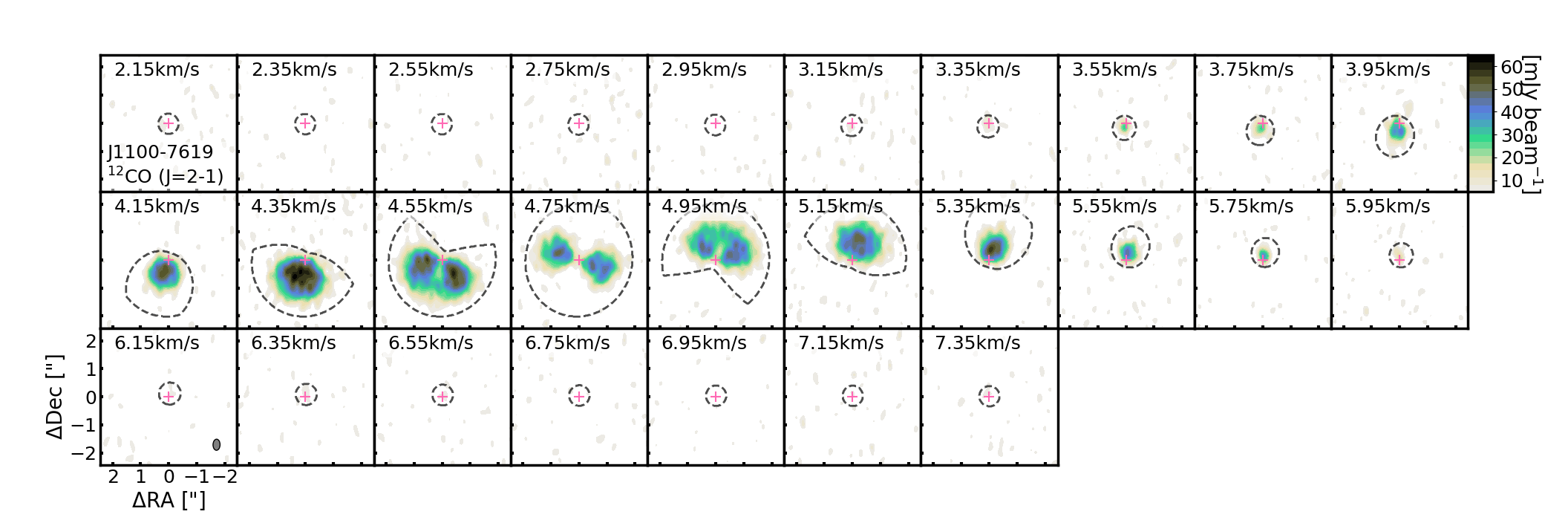}}
\caption{$^{12}$CO toward J1100-7619 above 2$\sigma$.
\label{figset_j1100_12CO}}
\end{figure*}

\begin{figure*}
\centering
\resizebox{0.99\hsize}{!}{
    \includegraphics{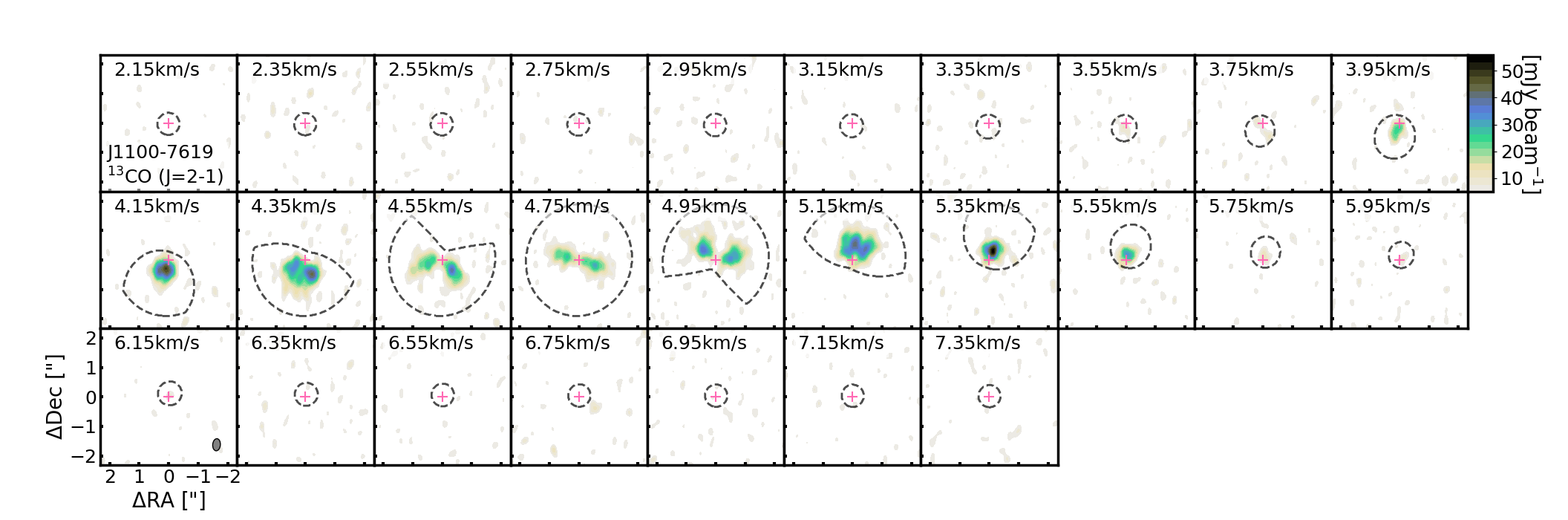}}
\caption{$^{13}$CO toward J1100-7619 above 2$\sigma$.
\label{figset_j1100_13CO}}
\end{figure*}

\begin{figure*}
\centering
\resizebox{0.99\hsize}{!}{
    \includegraphics{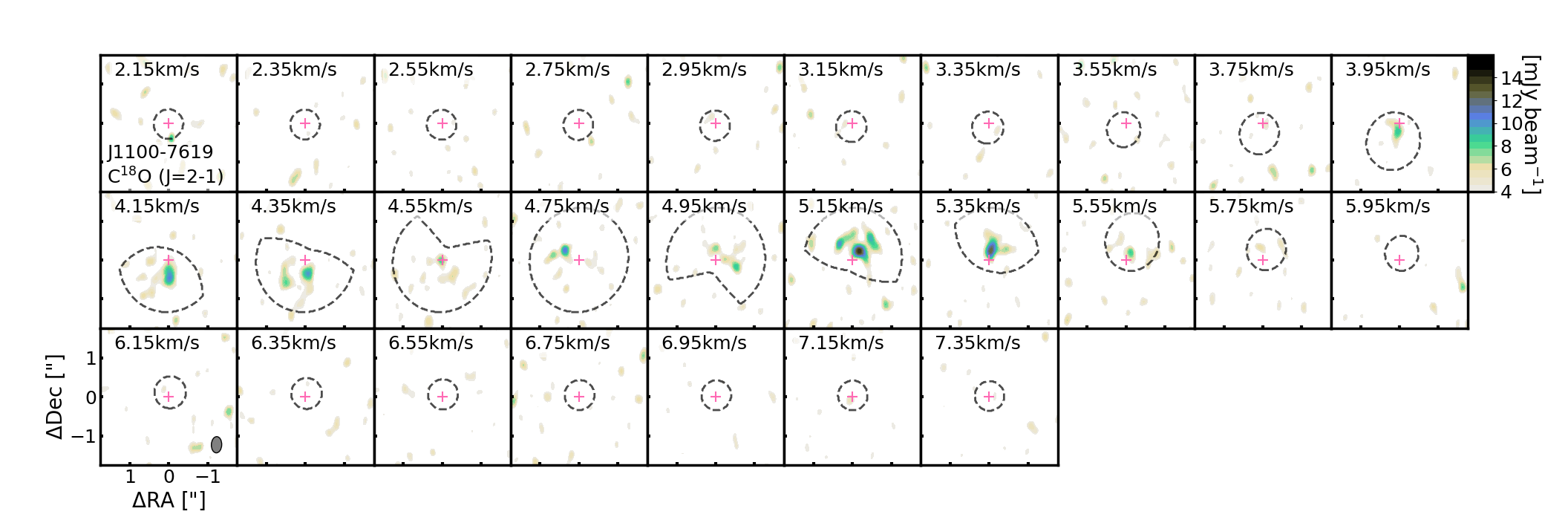}}
\caption{C$^{18}$O toward J1100-7619 above 2$\sigma$.
\label{figset_j1100_C18O}}
\end{figure*}

\begin{figure*}
\centering
\resizebox{0.99\hsize}{!}{
    \includegraphics{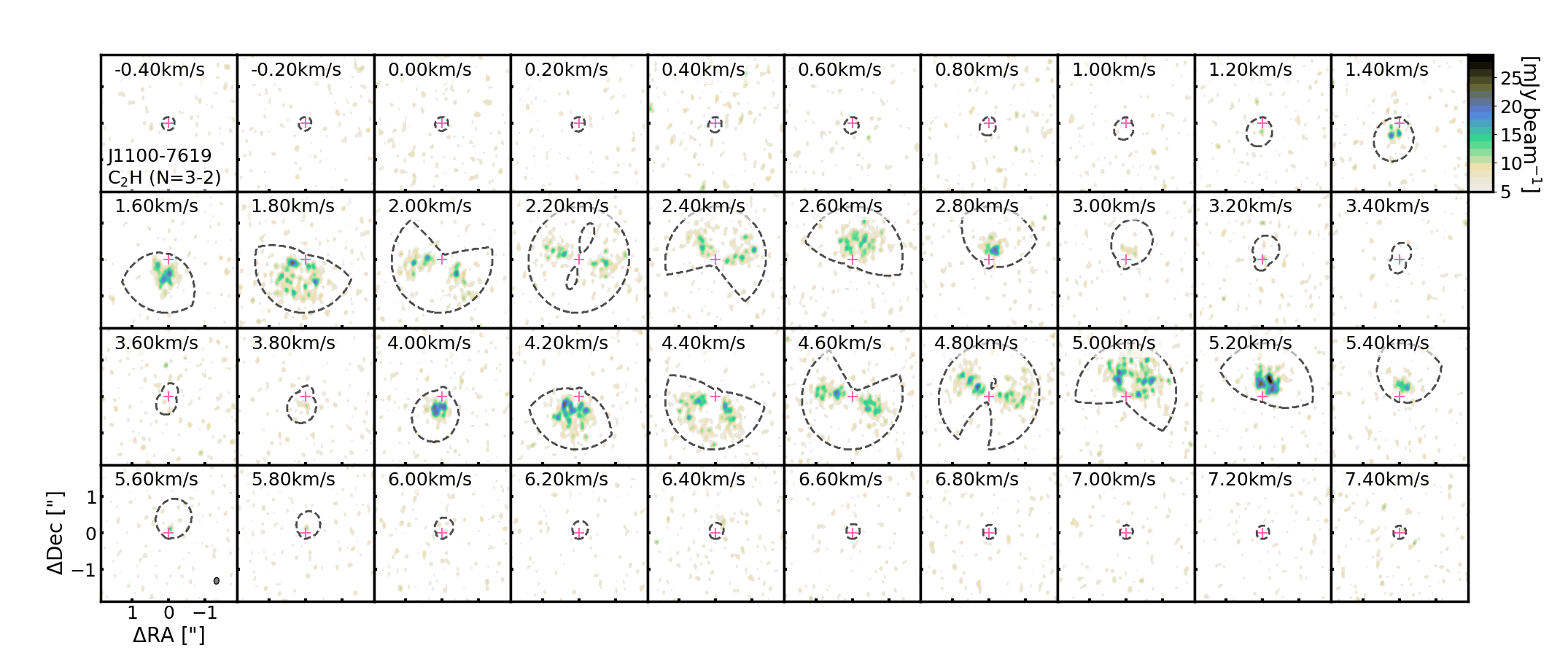}}
\caption{C$_2$H 3--2 toward J1100-7619 above 2$\sigma$.
\label{figset_j1100_C2H}}
\end{figure*}

\begin{figure*}
\centering
\resizebox{0.99\hsize}{!}{
    \includegraphics{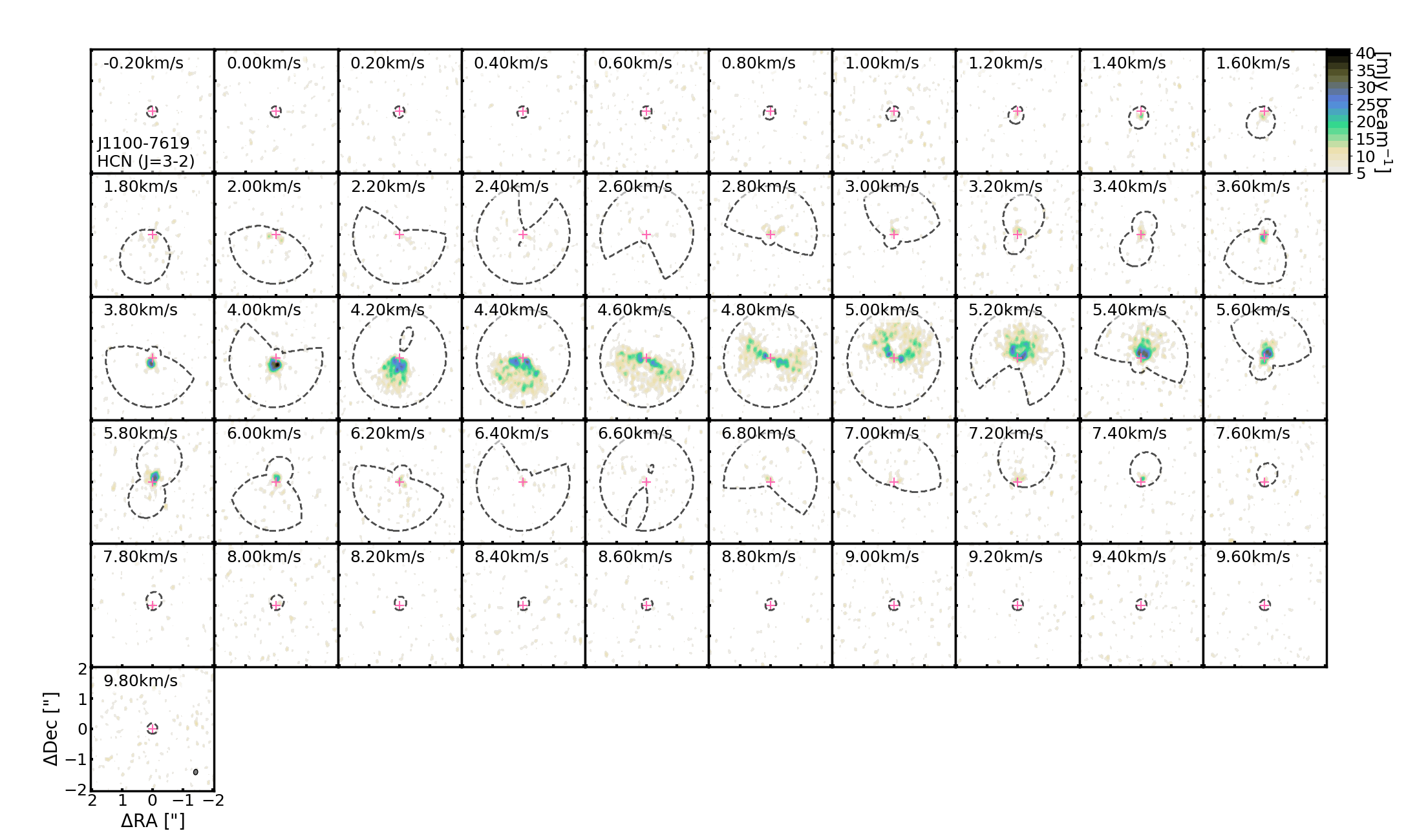}}
\caption{HCN 3--2 toward J1100-7619 above 2$\sigma$.
\label{figset_j1100_HCN}}
\end{figure*}

\begin{figure*}
\centering
\resizebox{0.99\hsize}{!}{
    \includegraphics{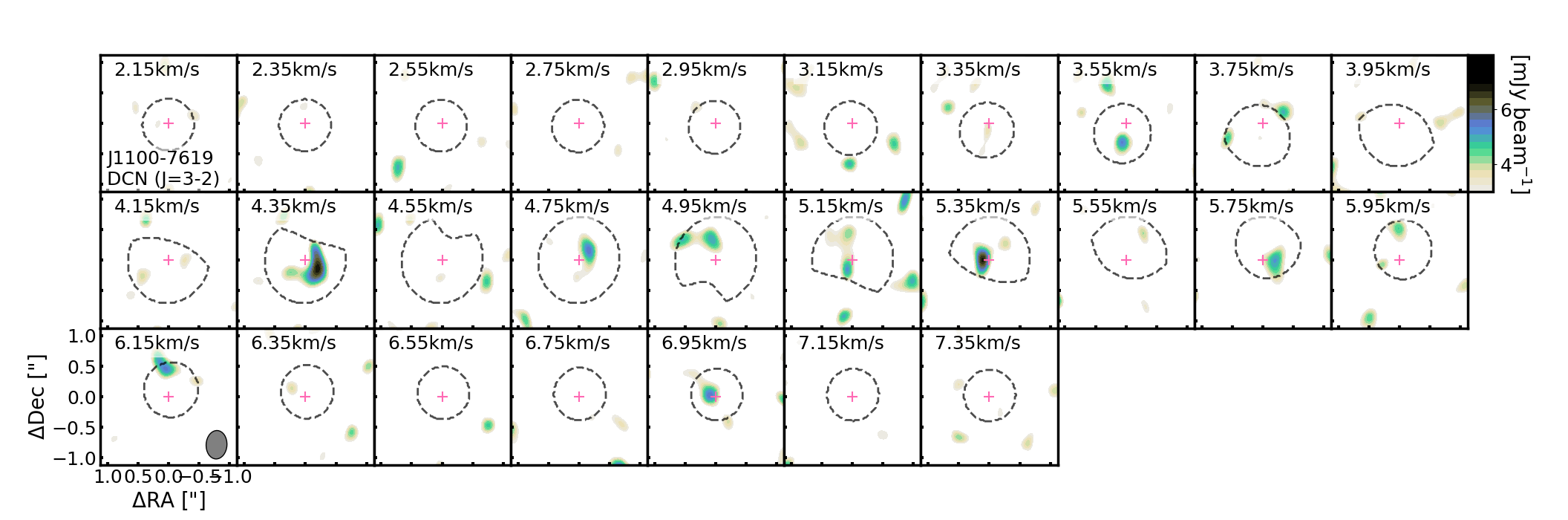}}
\caption{DCN 3--2 toward J1100-7619 above 2$\sigma$.
\label{figset_j1100_DCN}}
\end{figure*}

\begin{figure*}
\centering
\resizebox{0.99\hsize}{!}{
    \includegraphics{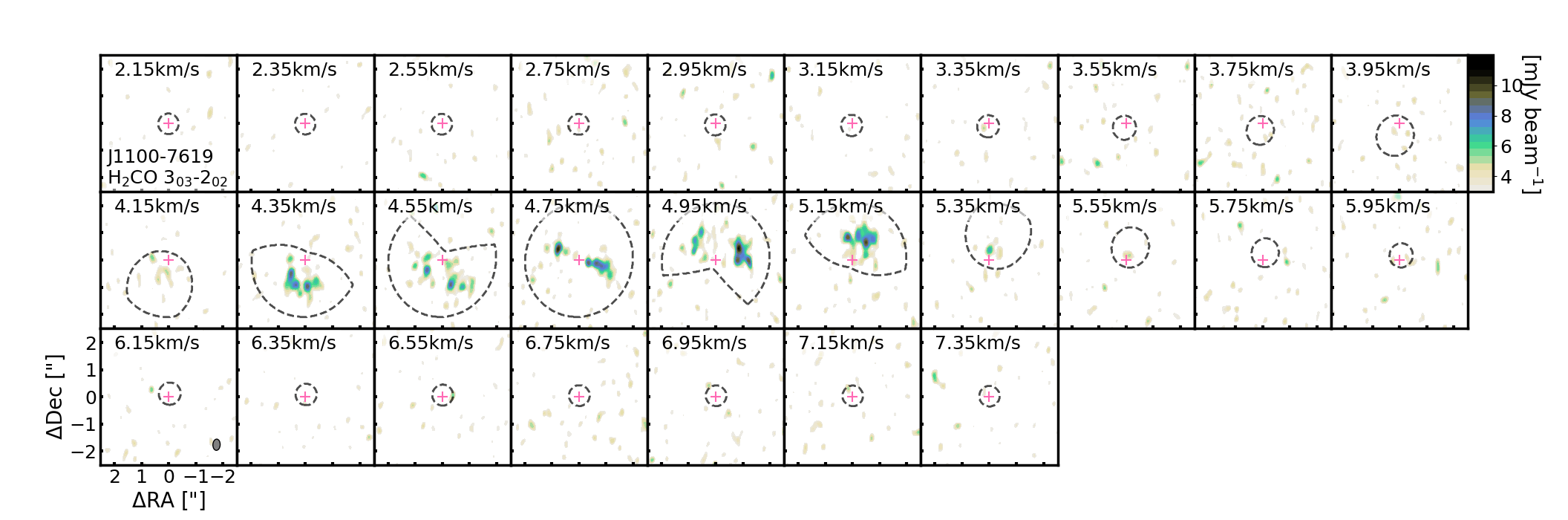}}
\caption{H$_2$CO 3--2 toward J1100-7619 above 2$\sigma$.
\label{figset_j1100_H2CO}}
\end{figure*}

Figures~\ref{figset_j1100_12CO} through~\ref{figset_j1100_H2CO} display channel maps of detected/tentatively detected emission toward J1100-7619.

\subsection{Channel Maps for J1545-3417}

\begin{figure*}
\centering
\resizebox{0.99\hsize}{!}{
    \includegraphics{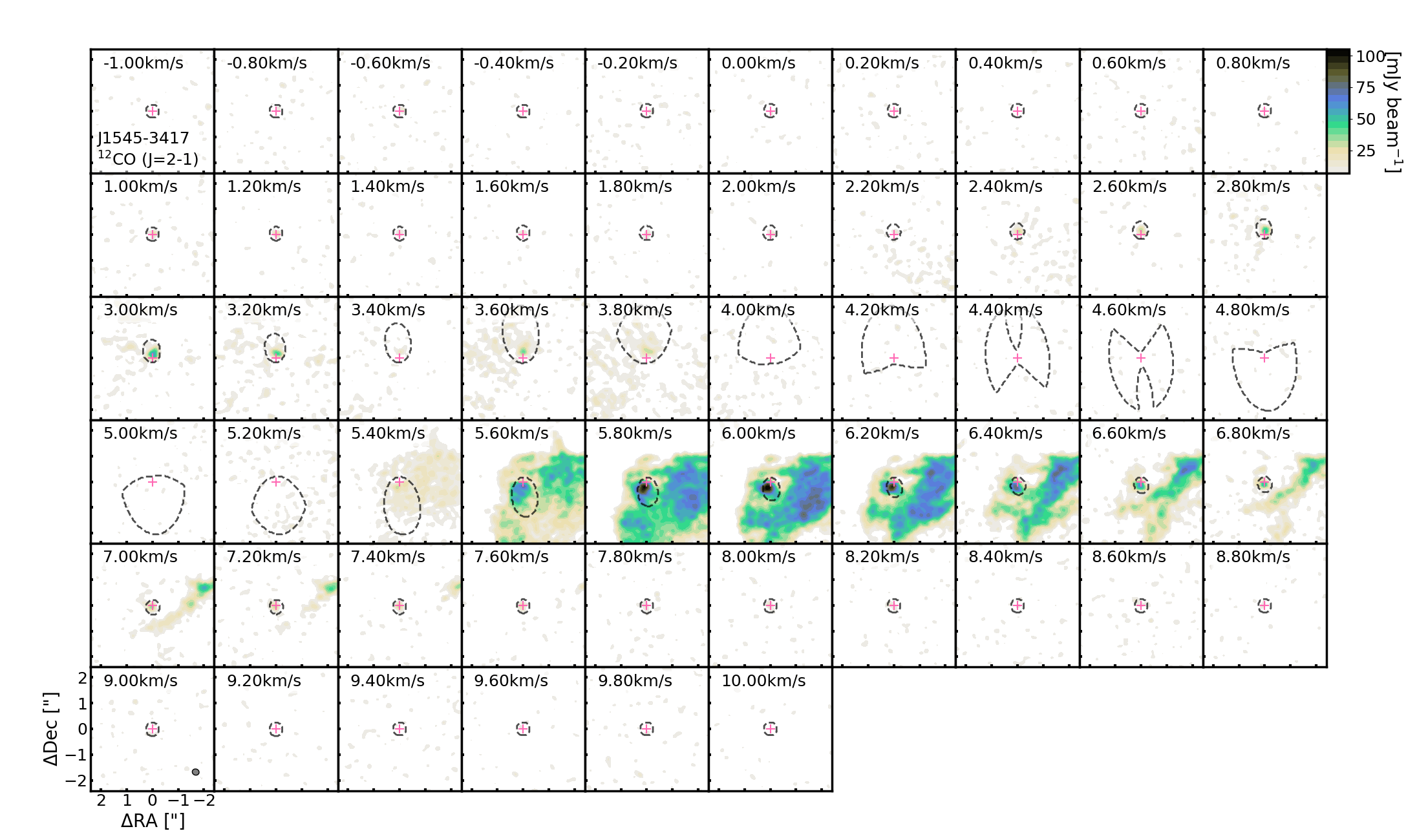}}
\caption{$^{12}$CO toward J1545-3417 above 2$\sigma$.
\label{figset_j1545_12CO}}
\end{figure*}

\begin{figure*}
\centering
\resizebox{0.99\hsize}{!}{
    \includegraphics{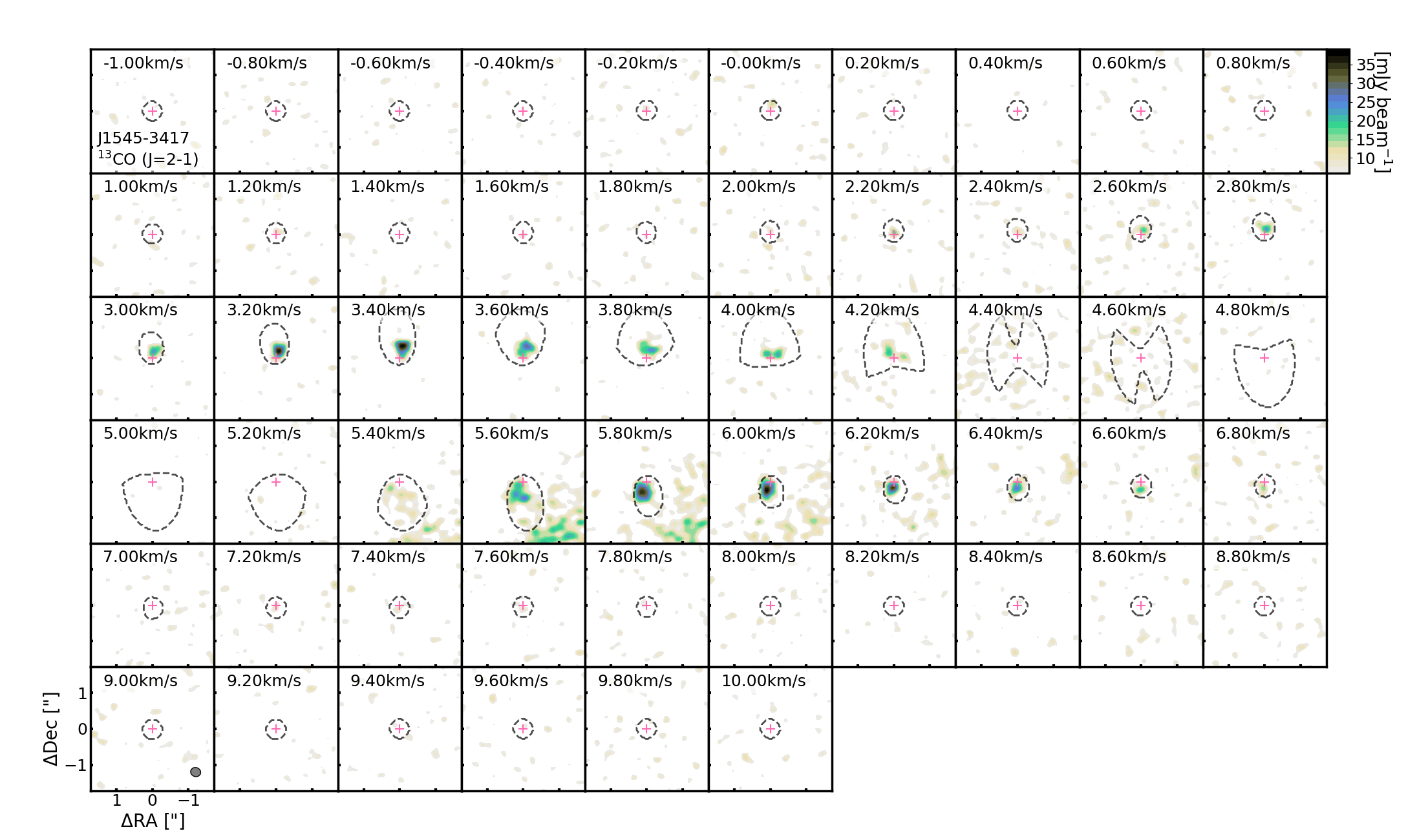}}
\caption{$^{13}$CO toward J1545-3417 above 2$\sigma$.
\label{figset_j1545_13CO}}
\end{figure*}

\begin{figure*}
\centering
\resizebox{0.99\hsize}{!}{
    \includegraphics{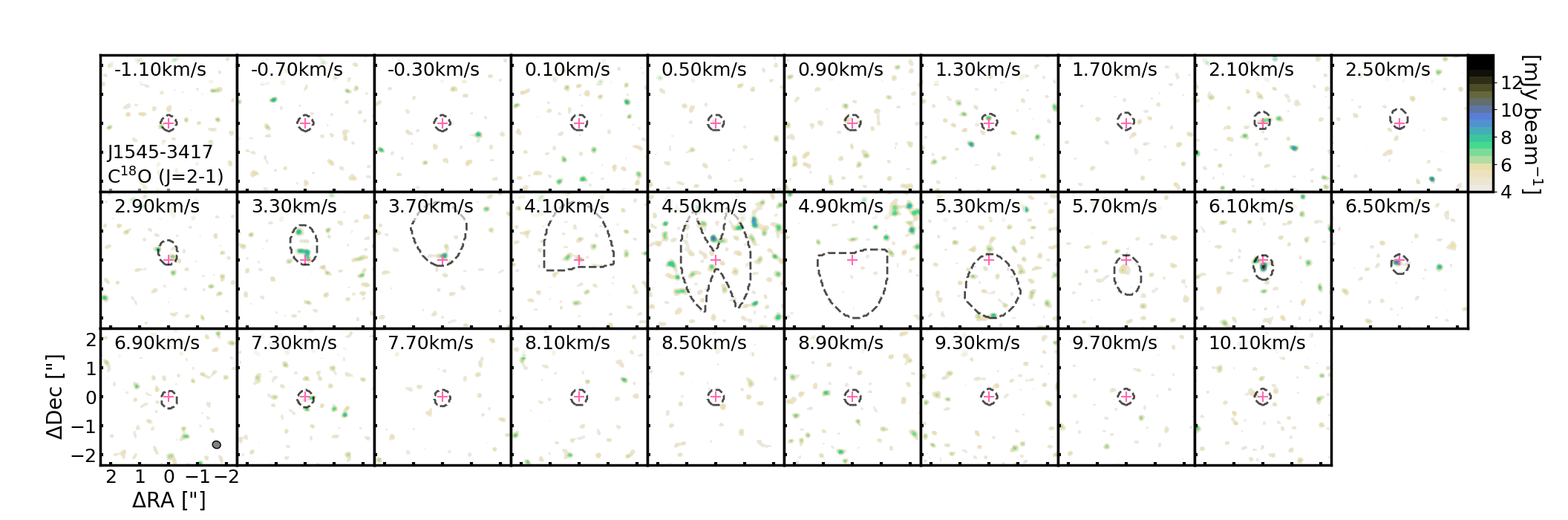}}
\caption{C$^{18}$O toward J1545-3417 above 2$\sigma$.
\label{figset_j1545_C18O}}
\end{figure*}

\begin{figure*}
\centering
\resizebox{0.99\hsize}{!}{
    \includegraphics{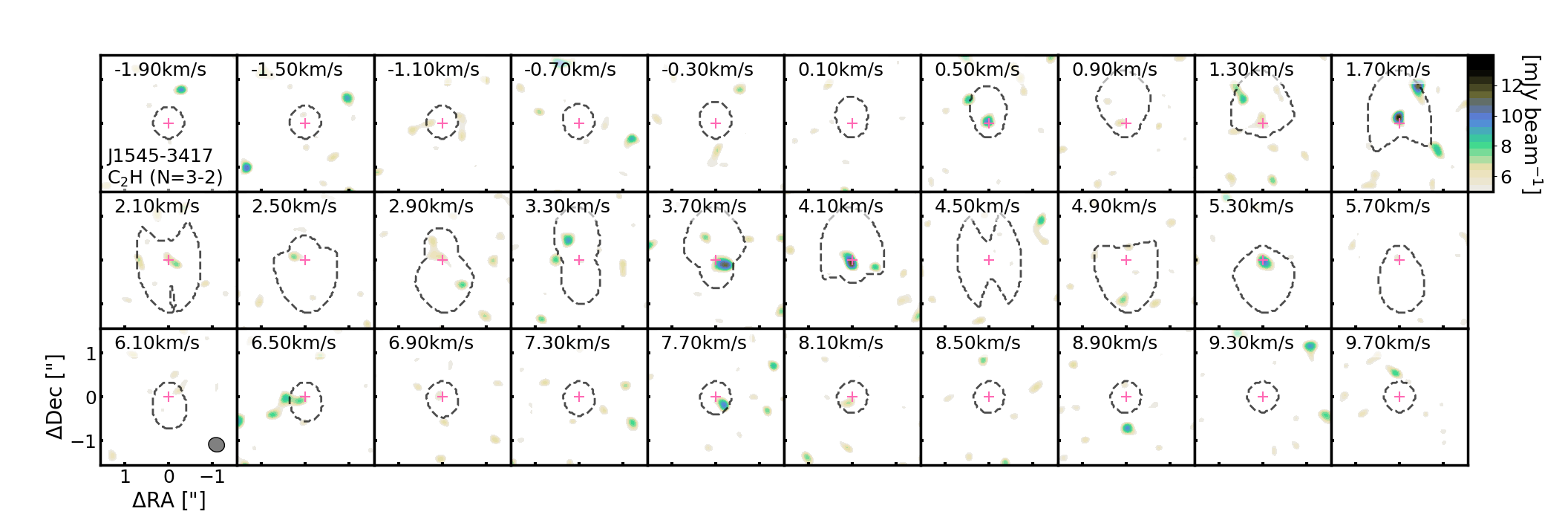}}
\caption{C$_2$H 3--2 toward J1545-3417 above 2$\sigma$.
\label{figset_j1545_C2H}}
\end{figure*}

\begin{figure*}
\centering
\resizebox{0.99\hsize}{!}{
    \includegraphics{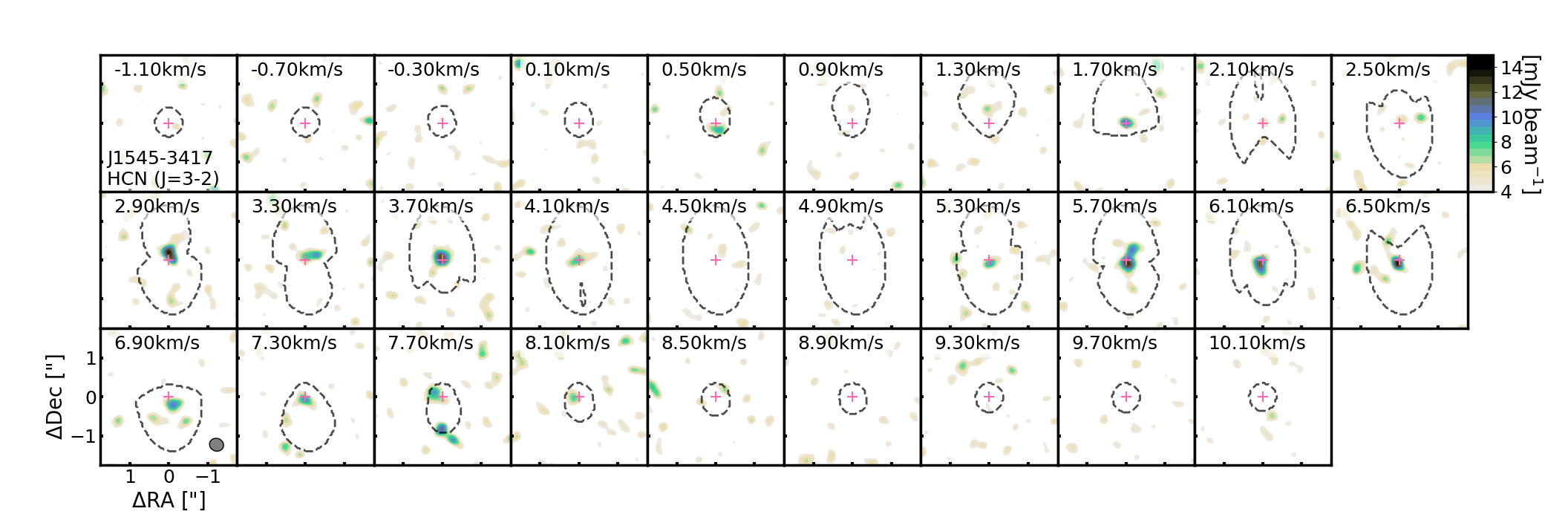}}
\caption{HCN 3--2 toward J1545-3417 above 2$\sigma$.
\label{figset_j1545_HCN}}
\end{figure*}

\begin{figure*}
\centering
\resizebox{0.99\hsize}{!}{
    \includegraphics{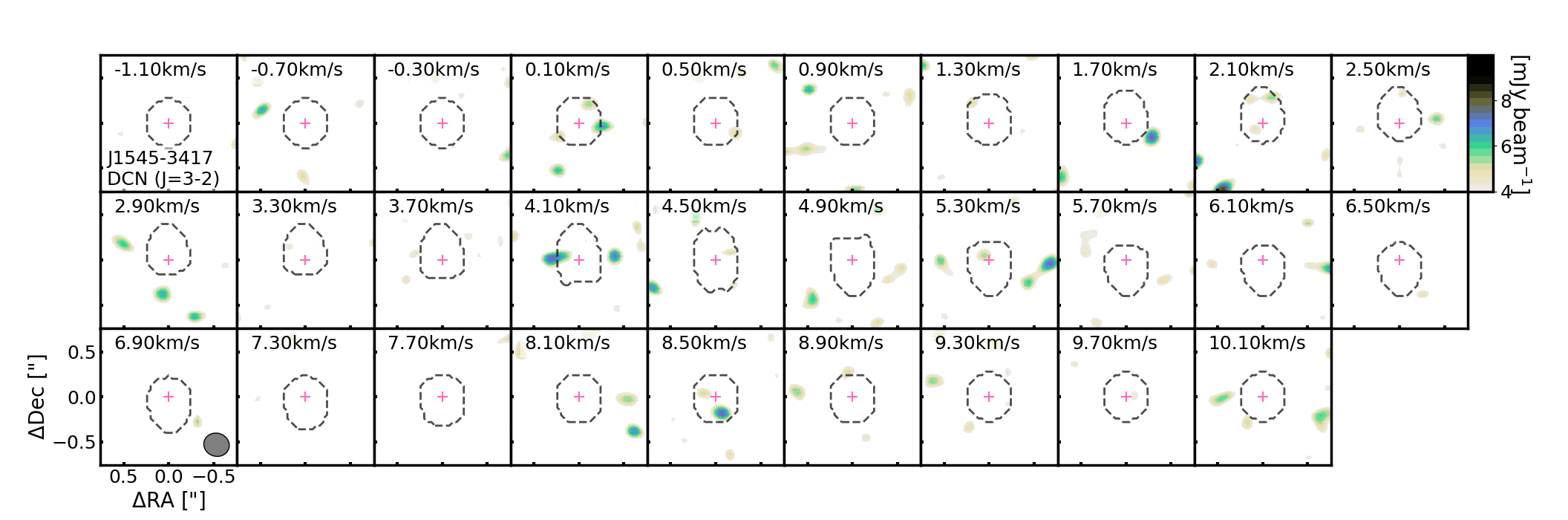}}
\caption{DCN 3--2 toward J1545-3417 above 2$\sigma$.
\label{figset_j1545_DCN}}
\end{figure*}

Figures~\ref{figset_j1545_12CO} through~\ref{figset_j1545_DCN} display channel maps of detected/tentatively detected emission toward J1545-3417.

\subsection{Channel Maps for Sz 69}

\begin{figure*}
\centering
\resizebox{0.99\hsize}{!}{
    \includegraphics{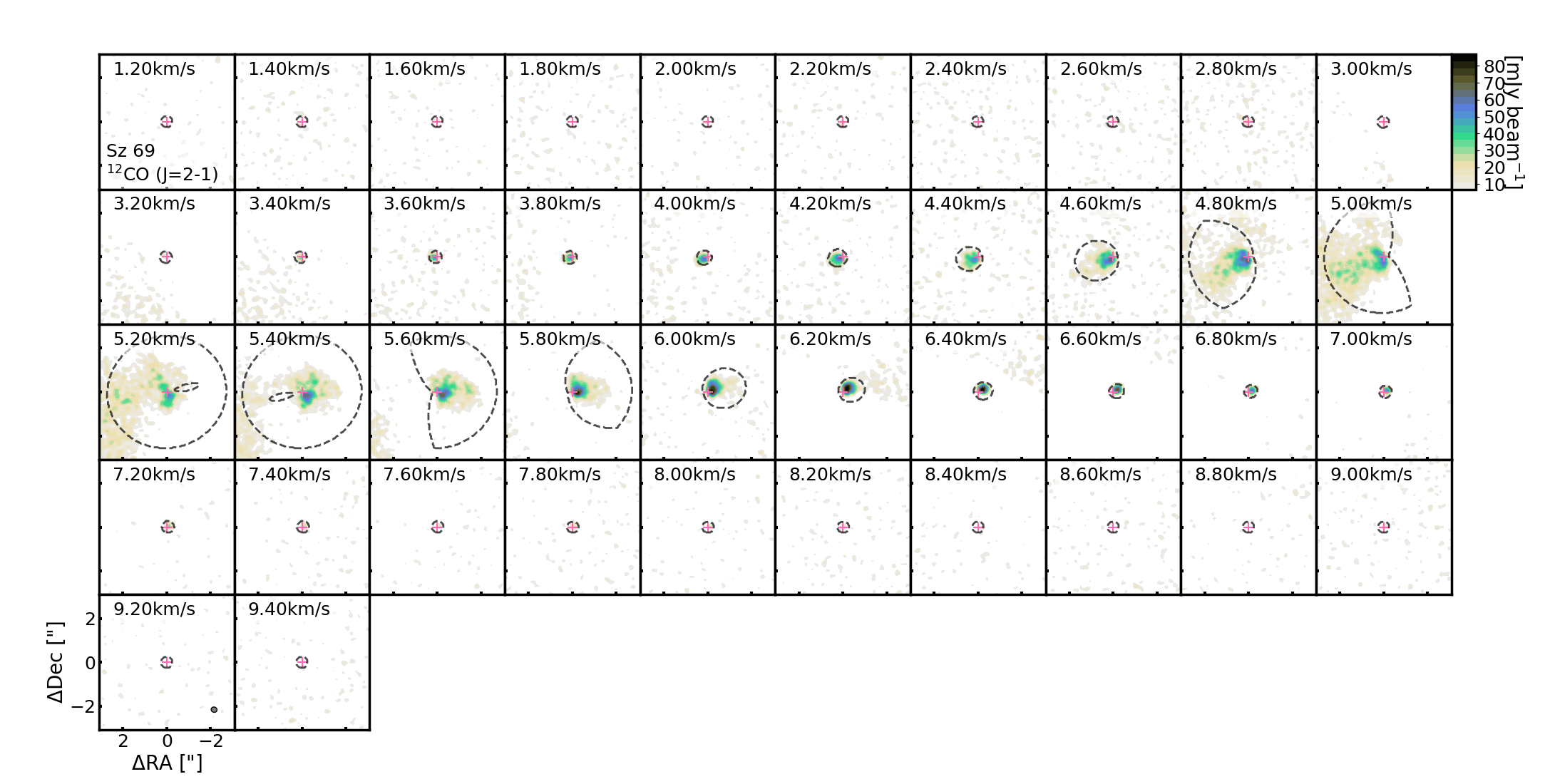}}
\caption{$^{12}$CO toward Sz 69 above 2$\sigma$.
\label{figset_sz69_12CO}}
\end{figure*}

\begin{figure*}
\centering
\resizebox{0.99\hsize}{!}{
    \includegraphics{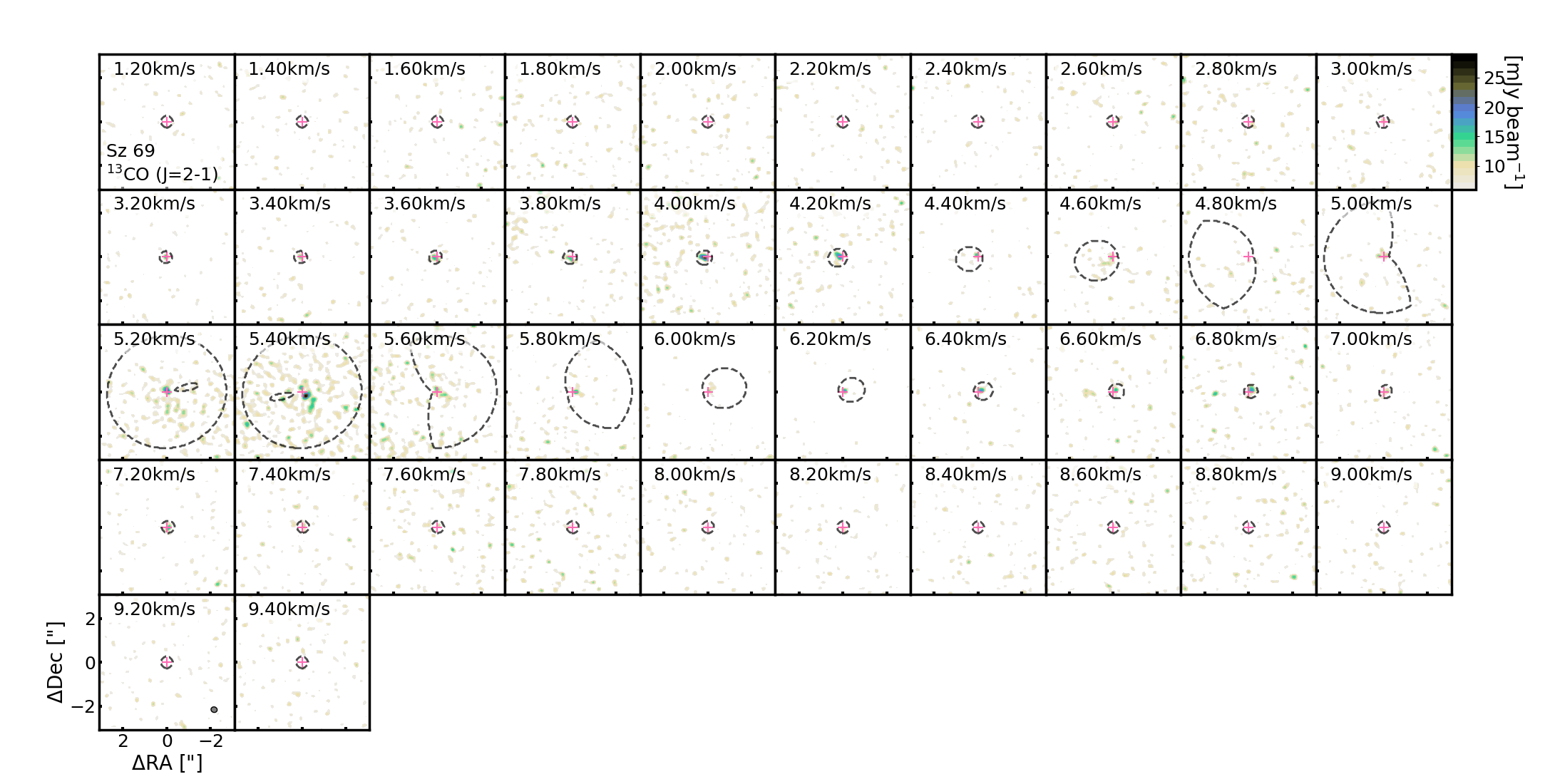}}
\caption{$^{13}$CO toward Sz 69 above 2$\sigma$.
\label{figset_sz69_13CO}}
\end{figure*}

\begin{figure*}
\centering
\resizebox{0.99\hsize}{!}{
    \includegraphics{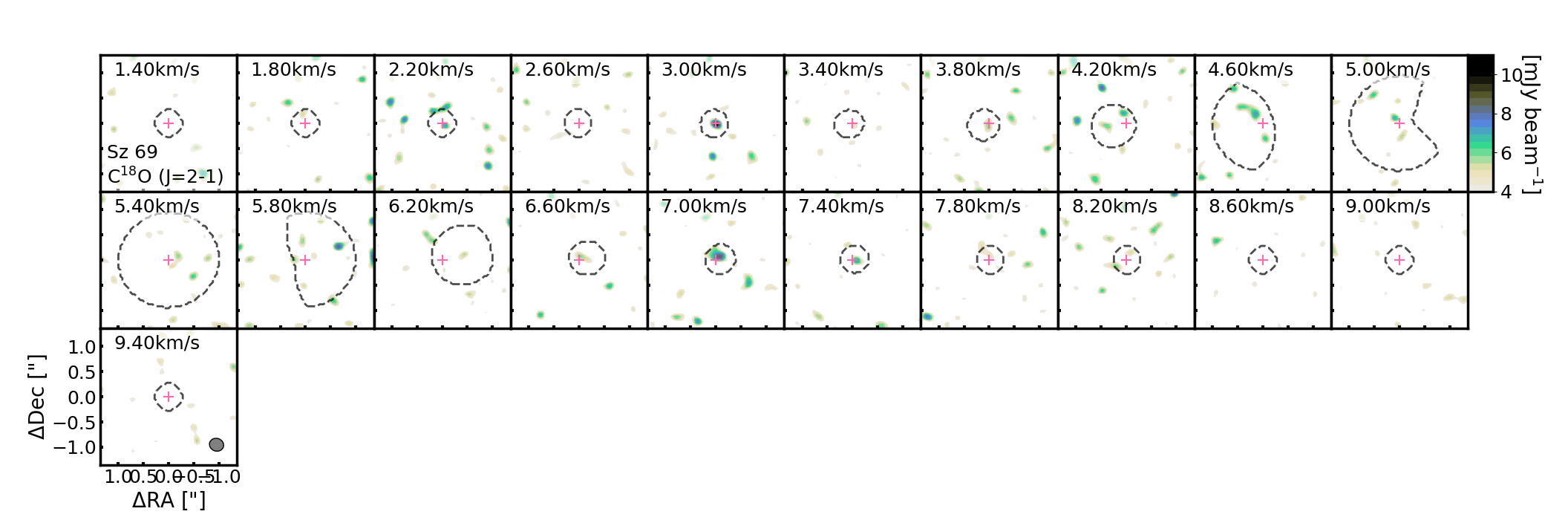}}
\caption{C$^{18}$O toward Sz 69 above 2$\sigma$.
\label{figset_sz69_C18O}}
\end{figure*}

\begin{figure*}
\centering
\resizebox{0.99\hsize}{!}{
    \includegraphics{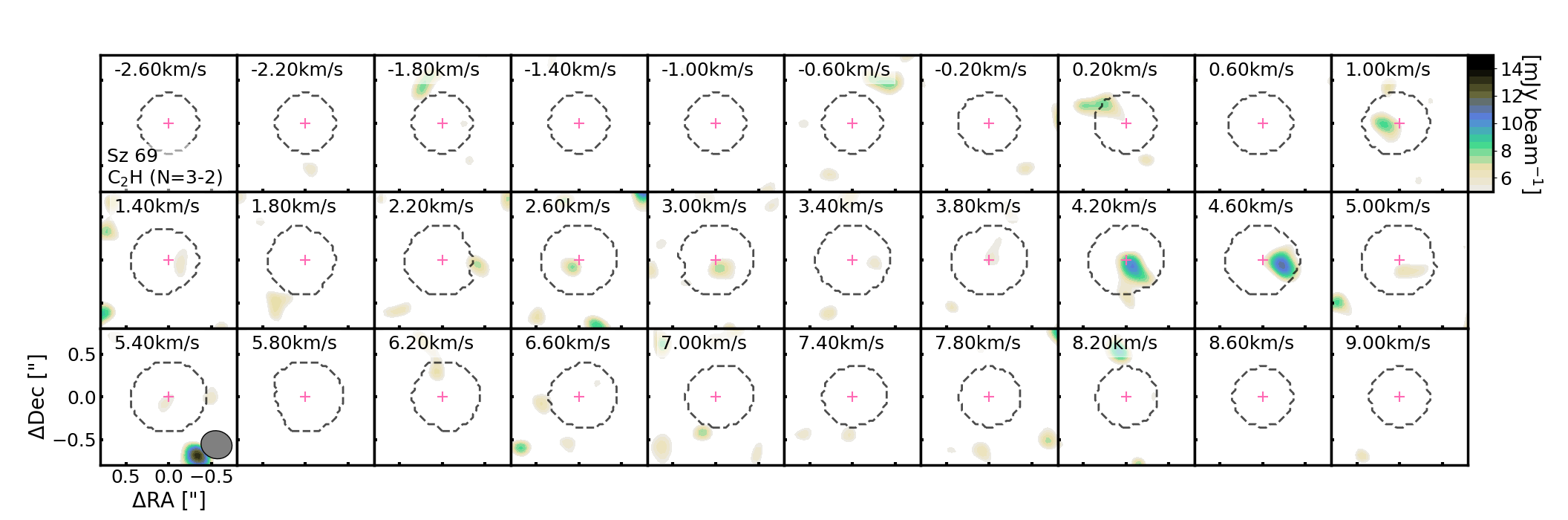}}
\caption{C$_2$H 3--2 toward Sz 69 above 2$\sigma$.
\label{figset_sz69_C2H}}
\end{figure*}

\begin{figure*}
\centering
\resizebox{0.99\hsize}{!}{
    \includegraphics{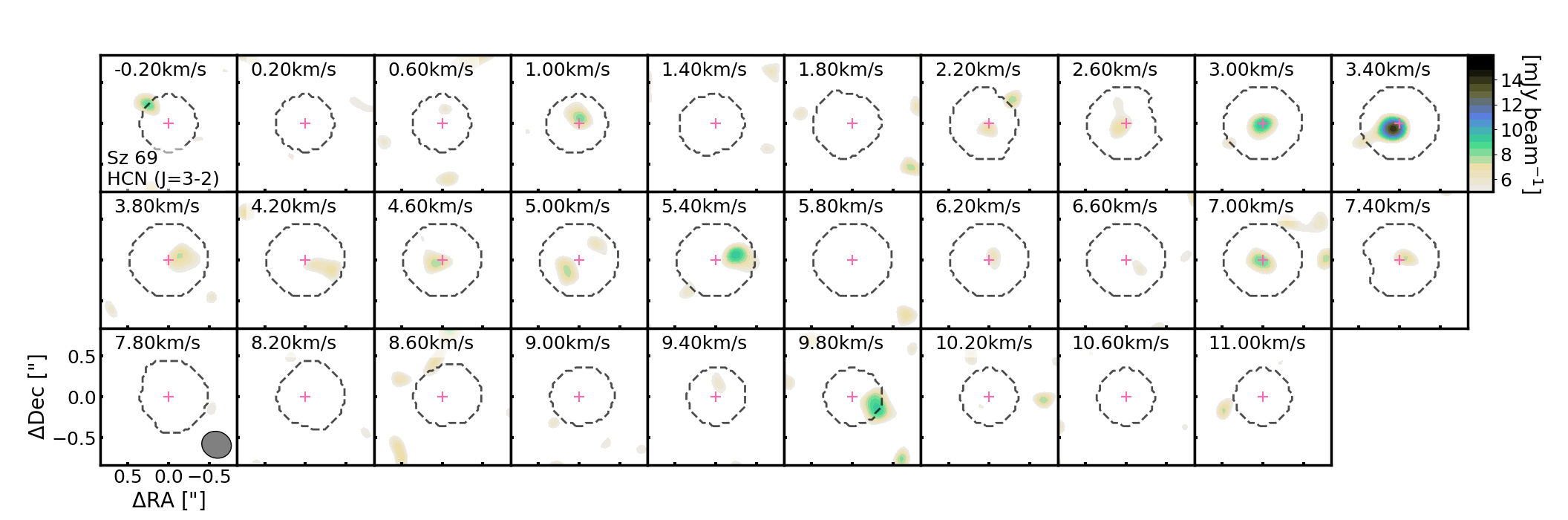}}
\caption{HCN 3--2 toward Sz 69 above 2$\sigma$.
\label{figset_sz69_HCN}}
\end{figure*}

\begin{figure*}
\centering
\resizebox{0.99\hsize}{!}{
    \includegraphics{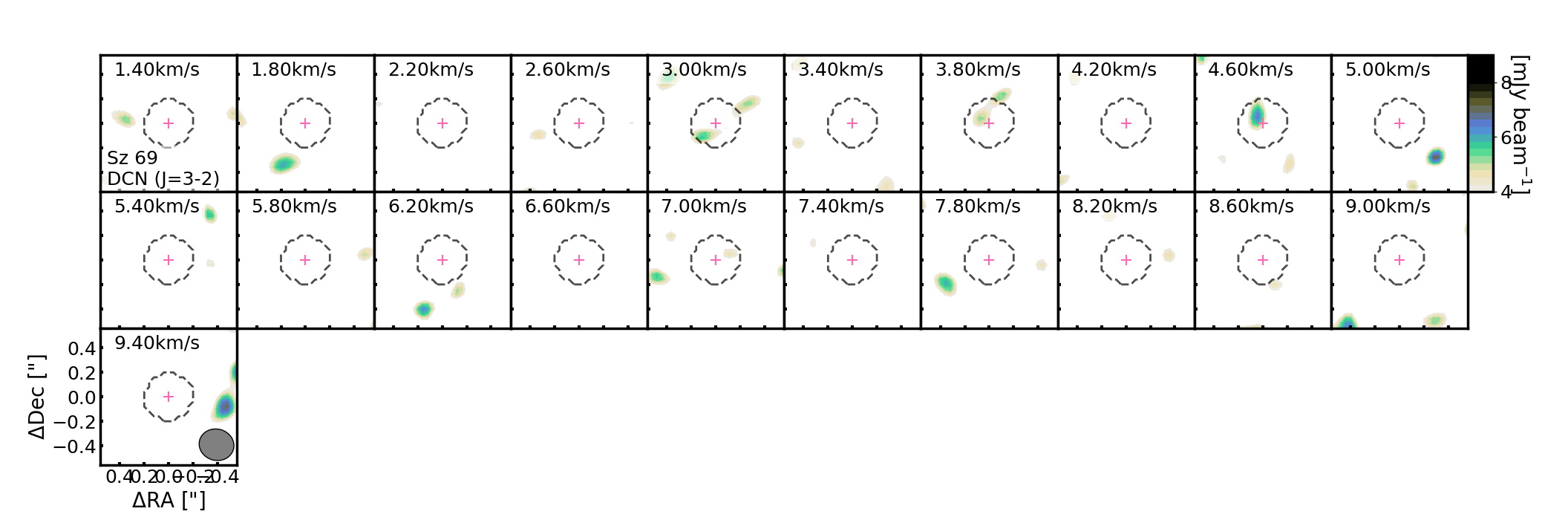}}
\caption{DCN 3--2 toward Sz 69 above 2$\sigma$.
\label{figset_sz69_DCN}}
\end{figure*}

\begin{figure*}
\centering
\resizebox{0.99\hsize}{!}{
    \includegraphics{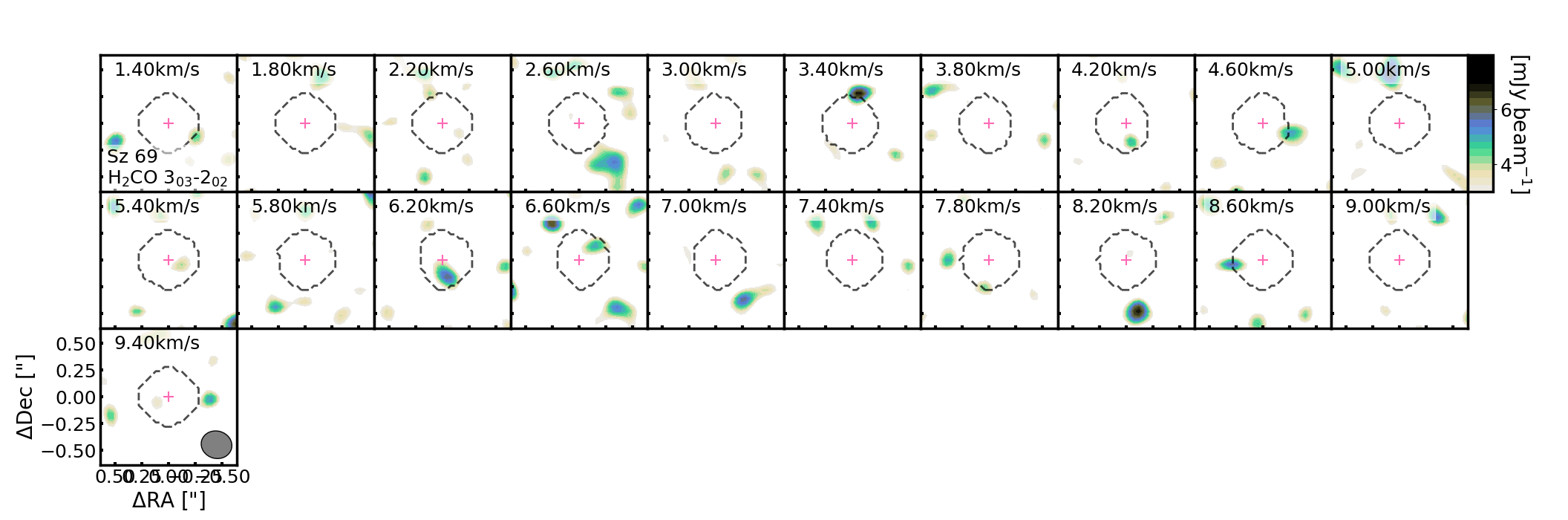}}
\caption{H$_2$CO 3--2 toward Sz 69 above 2$\sigma$.
\label{figset_sz69_H2CO}}
\end{figure*}

Figures~\ref{figset_sz69_12CO} through~\ref{figset_sz69_H2CO} display channel maps of detected/tentatively detected emission toward Sz 69.

\clearpage

\section{Correlation Coefficients}
\label{sec_appendix_linear}

Table~\ref{table_diskfluxfit} lists the Spearman correlation coefficients ($r_\mathrm{SCC}$) and their associated p-values for the log-log flux data presented in Figure~\ref{fig_fluxvsflux}.
The $r_\mathrm{SCC}$ value is a measure of how well the data can be described by a monotonic function.  The corresponding p-value is an estimate of statistical significance; i.e., a lower p-value indicates a lower probability that an uncorrelated system would have a correlation $\geq |r_\mathrm{SCC}|$.  Pearson correlation coefficients ($r_\mathrm{PCC}$) are included in Table~\ref{table_diskfluxfit} for completeness.  The $r_\mathrm{PCC}$ value is a measure of how well the data can be described by a linear function.  Note that the $r_\mathrm{PCC}$ value assumes the underlying datasets are Normally distributed.

Both correlation coefficients range from -1 to 1.  Data that is perfectly positively correlated will have an $r$ value of 1, while data that is perfectly negatively correlated will have an $r$ value of -1.  Completely uncorrelated data (i.e., pure scatter) will have an $r$ value of 0.  p-values range from 0 to 1.  Since our datasets are small, we treat $r_\mathrm{SCC}$ values as significant only if their corresponding p-values are $\leq$0.01.

\begin{deluxetable*}{lccccccccccccc}
\tablecaption{Correlation Coefficients for the Relative Disk Fluxes of M4-M5, Solar-Type, and Herbig Ae Disks. \label{table_diskfluxfit}}
\tablehead{
$\ln$($F_\mathrm{140pc} (x)$) & $\ln$($F_\mathrm{140pc} (y)$) & \multicolumn{4}{c}{M4-M5 Disks}                 & \multicolumn{4}{c}{Solar-Type + Herbig Ae Disks}                & \multicolumn{4}{c}{All Disks} \\
                         &                          & \# & $r_\mathrm{PCC}$         & $r_\mathrm{SCC}$     & p-val$_\mathrm{SCC}$      & \# & $r_\mathrm{PCC}$         & $r_\mathrm{SCC}$     & p-val$_\mathrm{SCC}$      & \# & $r_\mathrm{PCC}$         & $r_\mathrm{SCC}$     & p-val$_\mathrm{SCC}$   }%
 \colnumbers \startdata
\hline
C$^{18}$O (J=2--1) & C$_2$H (N=3--2) & 5           & 0.544     & 0.500     & 3.910e-01       & 10         & 0.739    & 0.600    & 6.669e-02      & 15          & 0.818     & 0.814     & 2.194e-04       \\
                   & DCN (J=3--2)    & 1           & ---       & ---       & ---             & 9          & -0.028   & -0.033   & 9.322e-01      & 10          & 0.339     & 0.248     & 4.888e-01       \\
                   & HCN (J=3--2)    & 5           & 0.637     & 0.500     & 3.910e-01       & 6          & 0.790    & 0.543    & 2.657e-01      & 11          & 0.822     & 0.882     & 3.302e-04       \\
                   & H$_2$CO         & 2           & ---       & ---       & ---             & 11         & 0.628    & 0.518    & 1.025e-01      & 13          & 0.756     & 0.687     & 9.509e-03       \\
HCN (J=3--2)       & C$_2$H (N=3--2) & 5           & 0.993     & 1.000     & 1.404e-24       & 7          & 0.803    & 0.750    & 5.218e-02      & 12          & 0.970     & 0.923     & 1.862e-05       \\
                   & DCN (J=3--2)    & 1           & ---       & ---       & ---             & 7          & 0.508    & 0.571    & 1.802e-01      & 8           & 0.530     & 0.643     & 8.556e-02       \\
                   & H$_2$CO         & 2           & ---       & ---       & ---             & 7          & 0.940    & 0.857    & 1.370e-02      & 9           & 0.917     & 0.917     & 5.066e-04       \\
C$_2$H (N=3--2)    & DCN (J=3--2)    & 1           & ---       & ---       & ---             & 10         & 0.565    & 0.588    & 7.388e-02      & 11          & 0.464     & 0.573     & 6.554e-02       \\
                   & H$_2$CO         & 2           & ---       & ---       & ---             & 11         & 0.755    & 0.555    & 7.665e-02      & 13          & 0.769     & 0.665     & 1.317e-02       \\
H$_2$CO            & DCN (J=3--2)    & 1           & ---       & ---       & ---             & 10         & 0.405    & 0.394    & 2.600e-01      & 11          & 0.509     & 0.509     & 1.097e-01 
\enddata
\tablecomments{The correlation coefficients describe the relationships between integrated fluxes, scaled to 140pc, for different molecular lines (columns 1 and 2) detected toward M4-M5, solar-type, and Herbig Ae disks, as depicted in Figure~\ref{fig_fluxvsflux} and described in Section~\ref{sec_results_relflux}.  Note that 1e+01 is shorthand for $1 \times 10^{+01}$.  All solar-type and Herbig Ae disk data was compiled from the ALMA detections of~\cite{cite_huangetal2017}, ~\cite{cite_bergneretal2019},~\cite{cite_bergneretal2020}, and~\cite{cite_peguesetal2020}.  All M4-M5 disk data is from this work.  Here we present the Spearman correlation coefficients $r_\mathrm{SCC}$ and the Pearson correlation coefficients $r_\mathrm{PCC}$.  We also present the p-values (p-val) for the $r_\mathrm{SCC}$ values as a measure of statistical significance.  Interpretations of the correlation coefficients and the p-values are described in the text of Appendix~\ref{sec_appendix_linear}.}
\end{deluxetable*}

\section{Investigation of Optical Depth as the Primary Contributor to Flux Correlations}
\label{sec_appendix_13CO}

\begin{figure*}
\centering
\resizebox{0.8\hsize}{!}{
    \includegraphics[trim=10pt 10.5pt 9.5pt 11.5pt, clip]{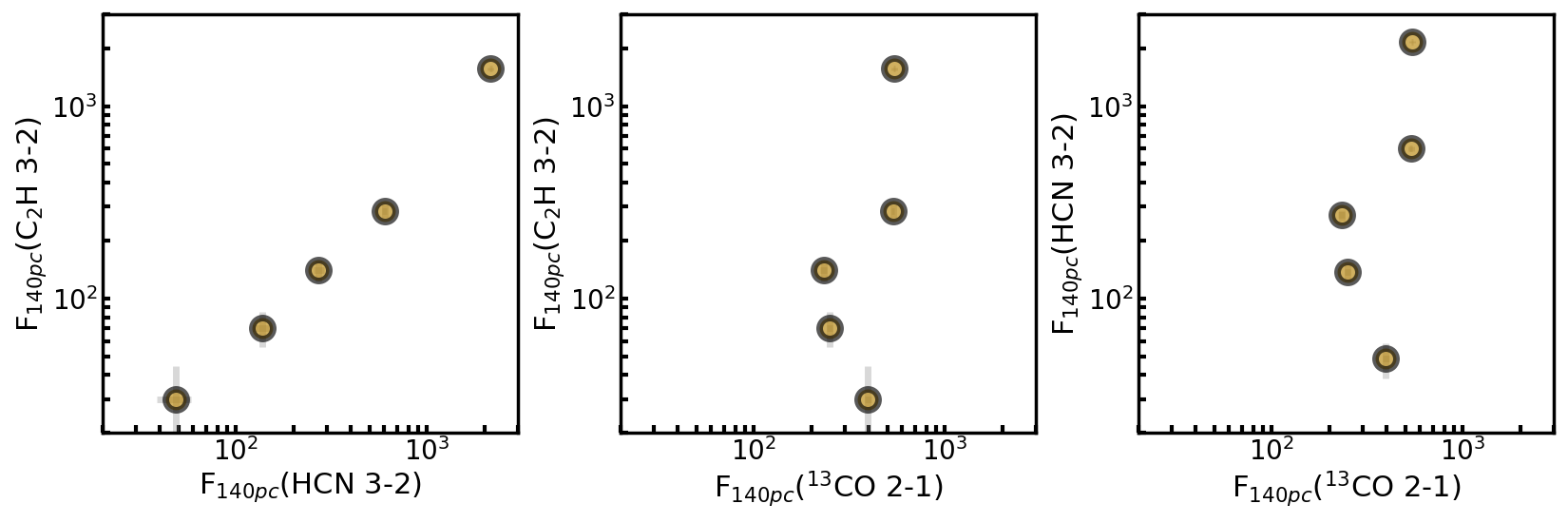}}
\caption{Comparisons of the total C$_2$H 3--2 and HCN 3--2 line fluxes to the total $^{13}$CO 2--1 line fluxes (all scaled to 140pc) for the M4-M5 disk sample.
\label{fig_13CO}}
\end{figure*}

The C$_2$H and HCN molecular line emission in this survey is still generally optically thick, which means that the emission we observe for these lines does not necessarily trace the underlying molecular abundances.  It is possible that the observed emission is instead purely tracing the size of the emission distribution, which would lead to a fixed dependence of emission on the distribution size.

We investigate this possibility here to determine if the correlations in Section~\ref{sec_results_relflux} are solely effects of optically thick emission.  We do so by comparing the C$_2$H 3--2 and HCN 3--2 line fluxes to $^{13}$CO 2--1 line fluxes for the M4-M5 disk sample.  The $^{13}$CO line is also expected to be optically thick.  A few ALMA studies have investigated the temperature of CO isotopologue emitting layers for solar-type disks.
\cite{cite_schwarzetal2016} derived an average $^{13}$CO gas temperature of 20-40K across the Tw Hya disk, with values between 20-25K beyond $\sim$30 AU.  \cite{cite_pinteetal2018} found the maximum brightness temperatures for $^{13}$CO in IM Lup to be $\sim$20K from ~140-300AU.  The TW Hya and IM Lup disks both host solar-type stars that are more massive than the host stars in our sample.  We expect that these $^{13}$CO emitting temperatures for solar-type disks are similar/upper limits to the $^{13}$CO emitting temperatures in our own M4-M5 disk sample.  We thus assume that the $^{13}$CO emitting layer is comparable to the C$_2$H and HCN emitting layers in the M4-M5 disks.

In the case where the C$_2$H 3--2, HCN 3--2, and $^{13}$CO 2--1 lines are uncoupled, we would expect to see the same level of correlation between the three pairs of lines.  Figure~\ref{fig_13CO} plots the total disk fluxes for all three lines against each other.  These fluxes are the same fluxes presented in Table~\ref{table_emflux} and are measured within the bounds of the Keplerian masks (Table~\ref{table_kepmask}).  We see again the clear correlation between the C$_2$H 3--2 and HCN 3--2 line fluxes.  However, we see that neither C$_2$H 3--2 nor HCN 3--2 show the same variation with $^{13}$CO 2--1.  This is evidence against a shared dependence of the relative line fluxes on optical depth.  We thus conclude that optically thick emission is not the primary reason for the C$_2$H 3--2 vs. HCN 3--2 correlation we find in Section~\ref{sec_results_relflux}.

\end{appendix}

\clearpage

\bibliography{projectbib}

\end{document}